\documentclass{aa}  

\usepackage{adjustbox}
\usepackage{threeparttable}
\usepackage{graphicx,url,twoopt,natbib}       
\usepackage{hyperref}   
\usepackage{longtable}
\usepackage{pifont}

\hypersetup{
  colorlinks=true,   
  urlcolor=blue,     
  linkcolor=blue,    
  citecolor=blue}

\usepackage{txfonts}
\usepackage{amsfonts}
\usepackage{caption}
\usepackage{subcaption} 
\usepackage{array}

\newcommand{\kms}{\mbox{km\,s$^{-1}$}}
\newcommand{\mum}{\mbox{$\mu$m}}
\newcommand{\mloss}{\mbox{$\dot{M}$}}

\newcommand{\msun}{\mbox{$M_{\odot}$}}
\newcommand{\my}{\mbox{$M_{\odot}$~yr$^{-1}$}}

\begin{document}

   \title{The dusty envelopes of asymptotic giant branch stars with ultraviolet excesses}

   \author{J. Alonso-Hernández \inst{1, 2} \and C. Sánchez Contreras \inst{1} \and R. Sahai \inst{3} \and J. Sanz-Forcada \inst{1}  }

   \institute{Centro de Astrobiología, CSIC-INTA, Camino bajo del Castillo s/n, 28692 Villanueva de la Cañada, Madrid, Spain \\ 
\email{jalonso@cab.inta-csic.es}  
\and Escuela de Doctorado UAM, Centro de Estudios de Posgrado, Universidad Autónoma de Madrid, E-28049 Madrid, Spain   \and Jet Propulsion Laboratory, California Institute of Technology, Pasadena, CA 91109, USA 
\\
}

\date{Received 20 February 2026 / Accepted 19 March 2026}

  \abstract
   {Roughly spherical envelopes around asymptotic giant branch (AGB) stars transform into the highly-asymmetric morphologies observed in planetary nebulae. The complex processes suffered in this metamorphosis are not yet completely understood. However, binarity emerges as a strong shaping factor, although the identification of binary companions in AGB stars is observationally challenging. The presence of ultraviolet (UV) excesses in AGB stars has been suggested as a potential indicator of binarity.}
   {In a first study, we characterised the properties of the gas component in the circumstellar envelopes surrounding a sample of 29 AGB stars with UV excesses. Now we intend to complement this information with an analysis of the dust component and compare the estimated parameters with those previously inferred from larger samples of AGB stars.}      
   {We modelled the spectral energy distributions of the sample using dust radiative transfer models. In some cases, we complemented the analysis with Herschel/PACS radial surface brightness profiles.} 
  {We derived mass-loss rates and gas-to-dust ratios, which are in the typical ranges for AGB stars. We found that the stellar and mass-loss parameters follow similar trends than those presented in the literature. There is an anticorrelation between the gas-to-dust ratio and the UV emission, although it is weaker than its correlations with pulsation and mass-loss. We also estimated the dust attenuation produced by the dust at UV wavelengths and describe its effects on the intrinsic UV emission.}
  {Stellar and mass-loss parameters of UV emitting AGB stars follow similar trends as found for larger samples of AGB stars. High-angular resolution observations are required to explore the dust forming regions and identify the presence of stellar companions. Circumstellar dust attenuation might play a dominant role in the observed UV emission, and needs to be accounted to estimate the intrinsic UV emission.}

   \keywords{Stars: AGB and post-AGB -- circumstellar matter -- mass-loss -- Ultraviolet: stars -- dust, extinction}

   \maketitle

\section{Introduction}\label{intro}

When stars with low and intermediate masses (i.e. with masses $\sim$1-8 \msun) leave the main-sequence, they evolve through the Hertzsprung–Russell diagram (HRD) increasing their luminosities while lowering their effective temperatures. They eventually reach the Asymptotic Giant Branch (AGB). At this stage their huge luminosities (from hundreds to thousands of solar luminosities) and intense stellar pulsation induce high mass-loss rates (\mloss$\sim$$10^{-8}$-$10^{-5}$ \my). The large amount of material transferred by these stars to the interstellar medium (ISM) causes AGB sources to become a major source of galactic chemical enrichment \citep[see][]{Tielens_2005, Ferrarotti_2006}.

During this process, the stellar material is expelled at low velocities ($\sim$5-30 \kms) and accumulates surrounding the stars, where the low temperatures allow the formation of molecular gas and dust in the vicinity of the star \citep[see][]{Hofner_2018}. This process leads to the formation of dense Circumstellar Envelopes (CSEs) that display (roughly) spherical shapes \cite[][]{Castro-Carrizo_2010}. In addition, AGB CSEs show a very rich molecular and dust chemistry \citep[for a recent overview see][]{agundez_2020}, which is mainly determined by the carbon-to-oxygen ratio (C/O). The chemistry in O-rich AGB stars (C/O$<$1) is dominated by O-bearing molecules (e.g. $\rm H_{2}O$, SiO) and related dust (silicates, oxides), whereas the chemistry in C-rich AGB stars (C/O$>$1) is dominated by C-bearing molecules (e.g. $\rm C_{2}H_{2}$, HCN) and related dust (carbon, carbides).

When these stars leave the AGB a dramatic transformation is observed in the shape of their CSEs, which might metamorphose into the large variety of shapes commonly found in Planetary Nebulae (PNe) and pre-PNe. They include highly elliptical, bipolar, or multipolar structures \citep{Sahai_1998,Ueta_2000,Sahai_2007,Sahai_2011a,Stanghellini_2016} that greatly contrast the shapes of AGB CSEs \citep[see][]{Balick_2002}. Moreover, during these post-AGB stages high-velocity outflows develop and CSEs become optically thinner allowing a high degree of ionisation by the central star \citep[see e.g.][]{Kwitter_2022}.

Even though it is still not well understood which processes lead to this morphological and kinematic transformation, binarity is expected to play a major role in shaping the CSEs \citep[see e.g.][]{Nordhaus_2006, De_Marco_2009}. Binarity should be common in AGB stars since it is found to be common on the main sequence \citep[see][]{Duquennoy_1991} as well as during the post-AGB phase\citep[see][]{Miszalski_2009}. However, the direct identification of stellar companions of AGB stars remains a major challenge due to their huge luminosity, the extinction produced by their CSEs, and the intrinsic variability (in both photometry and radial velocities) induced by stellar pulsations. 

Recently, high-angular resolution observations have identified the formation of binary-induced structures (e.g. spirals and rotating discs) on small scales in some AGB stars \citep[see e.g.][and references therein]{Decin_2020}. These structures, likely shaped by stellar companions, can serve as initial seeds for the development of the complex morphologies found in PNe.

On the other hand, \cite{Sahai_2008} discovered a subclass of AGB stars that show an intense ultraviolet (UV) emission (hereafter uvAGBs), orders of magnitude higher than the expected. These UV excesses can be explained by the presence of hot stellar companions and accretion processes \citep[see e.g.][]{Sahai_2008, Ortiz_2016}, although other phenomena like chromospheric activity cannot be ruled out \citep[see][]{Montez_2017}. Moreover, complementary observational efforts have found a large UV variability \citep{Sahai_2011b, Sahai_2016, Sahai_2018}, continuum-dominated UV spectra \citep{Ortiz_2019} and the detection of X-ray emission \citep{Ramstedt_2012, Sahai_2015, Ortiz_2021} in some of these stars. In these cases, the most likely explanation is the presence of interacting stellar companions and accretion processes.  

Recently, \cite{Sahai_2022} performed a series of modelling studies and proposed that the origin of UV excesses can be related with the ratio between FUV and NUV fluxes ($R_{\rm FUV/NUV}$). UV excesses with $R_{\rm FUV/NUV}$$\gtrsim$0.06 are likely produced by intense accretion, whereas $R_{\rm FUV/NUV}$$\lesssim$0.06 can be produced by chromospheric activity and/or less intense accretion processes.

The first characterisation of the CSEs of uvAGBs as a class was performed by \cite{alonso-hernandez_2024}, who studied their mass-loss properties based on CO rotational emission lines. One of their results was a statistically lower CO intensity against the 60 \mum\ flux in comparison with large samples of AGB stars, indicating lower amounts of molecular gas with respect to dust (i.e. lower gas-to-dust ratios).

In this paper we present the first characterisation of the dust component of the CSEs, based on the modelling of Spectral Energy Distributions (SEDs), in uvAGBs and explore the effects of dust opacity on the observed UV emission. In Sect.~\ref{obs} we describe our sample and observations. In Sect.~\ref{modelling} we describe the modelling procedure, the comparison with the observational data and the space of parameters explored. In Sect.~\ref{results} we describe the main results obtained in this study, which are later discussed in Sect.~\ref{dis}. Finally, we summarised our main conclusions in Sect.~\ref{summ}.

\section{Observational data sets}\label{obs}

In this study, we specifically focus on the sample of 29 UV emitting AGB stars presented in \cite{alonso-hernandez_2024}, which is based on a CO emission line survey of AGB stars with UV emission detected by the Galaxy Evolution Explorer \citep[GALEX,][]{GALEX}. The equatorial coordinates, distances and complementary information about the sources can be found in Table 1 of \cite{alonso-hernandez_2024}.

We based our analysis on three kinds of archival data, which are classified as follows: (i) photometric fluxes obtained from general catalogues and surveys, (ii) optical and infrared spectra, and (iii) {\it Herschel}/PACS images from which we extracted the photometric fluxes and radial surface brightness profiles.

\subsection{Photometry}

The SEDs were built using previously calibrated photometric fluxes that cover the optical to far infrared (FIR) range (0.5-200 \mum) with the most reliable fluxes using the VizieR Photometry viewer tool available at VizieR database \citep{vizier}. In the optical range, we selected photometry corresponding to the Johnson B and V, Gaia $\rm G_{BP}$, G and $\rm G_{RP}$ filters. In the infrared range, we selected the 2MASS J, H and K, Infrared Astronomical Satellite \citep[IRAS,][]{Neugebauer_1984} and AKARI \citep{Doi_2015} photometry (quality flag 2 or 3 for detections and 1 for upper limits respectively). The available WISE observations were not used due to saturation, as it is common in most galactic AGB stars \citep[see e.g.][]{Suh_2018}.

We noted the presence of near-field objects in the vicinity of some of our targets. However, in all the cases our AGB stars are significantly brighter than the surrounding objects and the effect of pollution was estimated, using the observations with highest angular resolution (e.g. Gaia in the optical and {\it Herschel}/PACS in the FIR, see also Sect.~\ref{PACS_obs}), to be around or less than 1\% of the fluxes (i.e. lower than total flux uncertainties).

When multi-epoch photometric values were available, we used the average fluxes. The flux uncertainties were estimated as the sum of their standard deviation and the average of their individual uncertainties. These uncertainties are dominated by the standard deviation, as they typically show a significant scattering related to the flux variability induced by the AGB pulsations.

\subsection{Spectra}

We gathered some large wavelength coverage spectra in order to obtain a more complete view of these objects. In particular, we mainly used Gaia DR3 XP \citep[see][]{Gaia_XP} as optical spectra (covering $\simeq$340-1020\,nm with R$\simeq$20-200) for 25 sources and IRAS Low Resolution Spectra (LRS) from the extended LRS atlas\footnote{Publicly available on \url{https://users.physics.unc.edu/~gcsloan/library/lrsatlas}} \citep[see][]{iras_atlas} as infrared spectra (covering $\simeq$7.7-22.7\,\mum\ with R$\simeq$20-60) for the whole sample.

Moreover, we used Infrared Space Observatory \citep[ISO,][]{Kessler_1996, Kessler_2003} Short Wavelength Spectrometer \citep[SWS;][]{Graauw_1996,Leech_2003} from the SWS atlas\footnote{Publicly available on \url{https://users.physics.unc.edu/~gcsloan/library/swsatlas}} \citep[see][]{sws_atlas} for 3 sources and Long Wavelength Spectrometer \citep[LWS;][]{Clegg_1996, Gry_2003} from the ISO archive for 1 source. We used the available ISO SWS (covering $\simeq$2.4-45.2\,\mum~with R$\simeq$1000-2000) and ISO LWS (covering $\simeq$43-197\,\mum~with R$\simeq$150-200) when available as infrared spectra instead of IRAS LRS due to their larger wavelength coverages and higher spectral resolutions.

\subsection{{\it Herschel}/PACS imaging}\label{PACS_obs}

We used archival observations performed with the Photodetector Array Camera and Spectrometer \citep[PACS;][]{Poglitsch_2010} on board {\it Herschel} Space Observatory \citep{Pilbratt_2010}, in particular photometric images in its three bands (``BLUE'' at 70 \mum, ``GREEN'' at 100 \mum, ``RED'' at 160 \mum) for seventeen sources. Four sources have available level 2.5 ``scan'' data, we used the images combined with the UNIMAP method \citep{Piazzo_2015} for VY\,UMa and with the JSCANAM method \citep{Gracia_carpio_2015} for V\,Eri, EY\,Hya and IN\,Hya. Moreover, thirteen sources have level 2.0 data, we combined the available ``scan'', as a first option, or ``chop-nod'', as a second option, images with standard {\it Herschel} Interactive Processing Environment \citep[HIPE,][]{HIPE} scripts. 

In first place, we checked whether the angular sizes of our targets are well-described by {\it Herschel}/PACS Point Spread Functions (PSFs). For this purpose, we estimated the PACS radial surface brightness profiles and compared them with those of the PSFs. We found this case in most of our targets (``point sources''), whereas in four sources (RU\,Her, SV\,Peg, T\,Dra, and V\,Eri) the angular size are slightly larger than the PSFs (``semi-extended sources'', see appendix~\ref{semi-extended}). Furthermore, VY\,UMa presents an extended detached shell with an extension of $\sim$35-70\arcsec\, \citep[previously reported by][]{Cox_2012, van_Marle_2014} that dominates the integrated FIR emission (see appendix~\ref{VY-UMA}).  

We performed open aperture photometry removing near-field sources in the {\it Herschel}/PACS image to estimate accurately their fluxes. For the ``point sources'', the photometric aperture radii were the standard values for point sources (12\arcsec, 20\arcsec\ and 25\arcsec\ in the BLUE and GREEN bands and 22\arcsec, 24\arcsec\ and 28\arcsec\ in the RED band), whereas for ``semi-extended sources'' we applied photometric aperture radii of 22\arcsec, 24\arcsec\ and 28\arcsec\ respectively.

In the case of VY\,UMa, we used aperture sizes of 12\arcsec\ for the BLUE and GREEN bands as well as 22\arcsec\ for the RED band, subtracted the surface brightness at the centre of the detached shell, and estimated the sky background in clear regions. We noted a large discrepancy with respect to the fluxes of IRAS and AKARI at similar wavelengths because their larger beams include, at least partially, the detached shell. We additionally estimated the emission of the detached shell, with aperture sizes of 70\arcsec\ in the three bands, and propose a model to fit the extended emission of this detached shell (see appendix~\ref{VY-UMA}).

The relative flux uncertainties from the individual and combined {\it Herschel}/PACS images can be lower than $1\%$. However, the flux variability between different epochs, even considering the scarcity of multi-epoch images and the sparse temporal coverage for most sources, is significantly larger (in some cases at least 5-20\%). Therefore, the instrumental uncertainties underestimate the intrinsic variability of AGB stars on the far-infrared (this also can apply to IRAS and AKARI photometry). The employed {\it Herschel}/PACS observations and their respective estimated fluxes are summarised in electronic format at the CDS. 

\section{Analysis: dust emission model}\label{modelling}

The circumstellar dust produces absorption, scattering, and thermal emission across the whole electromagnetic spectra, leading to changes in the overall shape of SEDs. Therefore, the comparison between the observed SEDs and those expected for the photospheres of AGB stars allows to characterise simultaneously the main physical parameters of both the stellar (e.g. the effective temperature, luminosity and roughly the C/O ratio) and dust (e.g. optical depth, dust density and temperature) components of the CSE. A broad wavelength coverage is particularly relevant to this analysis, although complementary infrared spectroscopy can be useful to constrain the dust composition by fitting the spectral features produced by the solid-state bands from the dust.

\subsection{Description of the DUSTY envelope models}\label{DUSTY}

We modelled the optical to far-infrared SED of our 29 targets with the 1D-radiative transfer code DUSTY \citep{Ivezic_1999}, which creates synthetic spectra based on the emission originated by a central source surrounded by spherical dust layers. The main six variables for DUSTY models are: (i) input stellar spectra, (ii) dust chemical composition, (iii) grain size distribution, (iv) temperature at the inner radius ($T_{\rm inn}$), (v) density distribution and (vi) optical depth.

We note that uvAGBs are binary candidates in which it is expected the presence of accreting stellar companions, which should be located inside the dust formation zone, that can produce shaping mechanisms, mass transfer and an anisotropic radiation field. These phenomena may lead to significant deviations from spherical symmetry in their CSEs, with a more pronounced effect in the innermost regions. Nevertheless, in the absence of spatially-resolved observations, we consider spherical symmetry to be a good approximation of the bulk of the CSE. We discuss possible effects of this approximation in Sect.~\ref{dis}.

The stellar component (input spectra) used as an input for DUSTY was included with the COMARCS stellar atmosphere library\footnote{Publicly available on \url{http://stev.oapd.inaf.it/atm/index.html}} (see \cite{Aringer_2016} for O-rich stars and \cite{Aringer_2019} for C-rich stars). These stellar models mainly depend on the stellar effective temperature ($T_{*}$) and the ratio between carbon and oxygen atomic abundances. 

Given the lack of accurate measurements, we selected COMARCS models with standard abundances of C/O$\simeq$0.55 for oxygen-rich stars and C/O$\simeq$1.40 for carbon-rich stars and solar metallicity \citep[e.g.][]{agundez_2020}. We remark that a variation of the C/O ratio only affects significantly the stellar spectra, in terms of SED modelling, at values near C/O$\simeq$1.00 due to the drastic change from O-rich to C-rich chemistry \citep[see][]{Aringer_2016, Aringer_2019}.

The dust composition is a key parameter that mainly affects the shape of the SED in the infrared, producing different and characteristic broad spectral features and modifying the shape of the continuum emission. For the O-rich AGB stars, we assumed a mixture of silicates \citep[from][]{Draine_1984}, alumina \citep[Al$_{2}$O$_{3}$, from ][]{Begemann_1997} and iron oxide \citep[FeO, from ][]{Henning_1995}. Whereas, for the C-rich AGB stars we assumed a mixture of amorphous carbon \citep[from][]{Hanner_1988}, silicon carbide \citep[SiC, from ][]{Pegourie_1988} and graphite \citep[from][]{Draine_1984}.

We performed a basic characterisation of the dust composition only for the three targets with available ISO SWS spectra. For this purpose, we performed an iterative process varying the relative dust abundances in 5\% steps until we found solutions that fitted the overall SEDs (see Sect.~\ref{SEDs}) and the infrared spectral features. We obtain the following best-fit grain compositions: RW\,Boo (60\% silicates, 30\% ${\rm Al}_{2}$O$_{3}$ and 10\% FeO), SV\,Peg (50\% silicates, 30\% ${\rm Al}_{2}$O$_{3}$ and 20\% FeO) and T\,Dra (80\% amorphous carbon, 15\% SiC and 5\% graphite), which reproduce the spectral features associated to the assumed dust compounds, although, we noted the presence of other small features.

We assumed similar compositions for the rest of targets based on their chemical type (for O-rich 50\% silicates, 40\% Al$_{2}$O$_{3}$ and 10\% FeO and for C-rich 80\% amorphous carbon, 5\% graphite and 15\% SiC). These compositions are in agreement with those expected according to the chemical type and mass-loss rates of our targets \citep[see e.g.][]{Waters_2011}. However, we acknowledge that dust composition is a major source of uncertainty due to the presence of spectral features with unknown carriers \citep[see e.g.][]{Sloan_2003}, featureless dust components \citep[e.g. metallic iron, see][]{Kemper_2002}, and large degeneracies between the different spectral features and with the overall shape of the SED \citep[see e.g.][]{Speck_2000, Jones_2014}.

On the other hand, the infrared region of the SEDs is also quite dependent of the grain size distribution \citep[see e.g.][]{Ysard_2018}, although it is very complicated to constrain empirically. In this study we used the standard Mathis-Rumpl-Nordsieck (MRN) distribution \citep{Mathis_1977}, which is the most common distribution in circumstellar and interstellar dust studies \citep[see references in][]{Mathis_1977}, described as:

\begin{equation}
    n(a) \propto a^{\rm -q} \hspace{1cm} \mathrm{for} \hspace{0.2cm} a_{\rm min}<a<a_{\rm max},
\end{equation}

\noindent where q$=$3.5, $\rm a_{min}$=0.005 \mum, and $\rm a_{max}$=0.25 \mum \, are respectively the exponent, the minimum and the maximum. These are the standard MNR values of the dust grain size distribution.

The temperature at the inner radius of the envelope indicates the maximum temperature of the dust  affecting, therefore, its thermal emission. This parameter is completely correlated with the inner radius of the envelope ($R_{\rm inn}$), which is also a DUSTY output, through the temperature distribution \citep{Ivezic_1999}.

The density distribution is assumed to be a power-law:

\begin{equation}
    \rho \propto \frac{1}{r^{n}} \hspace{1cm} \mathrm{for} \hspace{0.2cm} R_{\rm inn}<r<R_{\rm out},
\end{equation}

\noindent where $R_{\rm out}$ is the outer radius, and $n$, the index of the power-law, is a free parameter. In the case of a spherical wind expanding at a constant velocity, the density distribution is described by an index $n$=2 \citep[see][]{Ivezic_1995}. The ratio $Y$=$R_{\rm out}/R_{\rm inn}$ indicates the extension of the envelope.

Finally, the radial optical depth, $\tau_{550}$, at a fiducial wavelength of 550 nm, is used to parametrise the optical depth of the CSE across the entire wavelength range covered in the model. We used the optical depths at the central wavelengths of the two GALEX bands in order to estimate the effect of dust attenuation in the observed UV emission (see appendix~\ref{UV_opacities}).

\begin{figure*}[ h!]
     \centering
     \begin{subfigure}[b]{0.245\linewidth}
         \centering
         \includegraphics[width=\linewidth]{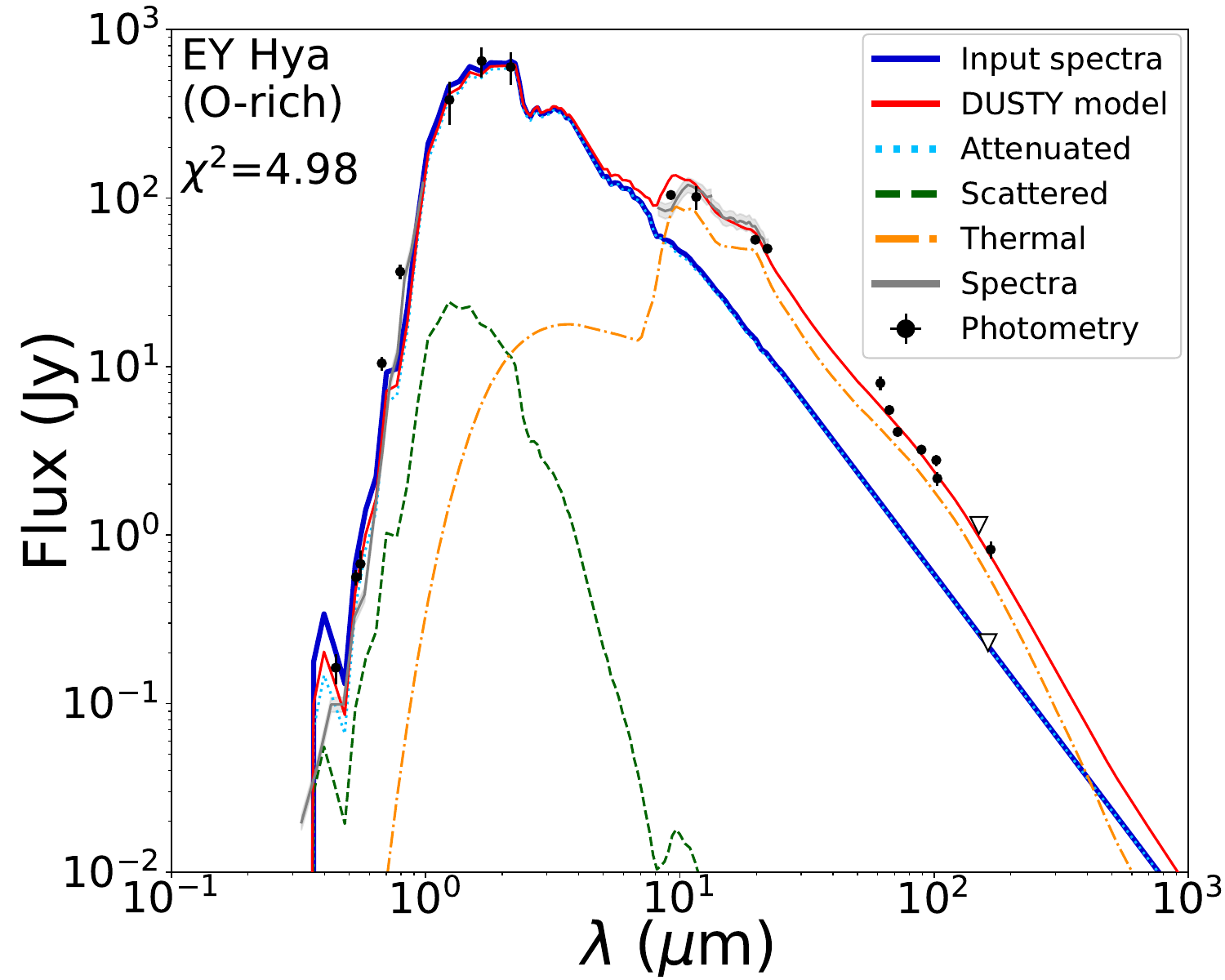}
     \end{subfigure}
     \begin{subfigure}[b]{0.245\linewidth}
         \centering
         \includegraphics[width=\linewidth]{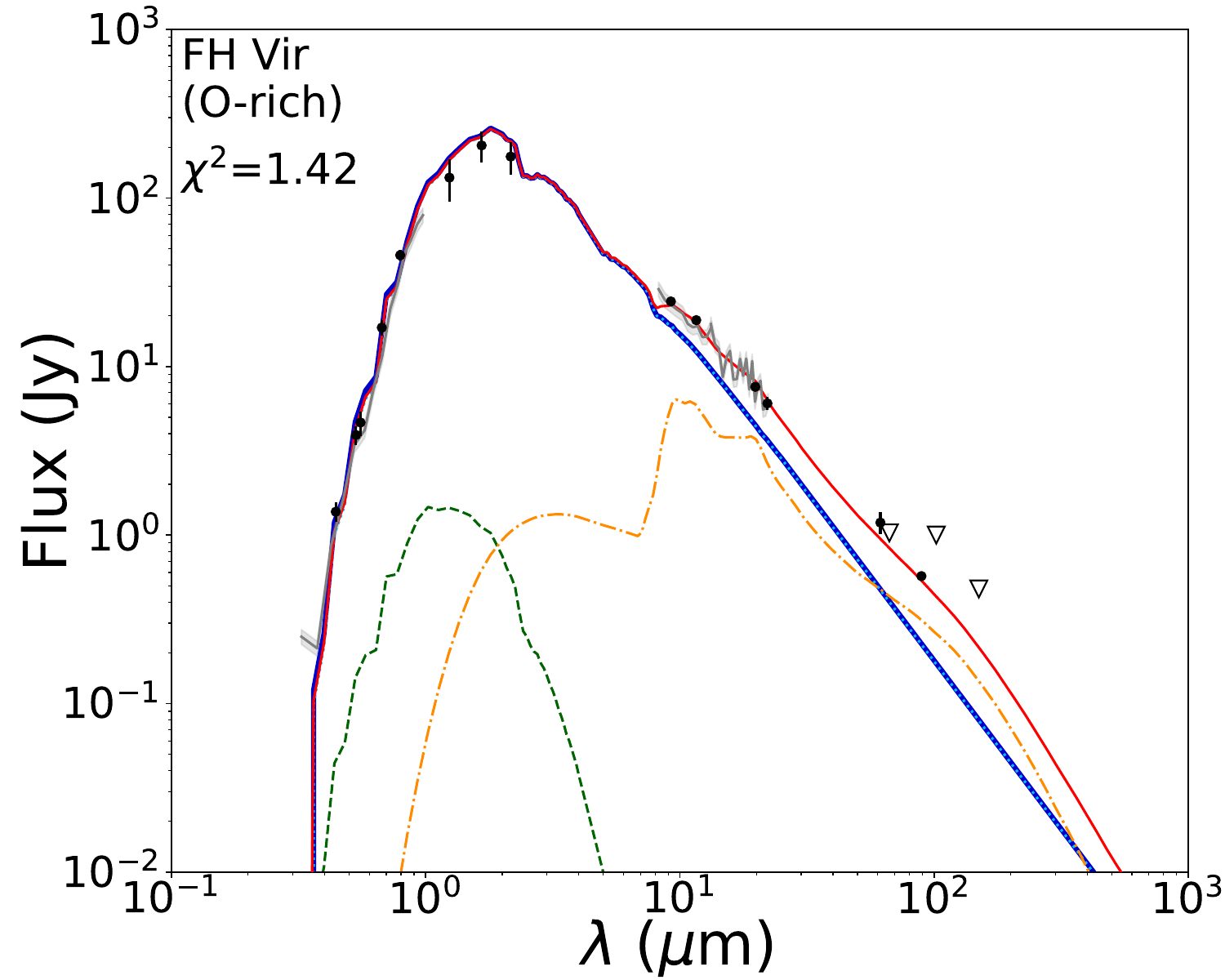}
     \end{subfigure}
     \begin{subfigure}[b]{0.245\linewidth}
         \centering
         \includegraphics[width=\linewidth]{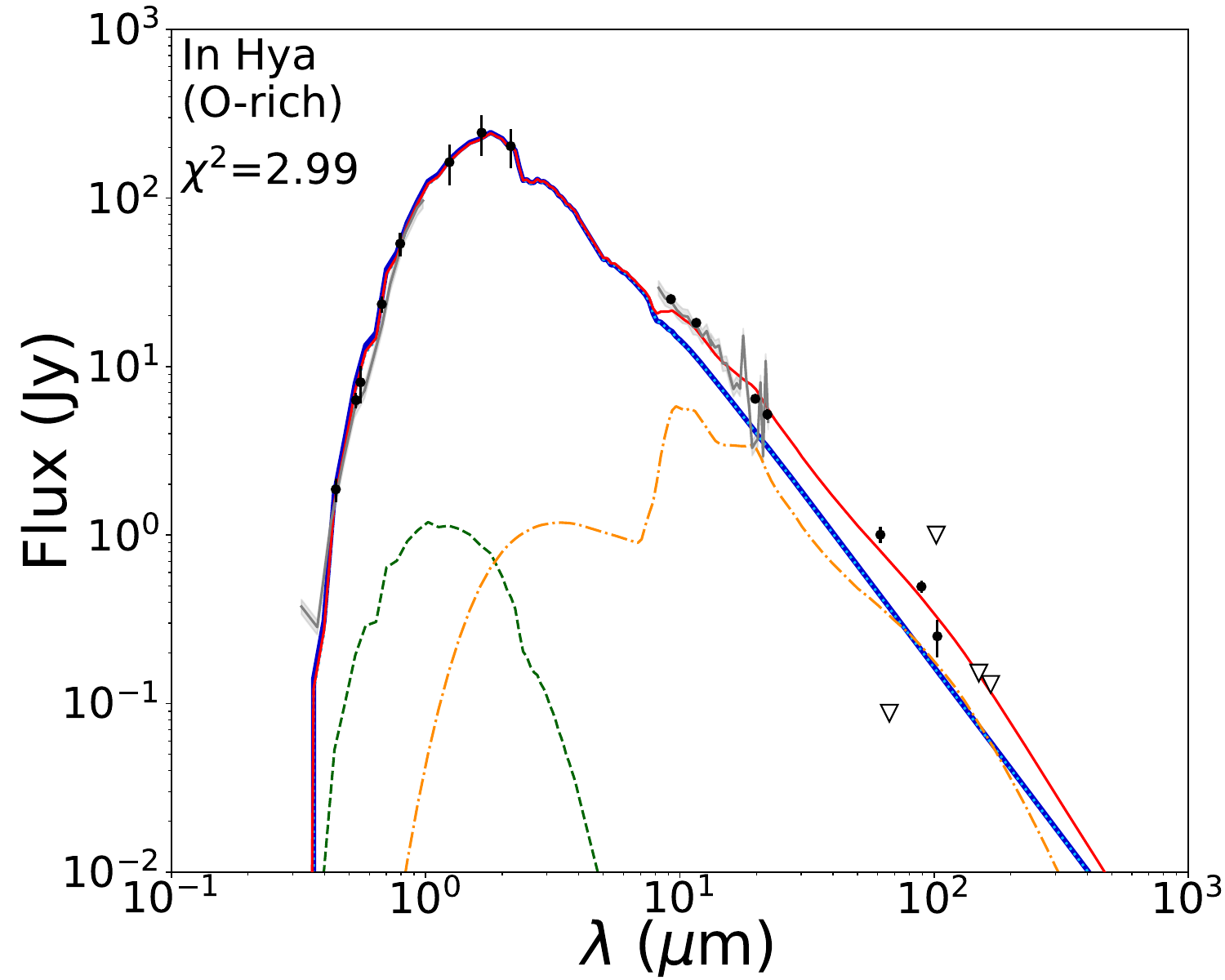}
     \end{subfigure}
     \begin{subfigure}[b]{0.245\linewidth}
         \centering
         \includegraphics[width=\linewidth]{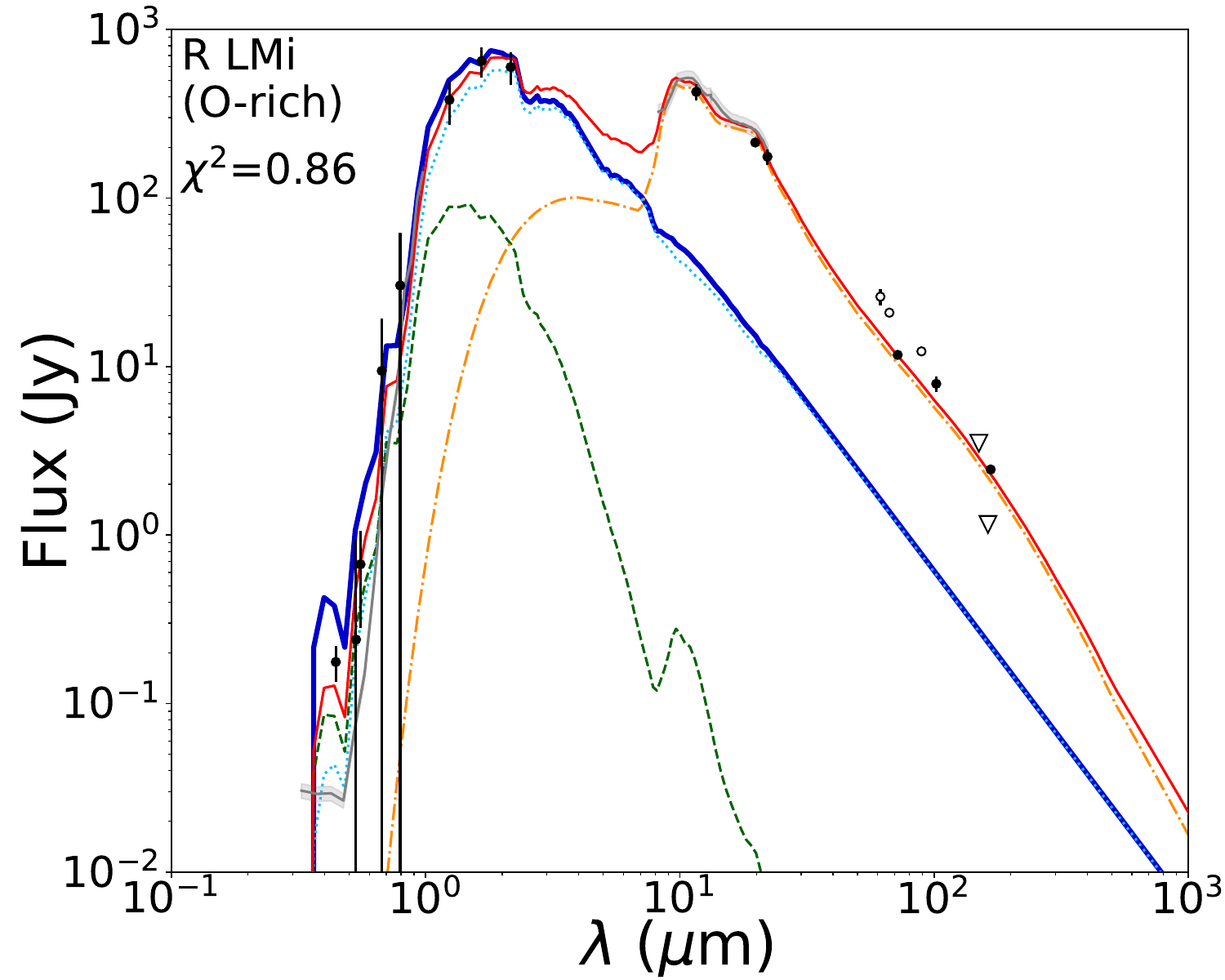}
     \end{subfigure}

     \begin{subfigure}[b]{0.245\linewidth}
         \centering
         \includegraphics[width=\linewidth]{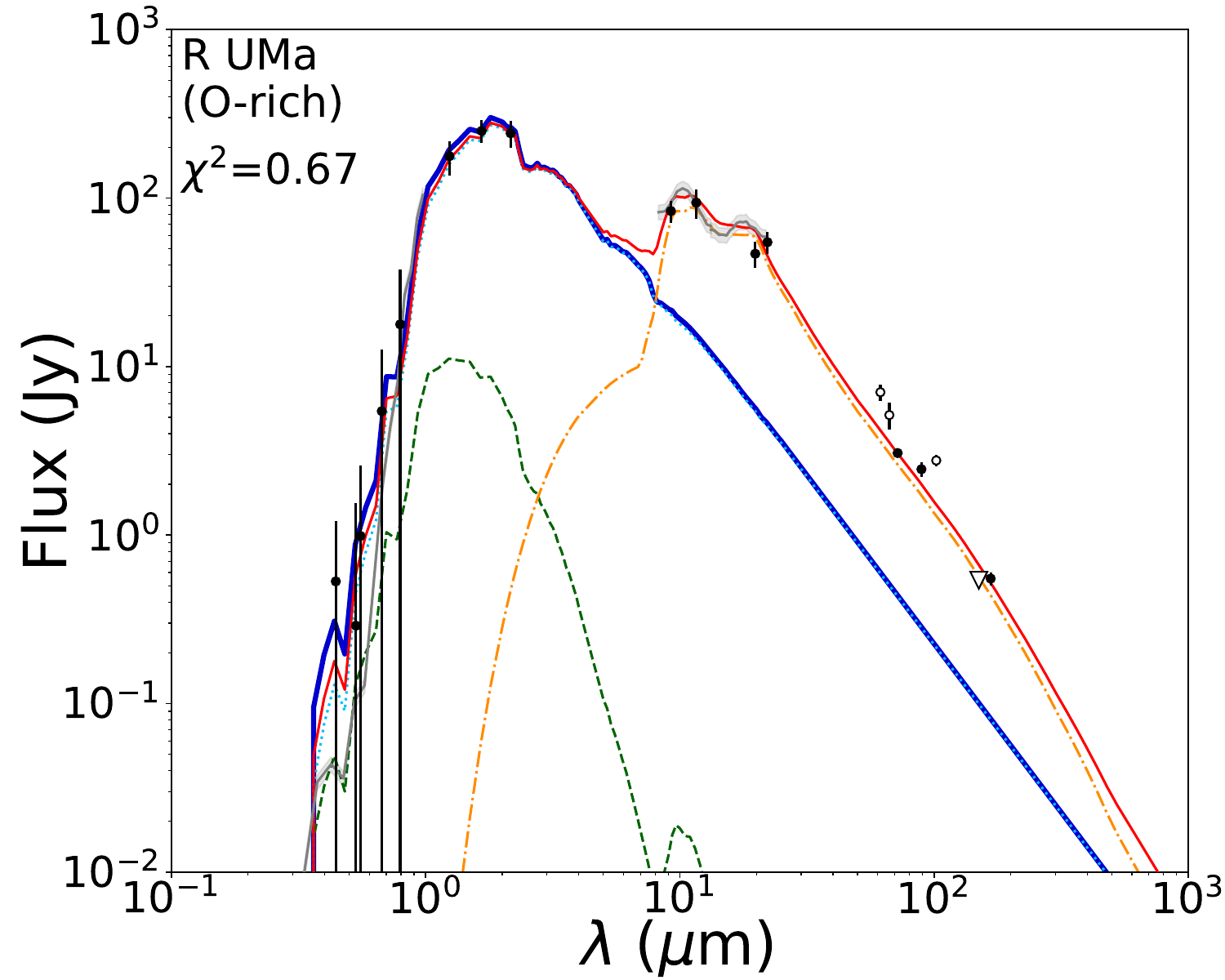}
     \end{subfigure}
     \begin{subfigure}[b]{0.245\linewidth}
         \centering
         \includegraphics[width=\linewidth]{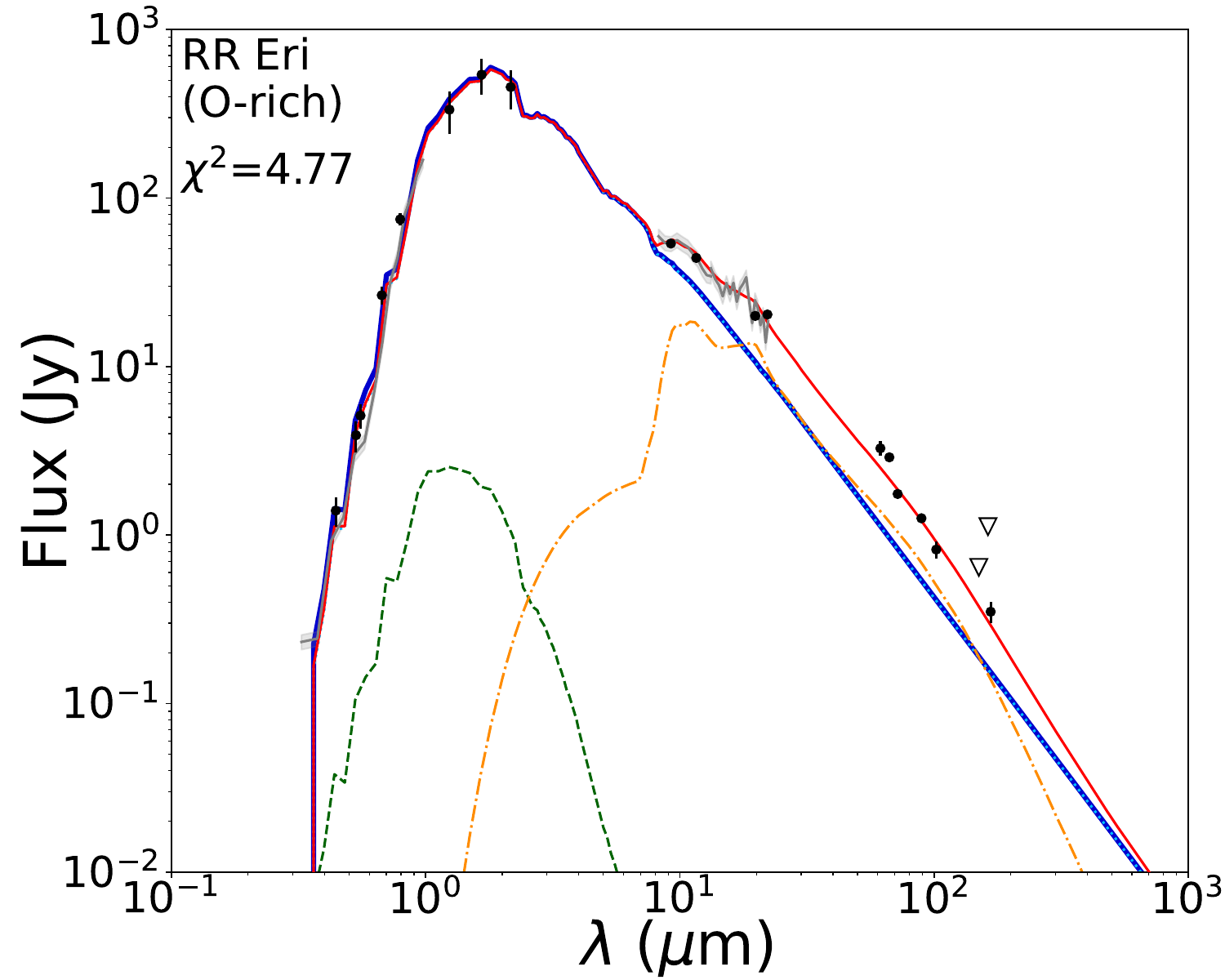}
     \end{subfigure}
     \begin{subfigure}[b]{0.245\linewidth}
         \centering
         \includegraphics[width=\linewidth]{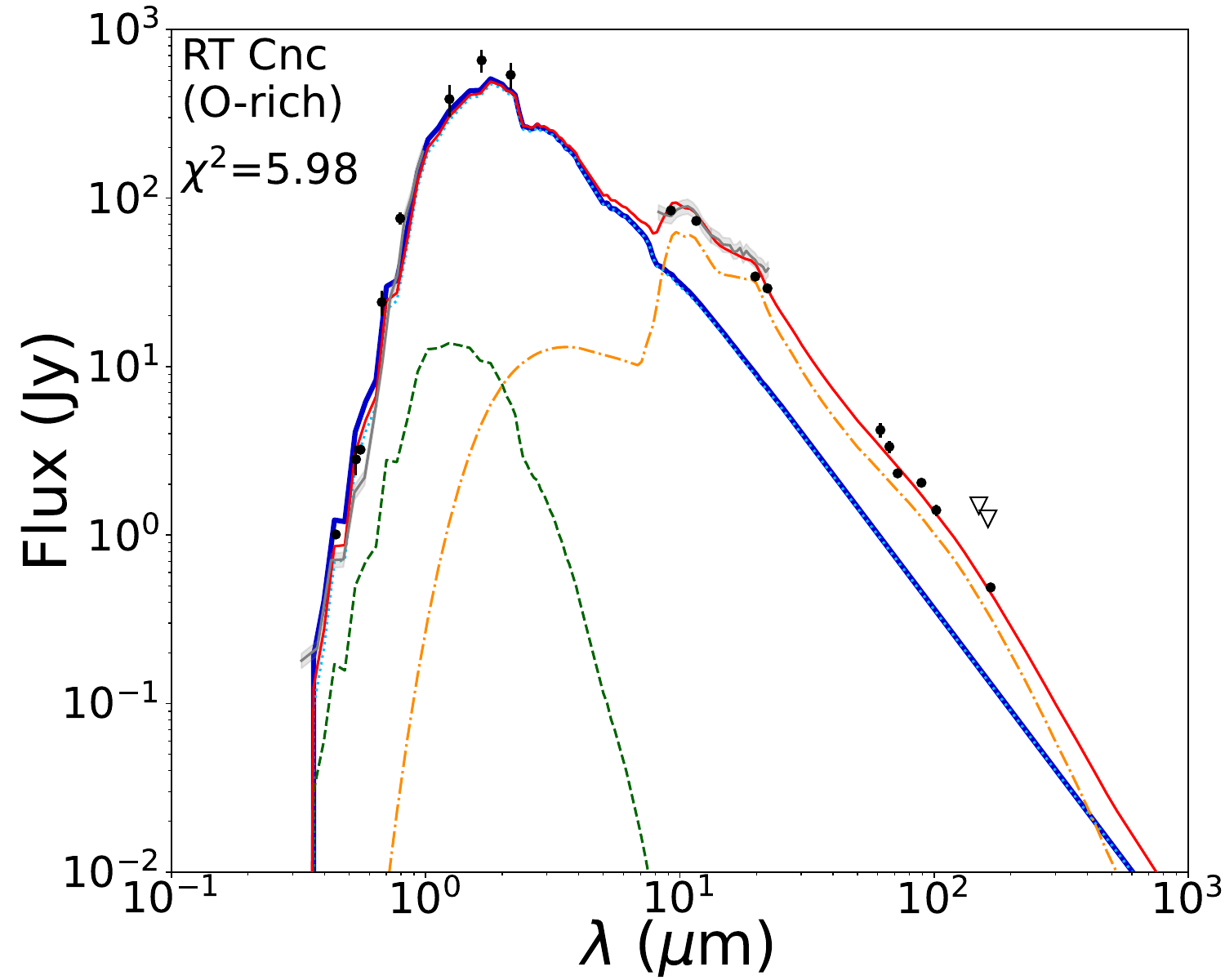}
     \end{subfigure}
     \begin{subfigure}[b]{0.245\linewidth}
         \centering
         \includegraphics[width=\linewidth]{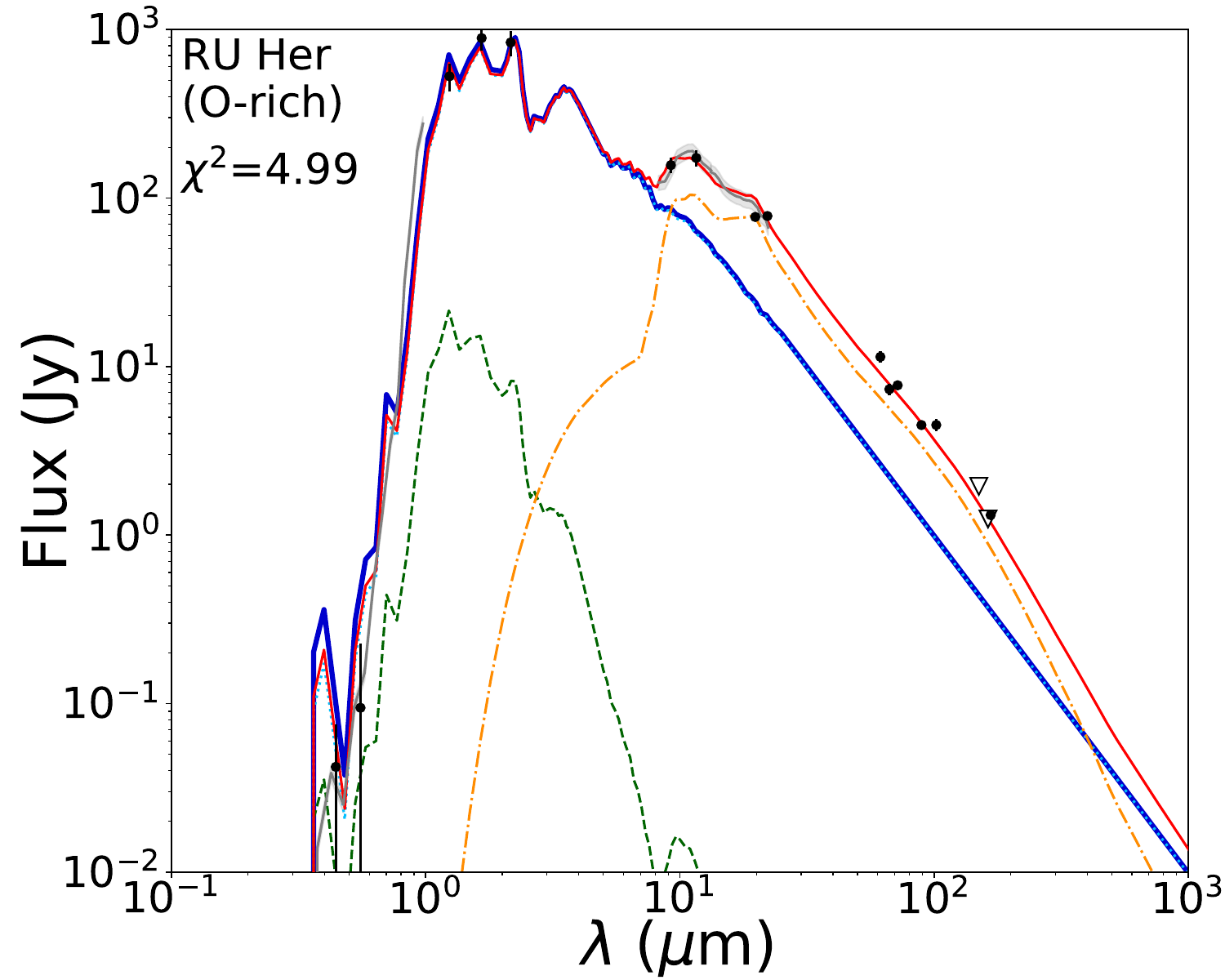}
     \end{subfigure}
        \caption{SEDs and best-fit DUSTY models for the dusty AGB stars. Solid blue line: input spectra, solid red line: total DUSTY model, dotted cyan line: Attenuated input spectra component, dashed green line: dust scattered component, dot-dashed orange line: dust thermal component. Photometric fluxes are indicated as filled circles (in a few cases the variability-induced uncertainties are larger than the average value), photometric data not included in the analysis as empty circles, and upper limits as empty triangles. Grey solid lines are GAIA, IRAS, and ISO spectra, shadowed areas indicate 10\% uncertainties.}
        \label{SED_fitting}
\end{figure*}

\subsection{Parameter space and SED fitting}\label{SEDs}

We performed a systematic search for the best-fit models with five free parameters: $T_{*}$, $T_{\rm inn}$,  $n$, $\tau_{550}$, and $Y$. To determine the best-fit parameters of the CSEs of the fit we perform a reduced $\chi^{2}$ minimisation, with $\chi^{2}$ described as:

\begin{equation}
    \chi^{2}= \frac{1}{N-p-1} \sum_{j}^{n}\frac{(O_{j}-M_{j})^{2}}{\sigma_{j}^{2}} ,
    \label{chi2}
\end{equation}

\noindent where $O_{j}$ is the photometric flux of each observation, $\sigma_{j}$ is its associated flux uncertainty, $M_{j}$ is the synthetic photometry estimated from the model, N is the number of observations and $p$ is the number of free parameters (in our case 5), and $N-p-1$ is the number of degrees of freedom of the fitting. Spectral observations were not used to estimate the $\chi^{2}_{\rm red}$ because they contain gas spectral lines that were not included in our SED modelling.

We estimated the synthetic photometry in each filter convolving the synthetic spectra produced by DUSTY with their respective transmission curves, which are provided by the SVO Filter Profile Service ``Carlos Rodrigo'' \citep{Rodrigo_2012, Rodrigo_2020, Rodrigo_2024}. In addition, we applied ISM extinction correction to the SEDs using the extinction law presented in \cite{Gordon_2023} and the E(B-V) values from \cite{alonso-hernandez_2024}. Finally, we scaled all the synthetic photometry by the coefficient that minimised the $\chi^{2}$.

Particular fluxes with very low relative uncertainties  might induce model overfitting in certain spectral regions, leading to wrong best-fit solutions. As mentioned in Sect.~\ref{obs}, far infrared single-epoch observations usually have very low flux uncertainties. However, a large dispersion of the photometric points due to the stellar pulsations is expected \citep[see e.g.][]{Onaka_1999, Groenewegen_2022, Tachibana_2023}, and also at larger wavelengths \citep[see e.g.][]{Jenness_2002, Dehaes_2007}.

Our simple study on the multi-epoch {\it Herschel}/PACS observations (see Sect.~\ref{PACS_obs}), indicates that variability-induced uncertainties should be at least around 5\%. Therefore, we impose a minimum of 5\% uncertainties in all photometric fluxes:

\begin{equation}
    \sigma_{j} = \mathrm{max}\{e_{j}, 0.05 \times O_{j} \},
\end{equation}

\noindent where $e_{j}$ is the uncertainty estimated as the sum of the instrumental error and the standard deviation when multi-epoch measurements are available. This conservative limit prevents from model over-fitting and provides reasonable best-fit solutions.

Moreover, we noted that a few sources (R\,LMi, R\,UMa and Z\,Cnc) show a large difference between the fluxes obtained from {\it Herschel}/PACS and those from IRAS and AKARI (see Fig.~\ref{SED_fitting} and appendix~\ref{SED_appendix}). We noted that, in these cases, IRAS and AKARI fluxes are systematically higher, indicating pollution from an irregular sky brightness due to their more limited angular resolution. This large difference in a short wavelength range produces a large increase of $\chi^{2}$ and complicates the overall reliability of the fitting. Therefore, we did not include these fluxes in the analysis. 

To explore the parameter space, we employed a simple fitting procedure, searching for an initial guest solution by visual inspection. Once we arrived to a good solution, we performed a numerical fitting with a grid of the five free parameters and chose the solution with the lowest $\chi^{2}$ as the best-fit solution.

We used linear spaces for $T_{*}$, $T_{\rm inn}$, n and $\tau$, and a logarithmic space for Y. The accuracy and ranges of the $T_{*}$ were limited by the availability of the stellar photospheric models, whereas for the rest of parameters they were determined considering the steepness of the $\chi^{2}$ distribution in each case. A discussion about the correlations between the different parameters and their associated uncertainties is provided in Sect.~\ref{fit_results}.

\section{Results}\label{results}

We present the main results from our characterisation of the dust component of the CSEs of our sample. We describe the best-fit models obtained from the SED modelling, derive additional parameters, and explore correlations between the parameters.

\subsection{Best-fit solutions and uncertainties}\label{fit_results}

\begin{figure*}[ h!]
     \centering
     \begin{subfigure}[b]{0.245\linewidth}
         \centering
         \includegraphics[width=\linewidth]{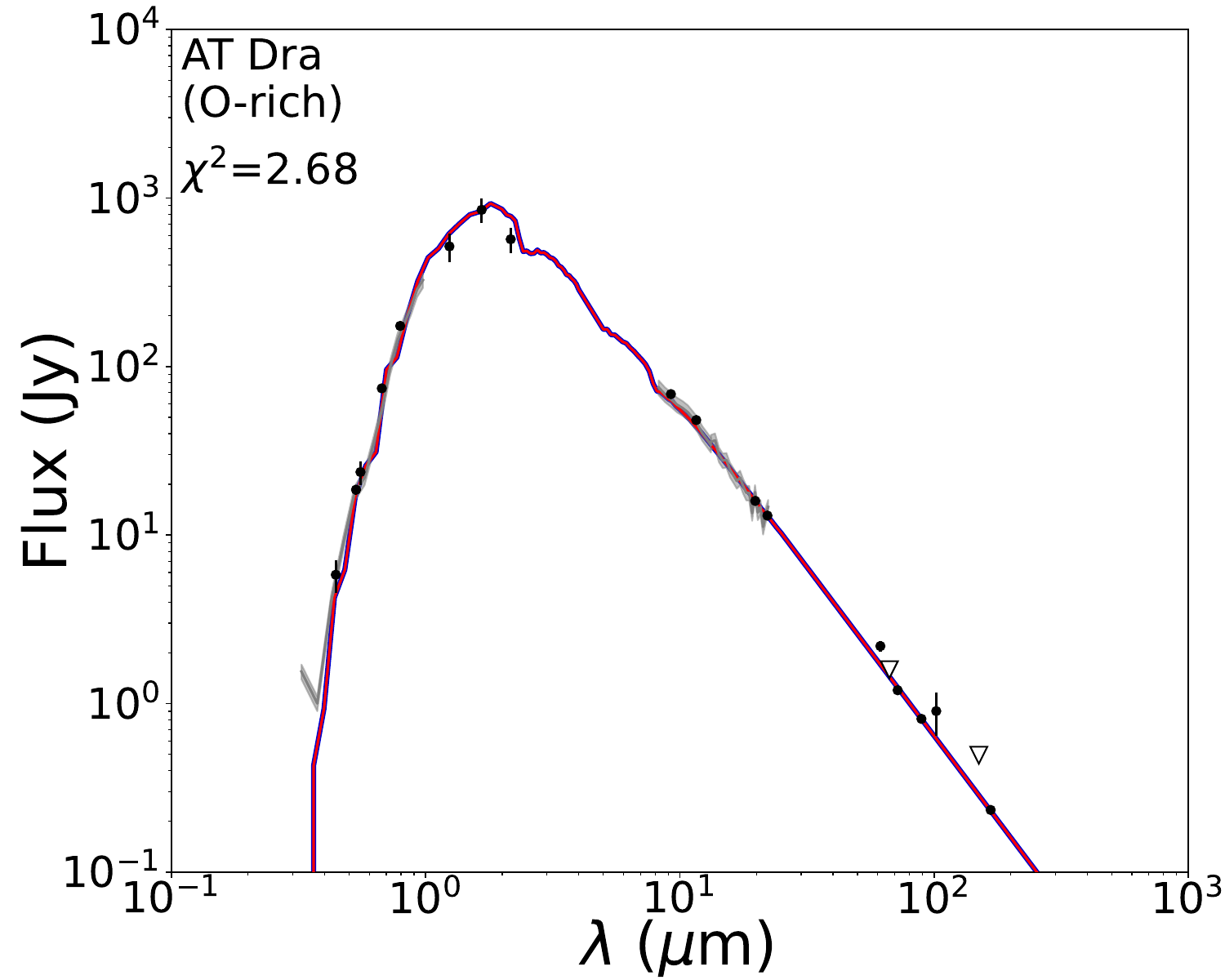}
     \end{subfigure}
     \begin{subfigure}[b]{0.245\linewidth}
         \centering
         \includegraphics[width=\linewidth]{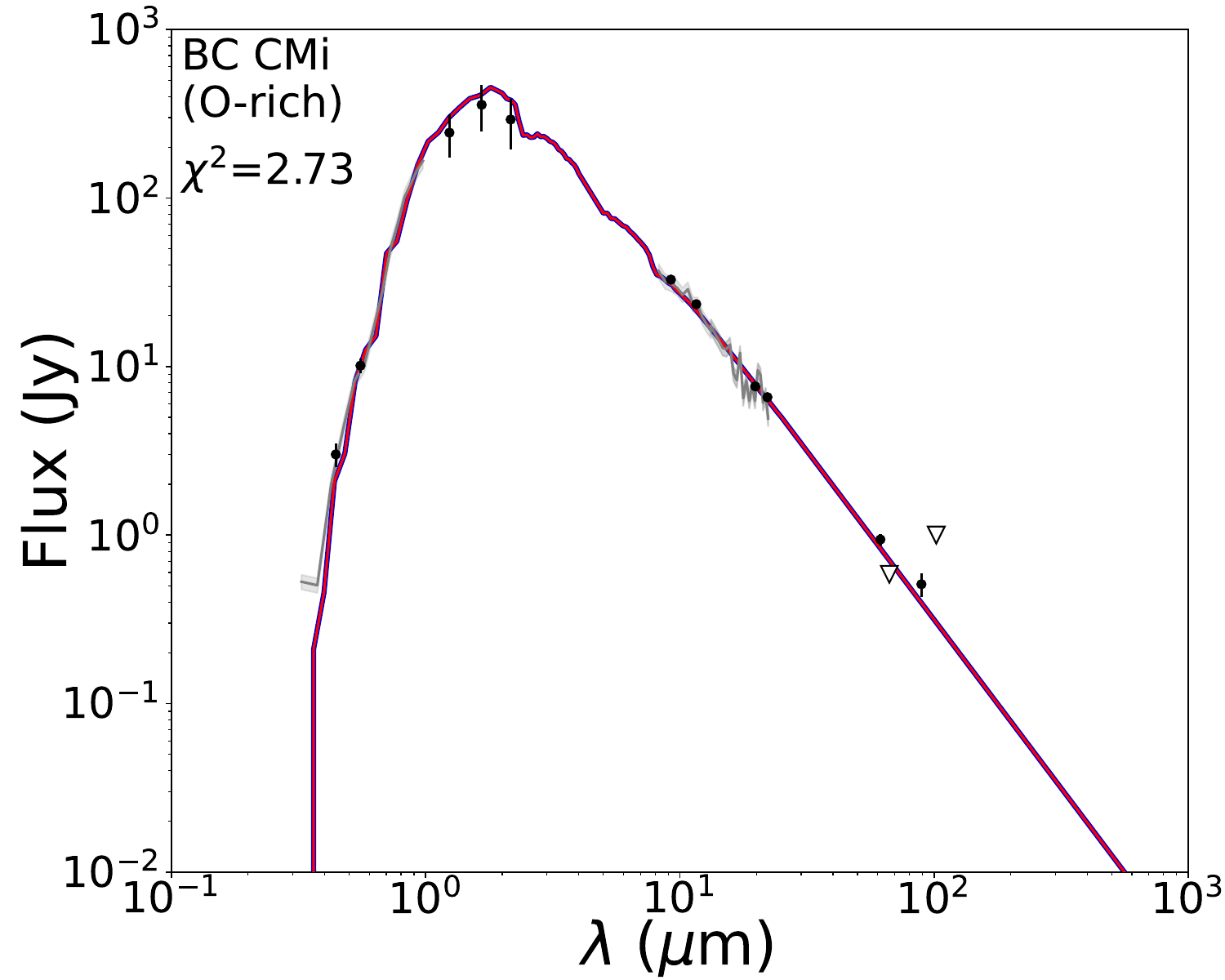}
     \end{subfigure}
     \begin{subfigure}[b]{0.245\linewidth}
         \centering
         \includegraphics[width=\linewidth]{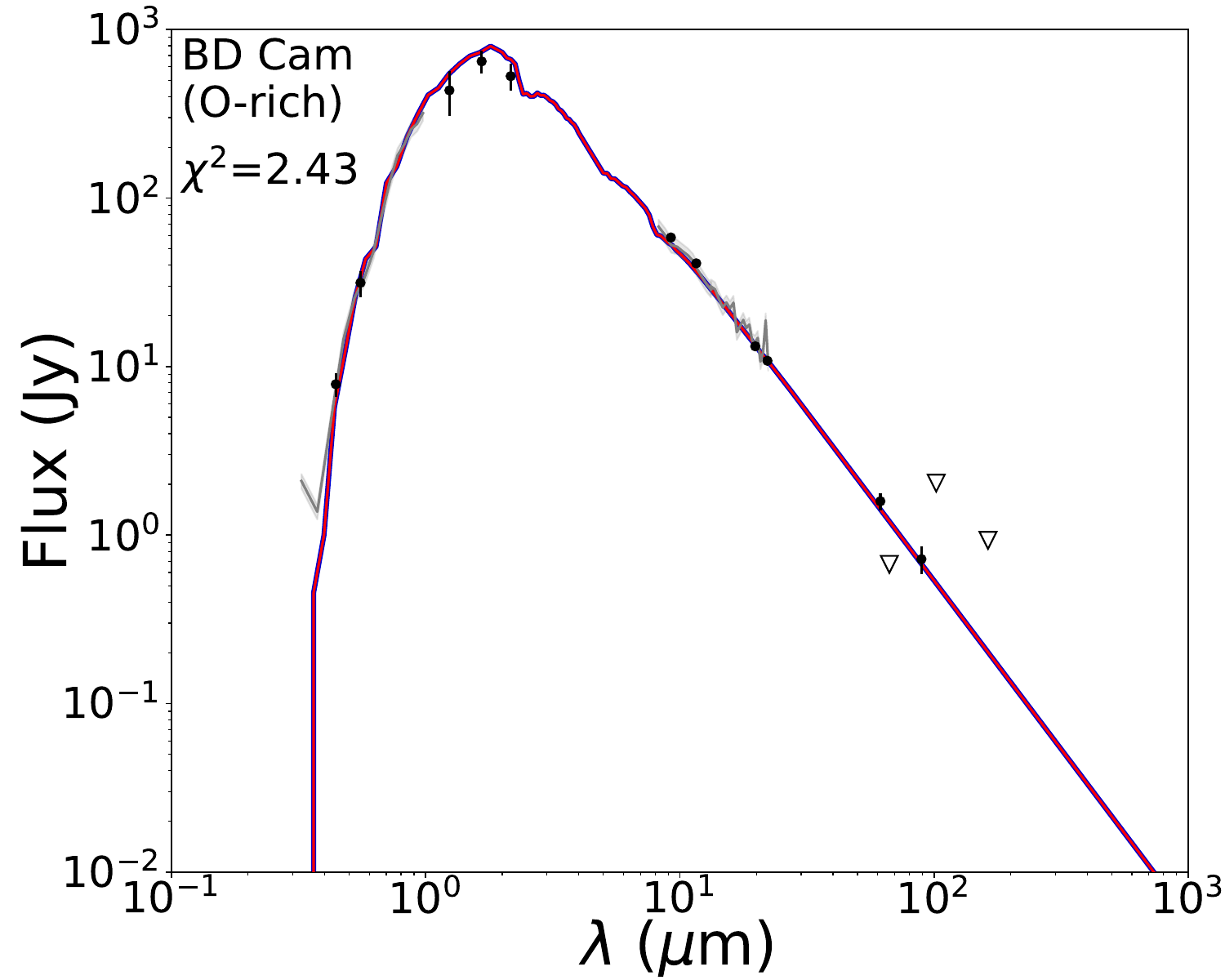}
     \end{subfigure}
     \begin{subfigure}[b]{0.245\linewidth}
         \centering
         \includegraphics[width=\linewidth]{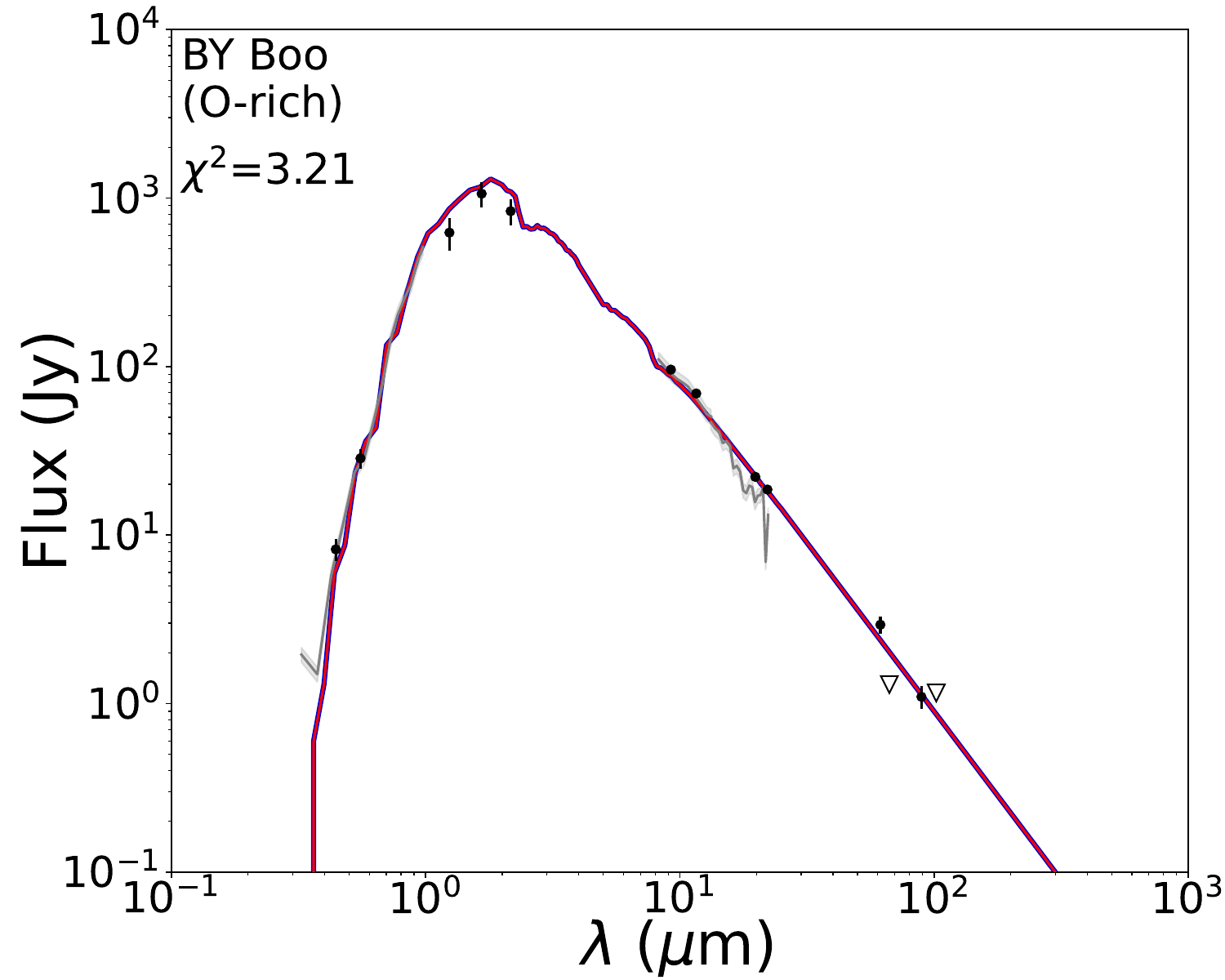}
     \end{subfigure}
     
     \begin{subfigure}[b]{0.245\linewidth}
         \centering
         \includegraphics[width=\linewidth]{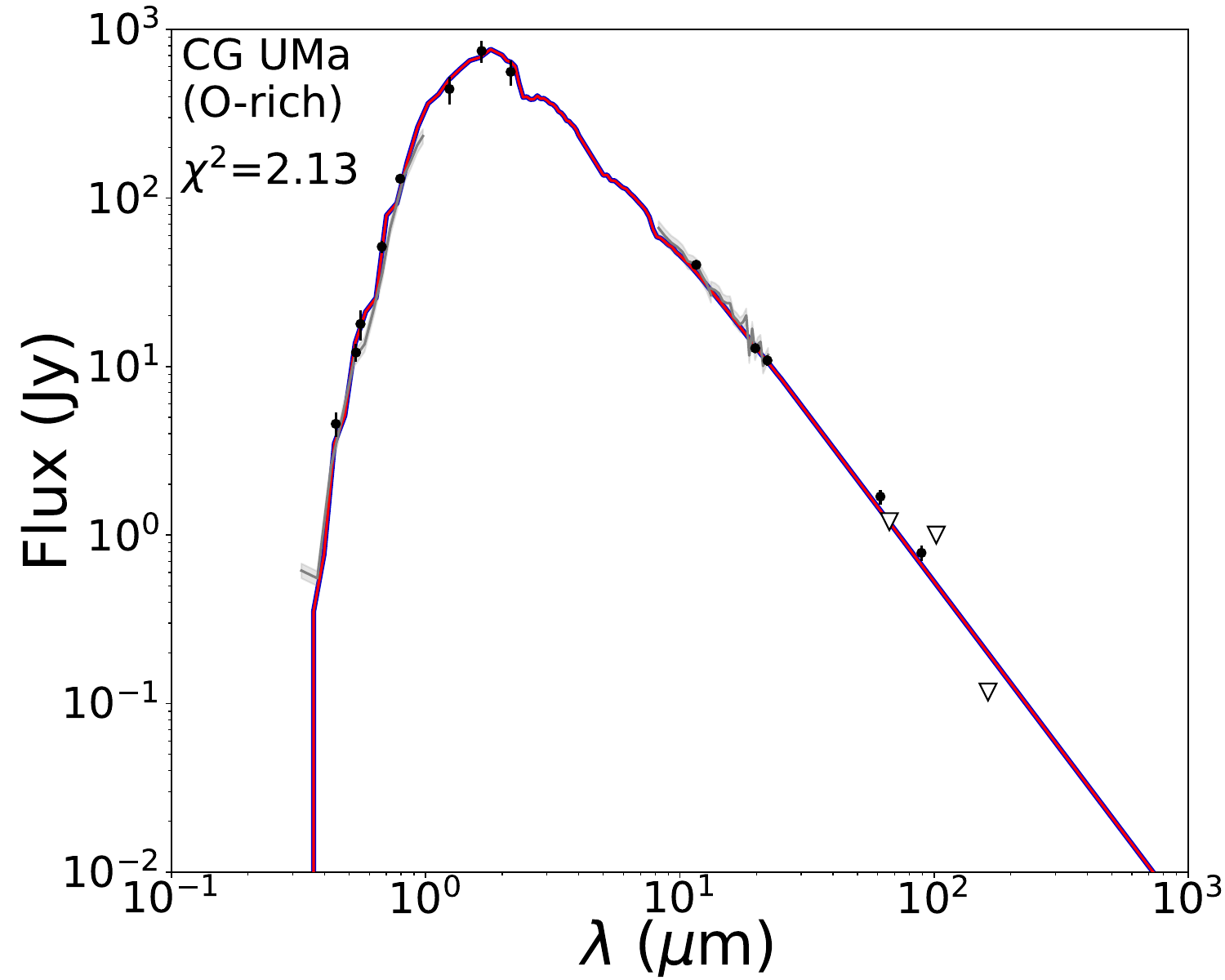}
     \end{subfigure}
     \begin{subfigure}[b]{0.245\linewidth}
         \centering
         \includegraphics[width=\linewidth]{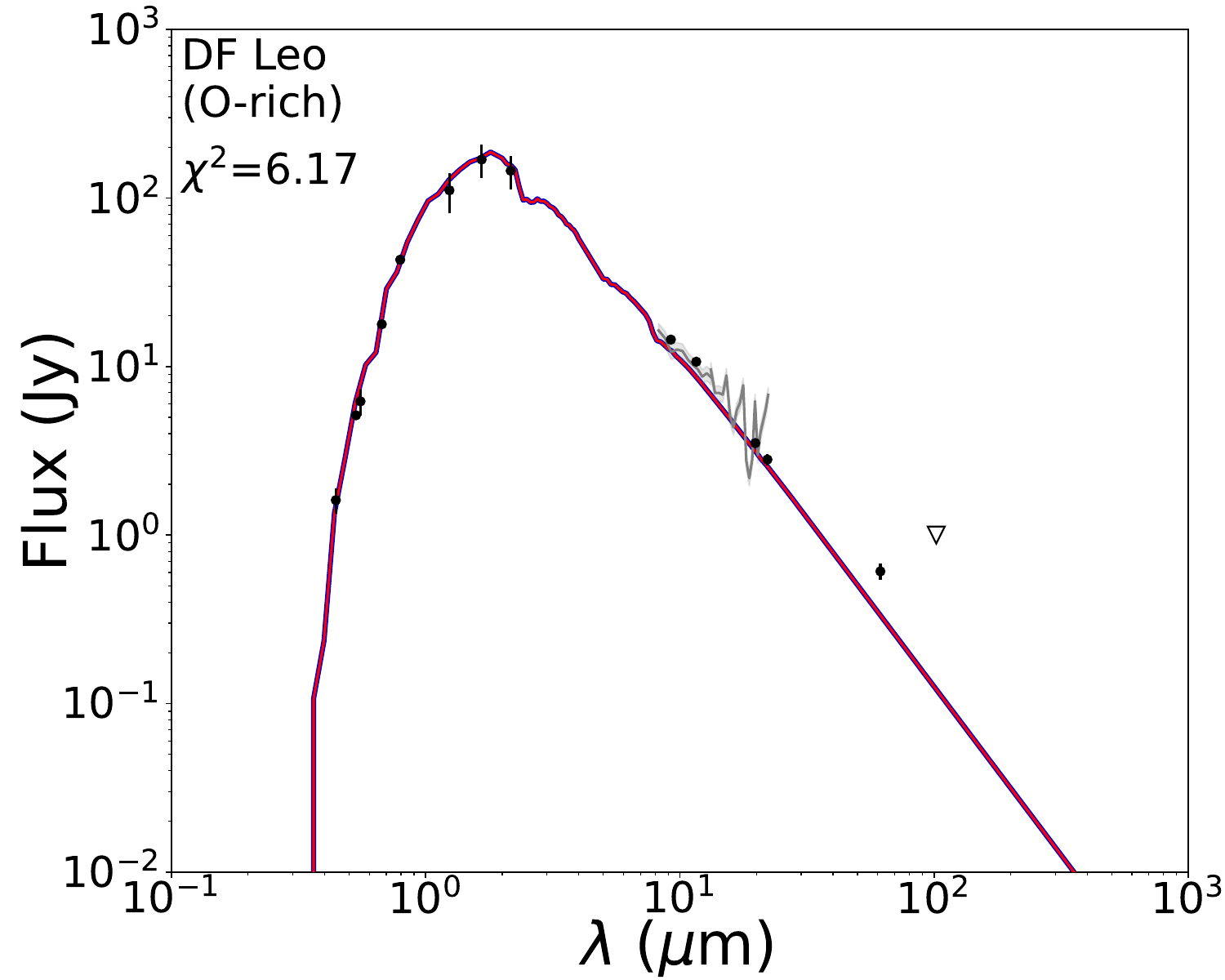}
     \end{subfigure}
     \begin{subfigure}[b]{0.245\linewidth}
         \centering
         \includegraphics[width=\linewidth]{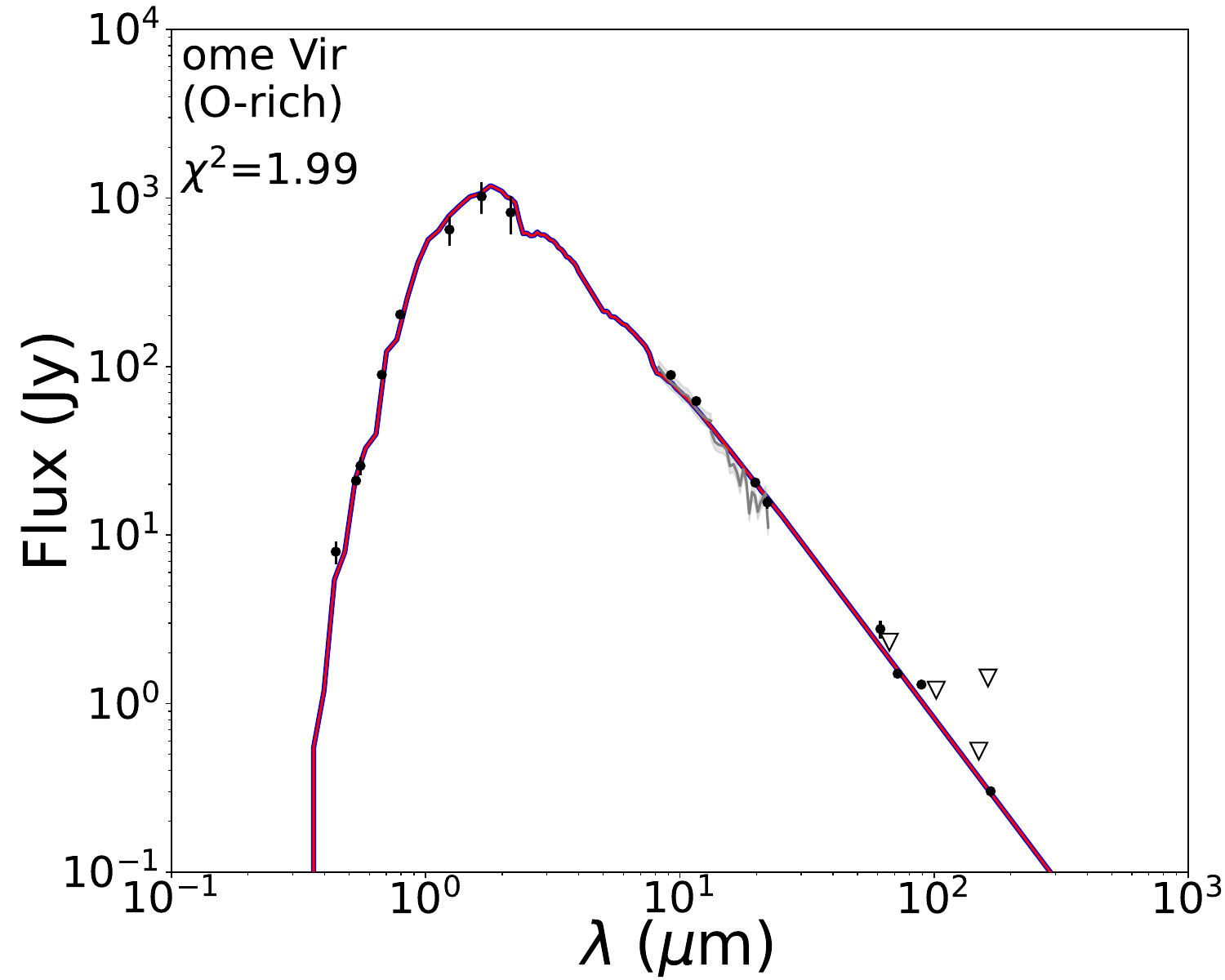}
     \end{subfigure}
     \begin{subfigure}[b]{0.245\linewidth}
         \centering
         \includegraphics[width=\linewidth]{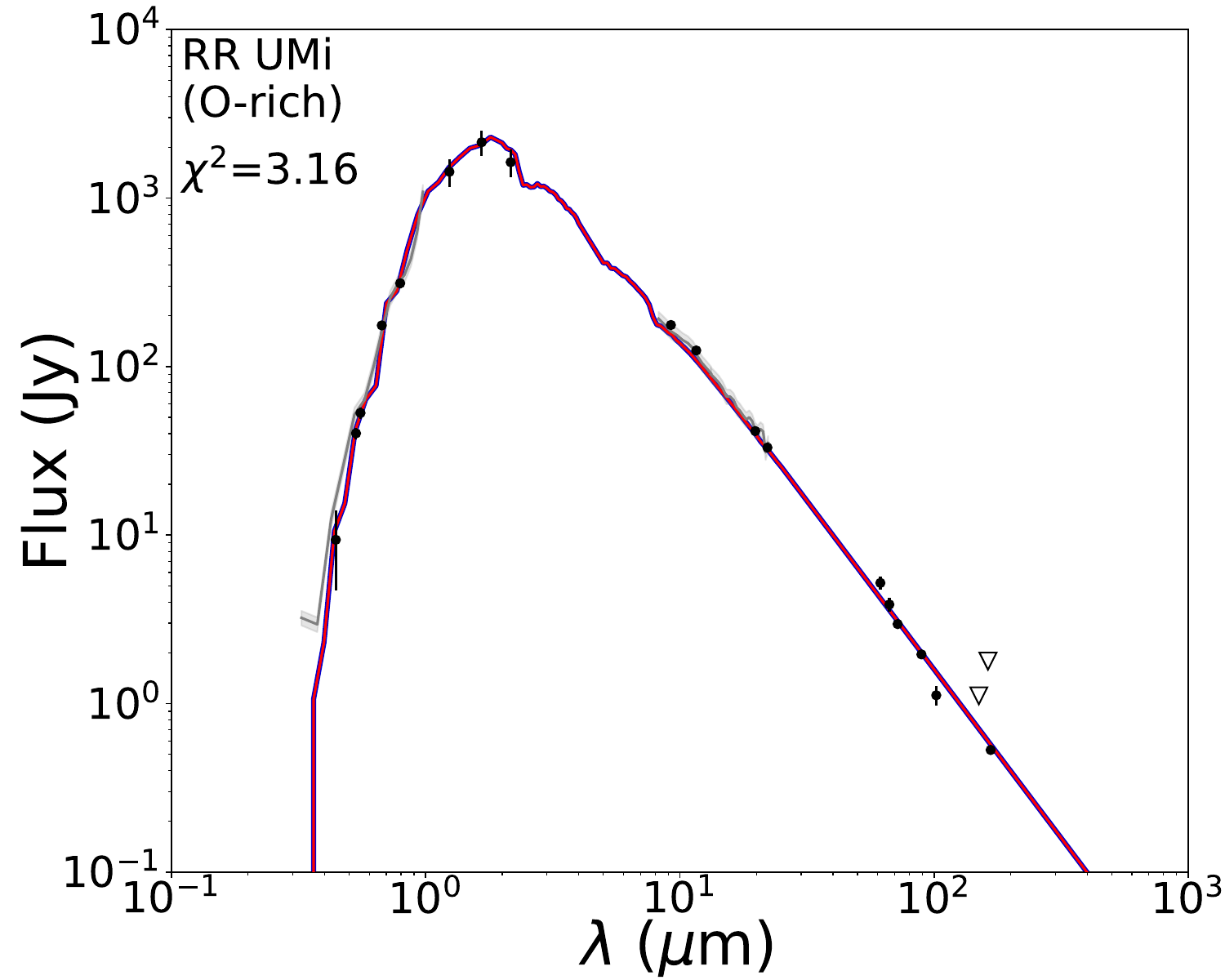}
     \end{subfigure}
     \par\bigskip
        \caption{Same as Fig.~\ref{SED_fitting} for ``naked'' sources.}
        \label{SED_nodust_fitting}
\end{figure*}

We find that the SEDs of most of our targets (21/29) show an infrared excess and require a dusty CSE (see Fig.~\ref{SED_fitting}, table~\ref{tab:DUSTY_models} and appendix~\ref{SED_appendix}); we derive additional parameters for these sources in Sects.~\ref{dust} and \ref{gas-to-dust}. The SEDs of some sources (8/29) can be fitted using only stellar photospheric models (see Fig.~\ref{SED_nodust_fitting} and table~\ref{tab:DUSTY_models_nodust} implying a negligible dust component, presumably because they have not yet experienced substantial mass-loss. We refer to such sources as ``naked'' \citep[as defined by][]{Sloan_1995}. 
Furthermore, we note that all the sources detected in CO by \cite{alonso-hernandez_2024} have infrared excess, whereas not all dusty AGB stars were detected in CO.

We highlight that all the models fit the SEDs with $\chi^{2}$$\gtrsim$1.0 (ranging $\sim$0.6-7.2) and show good agreements with the available spectra, for both dusty and naked AGB stars. Our best-fit DUSTY models reproduce satisfactorily the radial surface brightness profiles of semi-extended sources (see Appendix~\ref{semi-extended}), including the detached shell of VY\,UMa (see appendix~\ref{VY-UMA}).

Fig.\ref{fig:hist} shows the distributions of the best-fit values for $T_{*}$, $T_{\rm inn}$, $n$ and $\tau_{550}$. We find that dusty uvAGBs have $T_{*}$ in the range 2600-3500 K, with similar values for both O-rich and C-rich AGB stars, although we acknowledge that the sample of C-rich is very limited. The ``naked'' stars are located at the highest temperatures of the distribution (3400-3500 K) and also present low luminosities ($\sim$800-3000 $L_{\odot}$), indicating that they might be in the early-AGB phase or still on the Red Giant Branch \citep[see also][]{alonso-hernandez_2024}.

All the sources have values of $T_{\rm inn}$ lower than dust sublimation temperatures \citep[typically $\sim$1400-1500 K, see][]{agundez_2020}. In some cases, $T_{\rm inn}$ is rather low ($\sim$600-1000 K, see Fig.\ref{fig:hist} middle-left), which indicates a relative lack of warm dust in the immediate vicinity of the star.

The distribution of $n$ is around the expected values of 2 for radiatively driven winds \citep[see e.g.][]{Ivezic_1997}, although in some cases they are lower (see Fig.\ref{fig:hist} middle-right). These low values might indicate deviations from smooth outflows (we resume this point in Sect.~\ref{diagrams}). 

Most of the sources presented in this study have optically-thin envelopes at the fiducial wavelength $\tau_{550}$$\lesssim$1, although a few are optically-thick ($\tau_{550}$$\gtrsim$1, see Fig.\ref{fig:hist} right). This clustering at low $\tau_{550}$ can be expected because uvAGBs with low optical depths have a higher chance of being detected via their UV emission due to a low attenuation (see Sect.~\ref{dis} and Appendix~\ref{UV_opacities}).

We found that, in some cases, $Y$ is rather uncertain because the wavelength coverage of the observations does not extend to the sufficiently long wavelength that is required to probe the coldest dust located in the outermost regions of the envelopes \citep[see also][]{Heras_2005}.

We performed a sensitivity analysis to determine the uncertainties and correlations of the free parameters. The uncertainties were estimated from the log-likelihood confidence intervals and show asymmetric distributions. The uncertainties of the parameters are also affected by the spectral variability of AGB stars. In the case of targets without a dust component, $T_{*}$ is the only parameter correctly estimated as $\tau$ converges to zero in all cases ($\tau_{550}$$<$0.01) and the other parameters do not influence the fit.  

We found that the free parameters are degenerate and their uncertainties systematically underestimate second order effects, that is the parameters compensate the effects on the spectral shape between them. Some of these correlations display a characteristic ``banana'' shape indicating non-linearity (we present a detailed discussion about these correlations in appendix~\ref{correlations}).

\begin{figure*}[ h!]
    \centering
    \includegraphics[width=\linewidth]{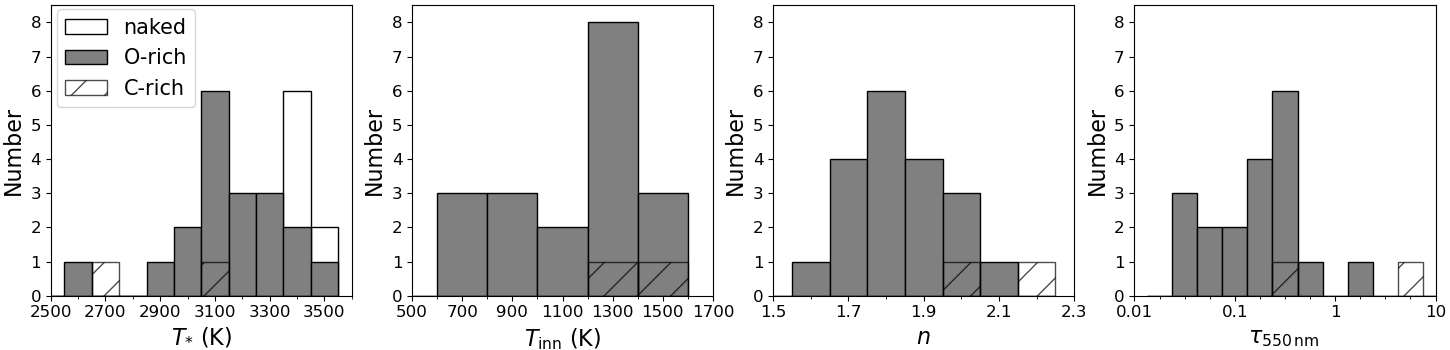}
    \caption{Histograms showing the distributions of the best-fit solution SED modelling parameters. From left to right: Stellar effective temperature, dust inner radius temperature, density power-law index, dust shell radial optical depth at 550 nm. White bins represent ``naked'' O-rich uvAGBs, grey bins represent O-rich uvAGBs with dust component and striped bins represent C-rich uvAGBs.}
    \label{fig:hist}
\end{figure*}

\begin{table*}[h!]

\renewcommand{\arraystretch}{1.2}
\small
\centering

\caption{Best-fit parameters from DUSTY models.} 
\label{tab:DUSTY_models}

\begin{adjustbox}{max width=\textwidth}
\begin{threeparttable}[b]

\begin{tabular}{l c c c c c c c c c c c }
\hline\hline 
Source    & $T_{*}$ & $T_{\rm inn}$ & $n$ & $\tau_{550}$ & log($Y$) & L & $R_{\rm inn}$ & $R_{\rm out}$ & M$_{ dust}$ & $\Dot{M}_{   \rm dust}$ & $\delta$ \\  
        & (K) & (K) &  &  &  & ($\mathrm{L_{\odot}}$) & ($10^{13}$cm) & ($10^{16}$cm) & ($10^{-6}$ $\rm  M_{\odot}$) &($10^{-10}$ $\rm  M_{\odot}\,yr^{-1}$) & \\  
\hline

EY\,Hya  & 3000$^{+100}_{-200}$  & 1350$^{+130}_{-160}$  & 1.86$^{+0.05}_{-0.06}$  & 0.35$^{+0.16}_{-0.09}$ & 3.8$^{+0.7}_{-0.4}$ & 5200$\pm$800 & 7.9 & 50 & 29 & 19 & 310 \\ 
FH\,Vir  &  3400$^{+100}_{-100}$  &  1500$^{+150}_{-150}$ &  1.75$^{+0.05}_{-0.05}$   & 0.05$^{+0.02}_{-0.02}$  &  4.4$^{+1.0}_{-0.2}$ & 2100$\pm$300 & 3.3 & 83 & 13 & -- & -- \\ 
IN\,Hya &  3500$^{+100}_{-200}$  &  1450$^{+170}_{-140}$ &  1.78$^{+0.07}_{-0.05}$  & 0.04$^{+0.03}_{-0.02}$  & 4.0$^{+1.0}_{-0.2}$  & 1700$\pm$200  & 3.3 & 33 & 2.4 & -- & -- \\ 
R\,LMi   &  3100$^{+100}_{-200}$ & 1250$^{+110}_{-130}$ &  2.03$^{+0.04}_{-0.04}$ &  1.6$^{+0.3}_{-0.2}$ &  4.6$^{+0.8}_{-0.2}$  &  3300$\pm$500 &  4.8 & 190 &  48 & 5.2 & 810 \\ 
R\,UMa   &  3200$^{+100}_{-400}$ & 700$^{+120}_{-80}$ &  2.13$^{+0.06}_{-0.04}$  &  0.40$^{+0.12}_{-0.10}$ & 4.0$^{+0.8}_{-0.4}$  &  5000$\pm$900 &  26 & 260 & 30 & 1.7 & 760 \\ 
RR\,Eri &   3300$^{+100}_{-200}$ &  800$^{+150}_{-120}$ &  1.84$^{+0.06}_{-0.06}$  & 0.04$^{+0.03}_{-0.02}$ &  3.0$^{+0.6}_{-0.5}$ & 4700$\pm$900  &  20 & 20 & 3.2 & 6.0 & 680 \\ 
RT\,Cnc  & 3300$^{+100}_{-100}$ &  1350$^{+140}_{-140}$ &  1.92$^{+0.06}_{-0.05}$   &  0.27$^{+0.09}_{-0.06}$  & 4.0$^{+0.7}_{-0.4}$ & 2100$\pm$300  & 4.2 & 42 & 5.5 & 3.0 & 230 \\ 
RU\,Her$^{*}$ &   $<$2700 &  700$^{+140}_{-130}$  &  1.98$^{+0.05}_{-0.06}$ & 0.20$^{+0.14}_{-0.07}$  & 3.4$^{+0.7}_{-0.4}$ & 20000$\pm$5000 &  50 & 130 & 72 & 13 & 1230 \\ 
RW\,Boo  &   3200$^{+100}_{-300}$ &  600$^{+170}_{-100}$ &  1.90$^{+0.14}_{-0.09}$   &  0.06$^{+0.06}_{-0.03}$ & 4.0$^{+1.0}_{-0.1}$  & 2200$\pm$300 & 6.6 & 66 & 3.5 & 2.1 & 1670 \\ 
RZ\,UMa  &   3100$^{+100}_{-200}$ &  1250$^{+150}_{-120}$ &  1.72$^{+0.05}_{-0.05}$   &  0.20$^{+0.06}_{-0.05}$  & 3.4$^{+0.5}_{-0.4}$   &  5100$\pm$1000 & 7.4 & 19 &  21 & 17 & 70 \\
ST\,UMa  &  3400$^{+100}_{-200}$  &  1150$^{+170}_{-110}$ &  1.86$^{+0.07}_{-0.04}$   &  0.09$^{+0.03}_{-0.03}$  & 3.6$^{+0.5}_{-0.1}$  & 2900$\pm$500 &  7.5 & 30 & 4.0 & --  & -- \\ 
SV\,Peg & 2900$^{+200}_{-200}$ & 950$^{+160}_{-100}$  &   1.72$^{+0.10}_{-0.06}$  & 0.20$^{+0.13}_{-0.07}$  &  3.0$^{+0.3}_{-0.5}$ &  10000$\pm$2000 & 21 & 21 & 41 & 45 & 330 \\ 
T\,Dra & 2700$^{+100}_{-200}$ & 1200$^{+110}_{-110}$  & 2.00$^{+0.03}_{-0.03}$  & 4.5$^{+1.0}_{-0.5}$ & 4.8$^{+0.7}_{-0.4}$ & 13000$\pm$3000 & 19 & 1200 & 1200 & 37 & 780 \\ 
TU\,And  &  3200$^{+100}_{-400}$  & 1500$^{+180}_{-120}$  &  1.84$^{+0.07}_{-0.05}$    &  0.35$^{+0.15}_{-0.09}$   & 4.4$^{+1.0}_{-0.1}$   &  7500$\pm$1700 &  5.9 & 150  & 127 & -- & -- \\ 
UY\,Leo  &  3300$^{+100}_{-200}$ & 1350$^{+160}_{-140}$  &   1.80$^{+0.06}_{-0.06}$   &  0.20$^{+0.05}_{-0.04}$ & 4.2$^{+1.0}_{-0.1}$   & 2000$\pm$300 &   4.0 & 64 & 23 & -- & -- \\
V\,Eri & 3000$^{+100}_{-300}$ & 1200$^{+140}_{-140}$ & 1.74$^{+0.06}_{-0.06}$   & 0.45$^{+0.19}_{-0.10}$ & 3.2$^{+0.4}_{-0.4}$ & 5100$\pm$1000 &  8.1 & 13 & 26  & 65 & 90 \\ 
VY\,UMa & 3100$^{+100}_{-300}$ & 1450$^{+160}_{-130}$ & 2.22$^{+0.12}_{-0.10}$   & 0.40$^{+0.15}_{-0.10}$ & 4.0$^{+1.0}_{-0.1}$ & 4700$\pm$900 & 6.1  & 24 & 0.15 & 0.050 & 40000 \\
VY\,UMa$^{**}$ & -- & -- & -- & -- & -- & -- & -- & -- & 5.5 & -- & --  \\
W\,Peg   & 3100$^{+100}_{-400}$   &  950$^{+150}_{-120}$  & 2.03$^{+0.05}_{-0.05}$  &   0.35$^{+0.16}_{-0.11}$ & 4.0$^{+0.8}_{-0.3}$ &5300$\pm$1000  &   15 & 150 & 26 & 4.1 & 1100 \\ 
Y\,CrB  & 3100$^{+100}_{-300}$ & 1300$^{+160}_{-120}$ &  1.76$^{+0.06}_{-0.05}$   &  0.25$^{+0.13}_{-0.08}$    &3.2$^{+0.5}_{-0.4}$ &  4500$\pm$800  & 6.4  & 64  & 7.5 & 8.1 & 160 \\ 
Y\,Gem  &  3100$^{+100}_{-200}$ & 1250$^{+160}_{-150}$  &   1.70$^{+0.06}_{-0.07}$  & 0.03$^{+0.03}_{-0.02}$  & 2.8$^{+0.5}_{-0.7}$ &   8500$\pm$1800 &  9.2  & 5.8 & 1.0 & -- & -- \\ 
Z\,Cnc   &  3200$^{+100}_{-200}$ &  1000$^{+200}_{-180}$ &  1.6$^{+0.2}_{-0.2}$  &  0.10$^{+0.05}_{-0.03}$ & 2.0$^{+0.3}_{-0.4}$   &  4200$\pm$600 & 12 & 1.2  & 1.1 & 11 & 55 \\

\hline
\end{tabular} 
\begin{tablenotes}
\item \textbf{Notes.} Column (1): name of the source, Col. (2): stellar effective temperature, Col (3): dust shell  inner radius temperature, Col (4): dust shell density power-law index, Col (5): dust shell radial optical depth at 550 nm (from $R_{\rm inn}$ to $R_{\rm out}$), Col. (6): logarithm of the dust shell outer-to-inner radii ratio, Col. (7): bolometric luminosity, Col. (8): dust shell inner radius, Col. (9): dust shell outer radius, Col. (10): dust shell mass, Col. (11): dust mas-loss rate, and Col. (12): gas-to-dust ratio. $^{*}$ In the case of RU\,Her, the sensitivity study indicates $\rm T_{*}$$<$2700\,K, although we noted that 2600\,K fits correctly the SED (see Fig.~\ref{SED_fitting}). $^{**}$ In the case of VY\,UMa, the dust mass estimated for the detached shell is included separately.
\end{tablenotes}

\end{threeparttable}
\end{adjustbox}

\renewcommand{\arraystretch}{1.0}

\end{table*}

\subsection{Derivation of related parameters}\label{dust}

We derived the bolometric luminosity, the inner and outer radii of the CSE ($R_{\rm \rm inn}$, $R_{\rm out}$), the radial optical depth as a function of wavelength (see also Appendix~\ref{UV_opacities}), and the dust mass ($M_{\rm dust}$) from the spatial integration of the density distribution \citep[see eqn. 1 of][]{Sahai_2023}.

\begin{equation}
   M_{\rm dust} = 4\pi \frac{n-1}{3-n} y(Y) R_{\rm inn}^{2} (\tau_{\lambda}/\kappa_{\lambda}),
\end{equation}

\noindent where $y(Y)$=($Y^{3-n}$-1)/(1-$Y^{1-n}$), and $\tau_{\lambda}$ and $\kappa_{\lambda}$ are, respectively, the radial optical depth of the shell and the dust mass absorption coefficient at a certain wavelength. In the case of a dust grain mixture, DUSTY considers only one type of grain whose properties are averaged from those of the mixture according to their abundances \citep[see][]{Ivezic_1995, Ivezic_1997}.

We considered 60 \mum\ as the reference wavelength, which is long enough with respect to the assumed grain sizes that grain size does not affect significantly the dust mass absorption coefficient \citep[see e.g.][]{Ysard_2018}. The mass absorption coefficients were estimated from their respective complex refractive indexes applying the Mie theory \citep[see e.g.][and references therein]{Draine_1984, Ivezic_1997}.

We obtained the values of $\kappa_{60\, \mum}$ for each dust composition used in this study. The different mixtures of O-rich dust has $\kappa_{60\, \mum}$$\simeq$100 cm$^{2}$\,g$^{-1}$ (with small differences between them) and the mixture of C-rich dust has $\kappa_{60\, \mum}$$\simeq$90  cm$^{2}$\,g$^{-1}$. These $\kappa_{60\, \mum}$ estimates are in good agreement with common values from the literature \citep[e.g. Fig. 3 from][]{Ysard_2018}. 

In the case of VY\,UMa, we estimated the dust shell masses for the two different components: the present-day CSE (point-source) and the detached shell (extended). This separation allows a more direct comparison with the rest of the sample. The present-day CSE has the lowest dust shell mass among the dusty sources (1.5$\times$$10^{-7}$ $\rm  M_{\odot}$), indicating that the mass-loss process was resumed recently.

The sample presents a wide range of masses, ranging from 1.5$\times$$10^{-7}$ to 1.2$\times$$10^{-3}$ $\rm  M_{\odot}$. These mass estimates have a strong dependence on $Y$ ($M_{\rm dust}$$\propto$$Y^{3-n}$, i.e. linearly when $n$=2), which is a relatively uncertain parameter.

\subsection{Dust mass-loss rates and gas-to-dust ratio.}\label{gas-to-dust}

We performed a comparison between the dust properties obtained in this study and those from the molecular gas component estimated by \cite{alonso-hernandez_2024}. For this, we assumed that the expansion velocity of the dust ($V_{\rm dust}$) is similar to that of CO and estimated the dust mass-loss rates by dividing the dust masses over the dust expansion times ($t_{\rm exp}$=$R_{\rm out} / V_{\rm dust}$):

\begin{equation}
    \Dot{M}_{\rm  dust}=M_{\rm dust}/t_{\rm exp}.
\end{equation}

Even though the dust masses estimated have a large dependence on $Y$, which is uncertain in some cases, the derived dust mass-loss rates are much less dependent of $Y$ because the expansion time compensates for this dependence (exactly when $n$=2) and our estimated values of $n$ are close to 2. We therefore estimate the average gas-to-dust ratios as the ratio between the gas and dust mass-loss rates: 

\begin{equation}
    \delta = \Dot{M}_{\rm gas}/\Dot{M}_{\rm dust},
\end{equation}

\noindent where we used the $\Dot{M}_{\rm gas}$ estimates from \cite{alonso-hernandez_2024}. Although the dust expansion velocities are expected to be somewhat greater than those of the gas because of the finite drift velocity of the dust grains relative to the gas, this difference is not significant for the dust mass-loss rates estimates and produces only slight differences
\citep[see][]{Groenewegen_1998}.

We obtained moderate dust mass-loss rates, in the range 0.05-65$\times$$10^{-10}$$\rm$ $M_{\odot}\,yr^{-1}$. The lowest estimate corresponds to the present-day CSE of VY\,UMa (5$\times$$10^{-12}$$\rm$ $M_{\odot}\,yr^{-1}$), in which the estimation of $\delta$ is not straightforward due to the presence of two components in the dust, whereas only one was detected in CO. Therefore, we assumed that the CO emission and gas mass-loss rate are related to the present-day dust CSE and used it to estimate $\delta$, which has a high value of 40000. The rest of $\delta$ values in the sample cover the range $\sim$50-1700.

\subsection{Trends and comparison with larger samples}\label{diagrams}

In this work, we characterised the dust component in CSEs of uvAGBs and estimate the gas-to-dust ratio by comparing the gas and dust mass-loss rates. The next step is to assess whether these parameters might be sensitive to the UV emission, and its underlying mechanism, or are determined by the stellar properties.

For this purpose, we performed a comparison with one independent stellar property and one extrinsic to the star. For the intrinsic stellar property, we used the variability classification and pulsation period ($P$) from \cite{Samus_2017}, summarised for this sample in \cite{alonso-hernandez_2024}. For the extrinsic parameter, we used $R_{\rm FUV/NUV}$, which is associated with binarity \citep[when $\gtrsim$0.06, see][]{Sahai_2022} and large UV excesses (see Appendix~\ref{UV_opacities}).

Fig.~\ref{fig:HR_diagram} shows the HRD for our sample, in which the dust shell masses and variability classification are also shown. We found that ``naked'' stars are located at the bottom-left corner (low luminosity and high temperature), which is consistent with an early evolutionary stage and indicates that they are not undergoing a significant mass-loss process yet. On the other hand, AGB stars with larger dust masses are located at the top-right corner (high luminosity and low temperature), these sources are in a more evolved stage and have more mass accumulated in their CSE. It can also be seen an evolutionary trend over the AGB, where irregular pulsators are located on the bottom-left corner, Miras pulsators are mostly located at the top-right, and Semi-regulars are spread over the other two groups.

Fig.~\ref{fig:correlations} shows the comparisons between the most relevant parameters estimated from the SED modelling with $P$ and $R^{\rm corr}_{\rm FUV/NUV}$ (defined as $R_{\rm FUV/NUV}$ corrected from dust attenuation, see appendix~\ref{UV_opacities}). We evaluate these comparisons qualitatively because the large dependence between these parameters and different methodologies employed in the literature prevents the derivation of simple relationships to compare with \citep[see discussion in Sect. 2 of][]{Hofner_2018}

We found a decreasing (non-linear) trend of $T_{*}$ with $P$ (see Fig.~\ref{fig:correlations} top-left). This relationship is a trend through the AGB, in which the stellar effective temperatures decrease while pulsation periods increase as the stars climb the AGB.

No clear correlation was found between $T_{\rm inn}$ and $P$ (see Fig.~\ref{fig:correlations} middle-top-left), indicating that $T_{\rm inn}$ does not depend directly on the pulsations and related stellar parameters. Low $T_{\rm inn}$ values indicate a lack of warm dust, which might result from a recent decrease in the mass-loss rate or from the accumulation of dust in structures at a certain distance from the star.

We also noted a dichotomy in the values of $n$ (see Fig.~\ref{fig:correlations} middle-bottom-left). Miras present standard values with $n$$\sim$2.0 on average, whereas semi-regulars systematically present lower values, with $n$$\sim$1.8 on average. Even though this result does not have a direct interpretation, the systematically lower values can be related with deviations from homogeneous outflows, in which pulsation driven mass-loss would have more important contribution in semi-regulars than in Miras.

We also found a positive (non-linear) correlation between $\tau_{550}$ and $P$ (see Fig.~\ref{fig:correlations} bottom-left). The optical depth is proportional to the mass-loss rate, which increases with $P$ \cite[see e.g.][]{McDonald_2016}, and indicates that the dust production rates are higher for more evolved stars.

\begin{table}[t]

\renewcommand{\arraystretch}{1.2}
\small
\centering

\caption{Properties of Stars for which DUSTY models show no dust component.} 
\label{tab:DUSTY_models_nodust}

\begin{adjustbox}{max width=\textwidth}
\begin{threeparttable}[b]

\begin{tabular}{l >{\centering\arraybackslash}p{2.70cm} >{\centering\arraybackslash}p{2.70cm} }
\hline\hline 
Source    & $T_{*}$ &  $L_{*}$  \\  
          & (K)  & L($L_{\odot}$) \\  
\hline 

AT Dra   &  3400$^{+100}_{-100}$    &   2500$\pm$400  \\
BC CMi   &  3400$^{+100}_{-100}$    &  820$\pm$90   \\
BD Cam   &  3500$^{+100}_{-200}$    &   2800$\pm$600 \\
BY Boo   &  3400$^{+100}_{-200}$    &   2100$\pm$300 \\
CG UMa   &  3400$^{+100}_{-100}$    &   2300$\pm$400 \\
DF Leo   &  3500$^{+100}_{-100}$    &  1300$\pm$200 \\
ome Vir  &  3400$^{+100}_{-100}$    &   2300$\pm$500 \\
RR UMi   &  3400$^{+100}_{-100}$    &   1400$\pm$300 \\
\hline
\end{tabular} 
\begin{tablenotes}
\item \textbf{Notes.} Column (1): name of the source, Column (2): stellar effective temperature, Column (3): bolometric luminosity.
\end{tablenotes}

\end{threeparttable}
\end{adjustbox}

\renewcommand{\arraystretch}{1.0}

\end{table}

On the other hand, no clear correlations were found between the parameters estimated from the SED modelling and $R^{\rm corr}_{\rm FUV/NUV}$ (see Fig.~\ref{fig:correlations} right panels). Only a slight decreasing trend of the stellar temperature with $R^{\rm corr}_{\rm FUV/NUV}$ might be seen. This correlation, although weak, is expected due in part to the contribution of the stellar photospheric emission in the NUV band increases with $T_{*}$, whereas its effect in the FUV band is negligible as it is dominated by accretion or chromospheric emission. Therefore, stars with larger $T_{*}$ present systematically lower $R_{\rm FUV/NUV}$.

Fig.~\ref{fig:gas_vs_dust} shows the comparison between the gas and dust mass-loss rates estimates for our sample of uvAGBs and those estimated by \cite{Wallstrom_2025} for the Nearby Evolved Stars Survey \citep[NESS,][]{Scicluna_2022} sample, which includes 485 AGB stars. Our CO detected uvAGBs are located at moderate gas and dust mass-loss rates (6$\times$$10^{-8}$$\rm  M_{\odot}\,yr^{-1}$$\lesssim$ $\Dot{M}_{\rm gas}$ $\lesssim$3$\times$$10^{-6}$$\rm  M_{\odot}\,yr^{-1}$ and 5$\times$$10^{-12}$$\rm  M_{\odot}\,yr^{-1}$$\lesssim$ $\Dot{M}_{\rm dust}$ $\lesssim$7$\times$$10^{-9}$$\rm  M_{\odot}\,yr^{-1}$) and, apart from VY\,UMa, show a good agreement with values derived from the NESS sample.

We found that our sources are in good agreement with those of NESS and are spread around the trend shown in \cite{Wallstrom_2025}. Furthermore, we noted that the two C-rich AGB stars and the four O-rich Miras in our sample are systematically above this trend. We expand the comparison with the NESS sample in Appendix~\ref{NESS_comparison}.

\begin{figure}[h!]
    \centering
    \includegraphics[width=\linewidth]{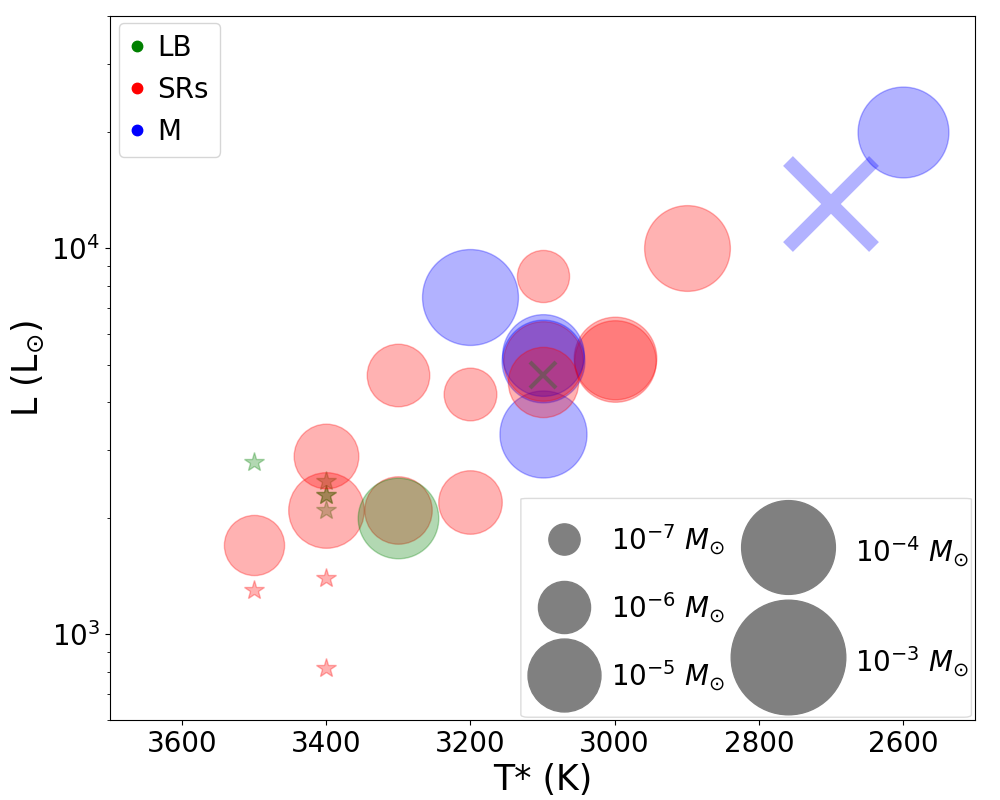}
    \caption{HRD of our the sample. The size of the markers is proportional to the dust shell masses as shown in the {\it bottom-right} legend. The markers colour represents the variability type. Red: semi-regulars (SRs), green: irregulars (LB), and blue: Miras (M). The shape of the markers represents the chemistry of the AGB stars. Circles: O-rich (``naked'' as asterisks), crosses: C-rich.}
    \label{fig:HR_diagram}
\end{figure}

Finally, we searched for correlations between the gas-to-dust ratio ($\delta$) and different parameters (see Fig.~\ref{fig:delta_vs_UV}). We use the linear correlation coefficient ($r$) as a simple statistical indicator to identify first-order trends (we refer to $|r|$$>$0.7 and 0.7$<$$|r|$$<$0.3 as strong and weak correlations respectively).

We checked the positive correlation between $\delta$ and the ratio between CO intensity and IRAS 60 \mum\, flux ($r$=0.37). When rejecting R\,UMa, which has a significant dispersion in the FIR photometry (see Sect.~\ref{SEDs}), this correlation becomes stronger ($r$=0.63). We also found weak positive correlations of $\delta$ with $\Dot{M}_{\rm gas}$ ($r$=0.51) and $P$ ($r$=0.64), whereas a weak anti-correlation was found with $R^{\rm corr}_{\rm FUV/NUV}$ ($r$=-0.44).

\begin{figure*}[ h!]
     \centering
     \begin{subfigure}[b]{\linewidth}
         \centering
         \includegraphics[width=\linewidth]{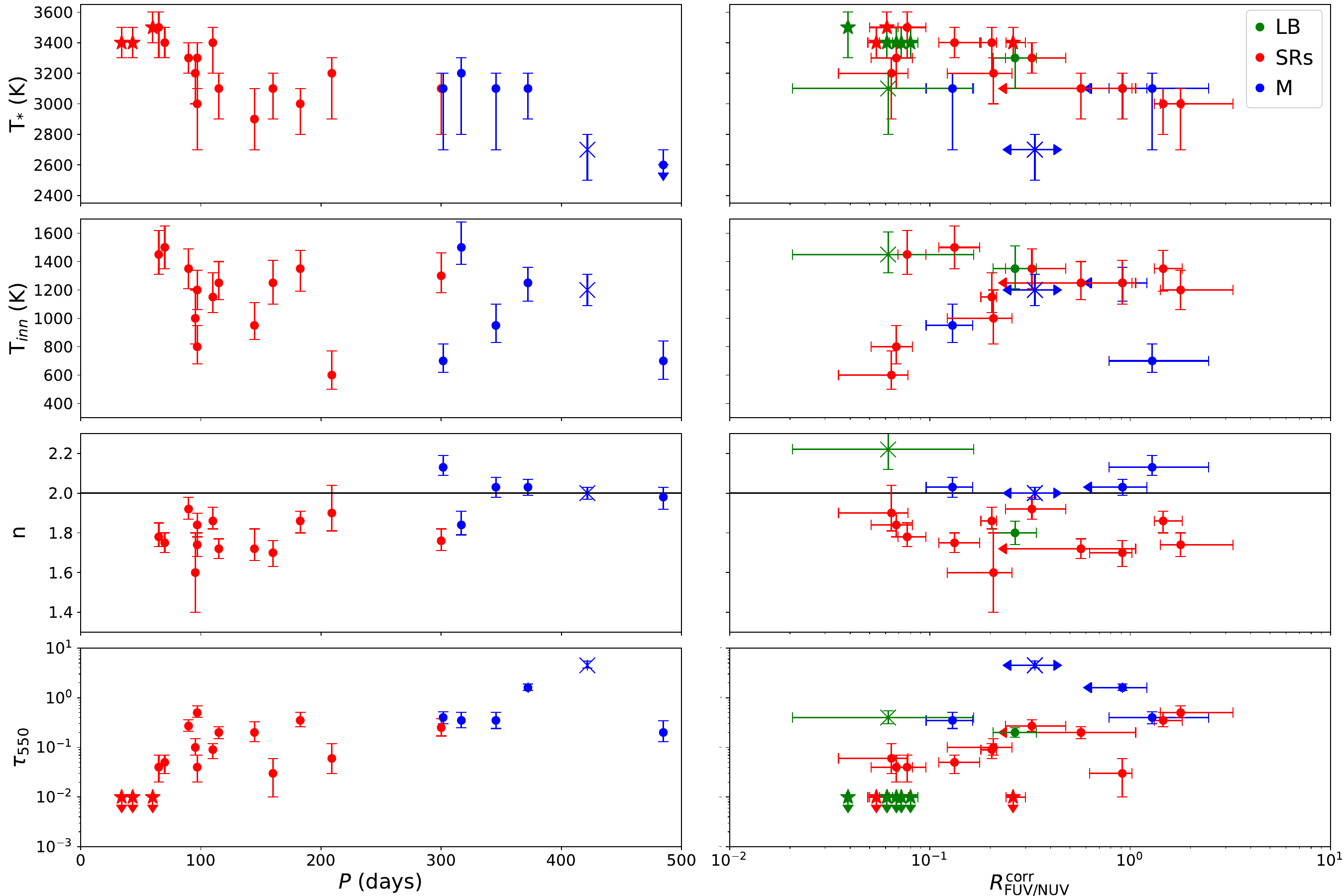}
     \end{subfigure}
        \caption{Comparison of the best-fit solution SED modelling parameters (from {\it top} to {\it bottom}: $T_{*}$, $T_{\rm inn}$, $n$ and $\tau_{\rm 550}$) with $P$ ({\it left}) and $R^{\rm corr}_{\rm FUV/NUV}$ ({\it right}). The colour and shape of the markers represent the stellar variability and chemistry type as in Fig.~\ref{fig:HR_diagram}. The horizontal line in the third row represents $n$=2. Error bars of $R^{\rm corr}_{\rm FUV/NUV}$ include multi-epoch variability of sources detected in both GALEX bands, arrows indicate that one observation was an upper limit.}
        \label{fig:correlations}
\end{figure*}

\section{Discussion}\label{dis}

In this study, we have characterised the dusty envelopes in a sample of 29 UV emitting AGB stars, which are binary candidates. In particular, we searched for trends between the parameters obtained from SED modelling, the pulsation properties, and the UV emission. We also performed a comparison with the NESS sample of AGB stars, to discern whether uvAGBs present systematic differences in their mass-loss rates or gas-to-dust ratios.

We found that UV emitting AGB stars cannot be clearly differentiated from the rest of AGB stars, on the basis of the their SED shapes, mass-loss rates or gas-to-dust ratios. This result is rather surprising considering the variety of binary-related physical phenomena that may affect the mass-loss, dust formation and enrichment, and gas-to-dust ratios in uvAGBs. In particular:

1. In uvAGBs, the internal UV/X-ray emission may photodissociate the abundance of CO in the innermost regions on the envelope. These abundance depletions can subtly modify the CO line profiles and intensities, remaining unperceived and affecting the gas mass-loss rates estimates \citep[e.g. a factor $\sim$2 for T\,Dra, see][]{alonso-hernandez_2025}.

2. Considering that uvAGBs are likely binary systems, they can suffer alterations in the dust formation process, especially within the wakes of the orbiting close stellar companions, where dust has been proposed to form abundantly \citep{Danilovich_2025}. Similarly, the impact of UV emission on the abundances of dust precursor molecular species \cite[e.g. SiO and C$_{2}$H$_{2}$, see][]{Van_de_Sande_2019, Van_de_Sande_2022} may affect dust formation.

3. The presence of binary-induced structures (e.g. spirals and discs), which usually cannot be directly identified by CO line profiles or SED modelling, may lead to incorrect estimates of the mass-loss rates for both dust \citep[see][]{Wiegert_2020} and gas \citep[see][]{Vermeulen_2025} when assuming spherical geometry.

The combination of these phenomena can potentially produce variations in the mass-loss rates of even orders of magnitude. Even though it might be expected that uvAGBs have lower gas-to-dust ratios as suggested by \cite{alonso-hernandez_2024}, these phenomena may each other in the overall result. On the other hand, some properties of dust grains (e.g. composition, size) in these systems may differ due to the UV irradiation or effects of stellar companions on the environment. 

We also remark that binarity is expected to be common in AGB stars. In particular, the fraction of uvAGBs is at least 10\% of the total population \citep[see][]{alonso-hernandez_2024}. Therefore, it is likely that the uvAGBs present in the NESS sample (see Appendix~\ref{UV_opacities}) are producing an overlap when comparing the samples, avoiding a clear identification of differences.

We also show that modelling dust attenuation at UV wavelengths is necessary for extracting intrinsic UV emission properties (see appendix~\ref{UV_opacities}). Optically thick CSEs can prevent even the detection of sources with intrinsically intense UV emission, resulting in an underestimate in the percentages of uvAGBs \citep[as suggested by][]{Montez_2017}. 

\begin{figure}[h!]
    \centering
    \includegraphics[width=\linewidth]{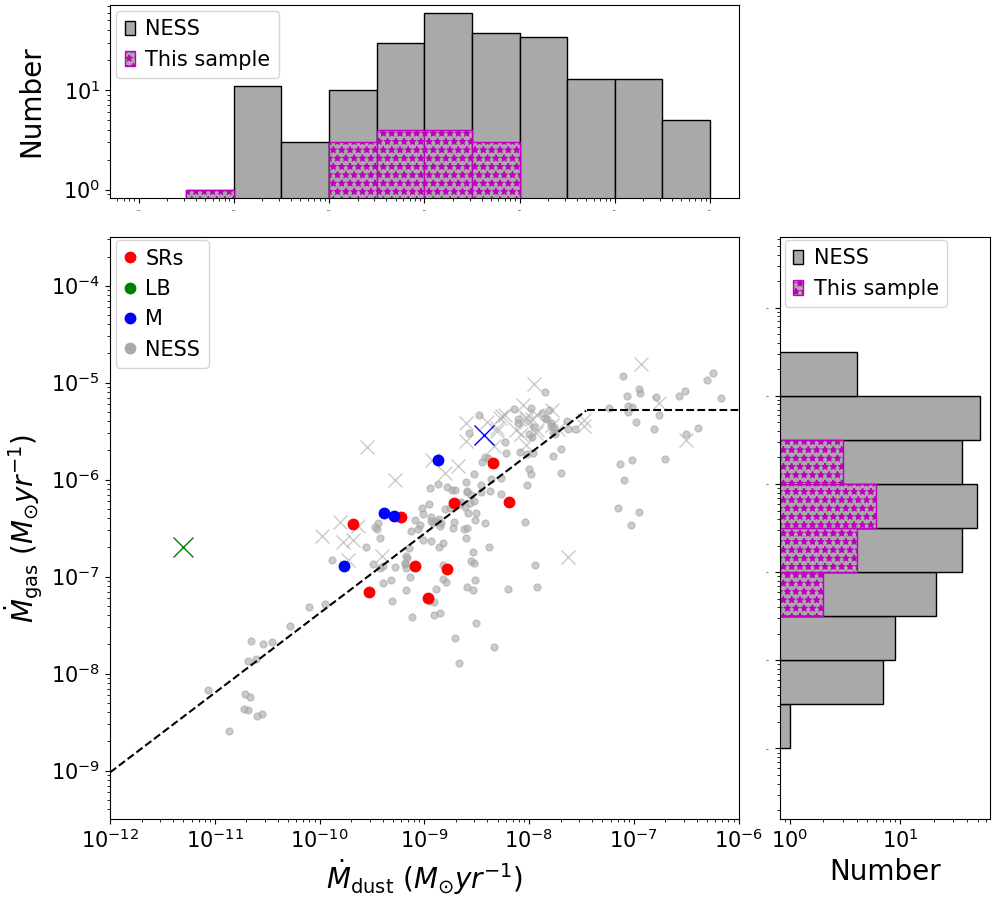}
    \caption{Comparison between the gas and dust mass-loss rates for our sample of uvAGBs. The colour and shape of the markers represent the stellar variability and chemistry type as in Fig.~\ref{fig:HR_diagram}. Grey markers represent sources from the NESS sample. The dashed line represents the relationship presented in \cite{Wallstrom_2025}. The histograms on the {\it top} and {\it right} of the figure indicate the distribution of $\Dot{M}_{\rm dust}$ and $\Dot{M}_{\rm gas}$ respectively.}
    \label{fig:gas_vs_dust}
\end{figure}

We also highlight the effect of dust attenuation on the $\rm R_{FUV/NUV}$$\gtrsim$0.06 threshold proposed by \cite{Sahai_2022} as a criteria to discern cases in which the UV excesses are clearly related with accretion processes. The observed $\rm R_{FUV/NUV}$ may well differ from the intrinsic ratio due to the difference in the optical depths between both bands. The optical depth in the FUV band is larger than in the NUV band, leading to intrinsic $\rm R_{FUV/NUV}$ larger than observed. O-rich AGB stars present a larger difference in the optical depth between both bands (see appendix~\ref{UV_opacities}).

The wavelength dependence of dust attenuation also affects the overall shape of the UV spectra, affecting both continuum measurements and line intensity ratios. Although this effect is small when comparing spectral lines to their adjoining continuum emission \citep[e.g.][]{Guerrero_2020}, it can be important when measuring the continuum over a broad spectral range or when estimating ratios between distant spectral lines due to the difference in attenuation \citep[e.g.][]{Ortiz_2019}.

\begin{figure*}[h!]
     \centering
     \includegraphics[width=\linewidth]{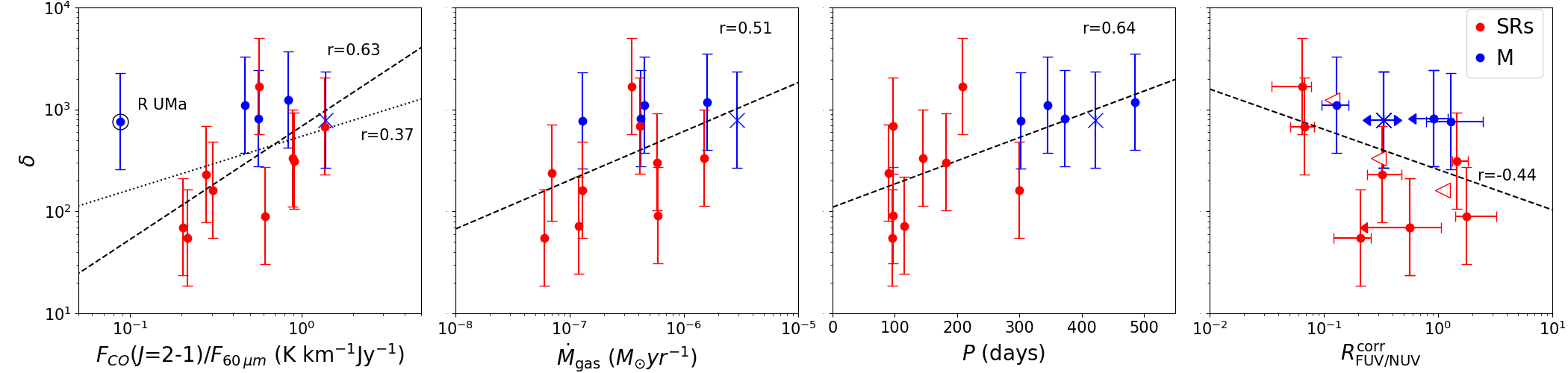}
     \caption{Comparison between the gas-to-dust ratio $\delta$ and different parameters (from {\it left} to {\it right}: ratio of CO J=2-1 velocity-integrated line flux to the IRAS 60\,\mum~flux, $\Dot{M}_{\rm gas}$, $P$, and $R^{\rm corr}_{\rm FUV/NUV}$). The colour and shape of the markers represent the stellar variability and chemistry type as in Fig.~\ref{fig:HR_diagram}. The dashed lines represents linear fits to the points, the dotted line in the first panel is the best fit excluding R\,UMa. Error bars of $R^{\rm corr}_{\rm FUV/NUV}$ are estimated as in Fig.~\ref{fig:correlations}, empty arrows are upper limits for non detections in the FUV.}
        \label{fig:delta_vs_UV}
\end{figure*}

We found a weak anticorrelation between the gas-to-dust ratio and $R^{\rm corr}_{\rm FUV/NUV}$, although we acknowledge that the uncertainties associated with the dust attenuation correction and the intrinsic variability of the UV emission in these sources may well be hiding a stronger anticorrelation with $R^{\rm corr}_{\rm FUV/NUV}$. Furthermore, the correlations between the gas-to-dust ratio with the mass-loss rate and stellar pulsations are stronger (see Fig.~\ref{fig:delta_vs_UV}).

In this context, we remark that uvAGBs are predominantly semi-regulars \citep[see][]{alonso-hernandez_2024} and have moderate or low mass-loss rates (the dust attenuation produces an observational bias to detect UV emission in AGB stars with low opacities), which have systematically lower gas-to-dust ratios. Future studies are required to better understand the effect of UV emission on gas-to-dust ratios.

\section{Summary and conclusions}\label{summ}

In this paper we studied the dust properties of UV emitting AGB stars estimated from SED modelling. We included the reduction of {\it Herschel}/PACS images and performed aperture photometry, which was useful to check the angular sizes and density distributions of the CSEs and to model the extended detached shell of the C-rich AGB star VY\,UMa.

We obtained moderate dust mass-loss rates in the range $10^{-10}$--$10^{-8}$$\rm  M_{\odot}\,yr^{-1}$ and gas-to-dust ratios ranging from 50 to 1700 (excluding VY\,UMa, see Table~\ref{tab:DUSTY_models}), which are intermediate values in comparison with those estimated for the NESS sample (see Fig.~\ref{gas-to-dust} and Appendix~\ref{NESS_comparison}). Our sample of uvAGBs presents similar mass-loss rates and shows, at least qualitatively, the same trends as larger samples of AGB stars. However, it is expected that larger samples also contain some uvAGBs as they are at least 10\% of the population\citep[see][]{alonso-hernandez_2024}, which may prevent us from identifying clear differences.

We found a weak anticorrelation ($r$=-0.44) between the gas-to-dust ratio and $R^{\rm corr}_{\rm FUV/NUV}$. This result indicates that the gas-to-dust ratio of uvAGBs with larger $R^{\rm corr}_{\rm FUV/NUV}$ might be more affected by UV emission and/or by the presence of a stellar companions. However, we note that $R^{\rm corr}_{\rm FUV/NUV}$ is affected by the uncertainties on the dust attenuation correction as well as the intrinsic UV variability found in uvAGBs. On the other hand, the gas-to-dust ratio is also affected by pulsations and mass-loss rates. Therefore, the relationship between the gas-to-dust ratio and the UV emission is not simple.

It is important to obtain not only samples with confirmed stellar companions, but also of single (or not interacting) stellar systems, to better understand the effect of binarity on mass-loss processes. Future observations with high angular resolution are required to explore the dust forming regions surrounding these stars, identify the presence of stellar companions, and constrain their effect on dust formation and enrichment. 

Infrared interferometry is useful to observe the dust distribution in the stellar vicinity, which is key to identifying whether dust formation can be correctly described by radiatively driven winds \citep[][]{RASTAU_2025}, or to indicate the presence of stellar companions \citep[see e.g.][]{Planquart_2024}. Sub-mm interferometry has already shown large capabilities to identify binary-induced structure and modified dynamic in the interior of CSEs \citep[see e.g.][]{Decin_2020}. Moreover, larger spectral coverage is required for robust SED modelling, especially extending into the far-infrared, which traces the outermost regions of the CSEs where a significant mass of cool dust may reside.

We also described the effects of dust attenuation on the observed UV emission and the importance of correct modelling when analysing photometric or spectroscopic observations. Correct modelling of dust attenuation can provide a better understanding of the nature of the UV source (e.g. UV excesses and $R_{\rm FUV/NUV}$) as it is not negligible in some cases. Furthermore, dust attenuation can prevent the detection of uvAGBs.

The detectability of UV emission is biased towards AGB stars with low mass-loss rates (and low opacities), which have systematically lower gas-to-dust ratios. A systematic study, using a larger sample, in which binary and single AGB stars are identified, and covering wider ranges of parameters, would help to refine the observed trends with the mass-loss properties.

\section*{Data availability}\label{additional_material}

Tables containing the employed photometry for the SEDs and the photometric fluxes estimated from the individual {\it Herschel}/PACS observations are only available in electronic form at the CDS via anonymous ftp to cdsarc.u-strasbg.fr (130.79.128.5) or via http://cdsweb.u-strasbg.fr/cgi-bin/qcat?J/A+A/.

\begin{acknowledgements}
We thank the referee, Roberto Ortiz, whose valuable comments helped us improve the quality of the manuscript. This work is part of the I+D+i projects PID2019-105203GB-C22, PID2022-137241NB-C42 and PID2023-146056NB-C22 funded by Spanish MCIN/AEI/10.13039/501100011033 and by “ERDF A way of making Europe". J.A.H. is supported by INTA grant PRE\_MDM\_05 and acknowledges CSIC grant iMOVE 23023. R.S.'s contribution to the research described here was carried out at the Jet Propulsion Laboratory, California Institute of Technology, under a contract with NASA (80NM0018D0004), and funded in part by NASA via various ROSES/ADAP awards and HST/GO awards (administered by STScI). {\it Herschel} is an ESA space observatory with science instruments provided by European-led Principal Investigator consortia and with important participation from NASA. ISO is an ESA project with instruments funded by ESA Member States and with the participation of ISAS and NASA. Some of the data presented in this paper were obtained from the Multimission Archive at the Space Telescope Science Institute (MAST). STScI is operated by the Association of Universities for Research in Astronomy, Inc., under NASA contract NAS5-26555. Support for MAST for non-HST data is provided by the NASA Office of Space Science via grant NAG5-7584 and by other grants and contracts.
\end{acknowledgements}

\bibliographystyle{aa} 
\bibliography{main.bib}

\begin{appendix}

\section{Semi-extended sources}\label{semi-extended}

\begin{figure}[h!]
     \centering
     \begin{subfigure}[b]{0.49\linewidth}
         \centering
         \includegraphics[width=\linewidth]{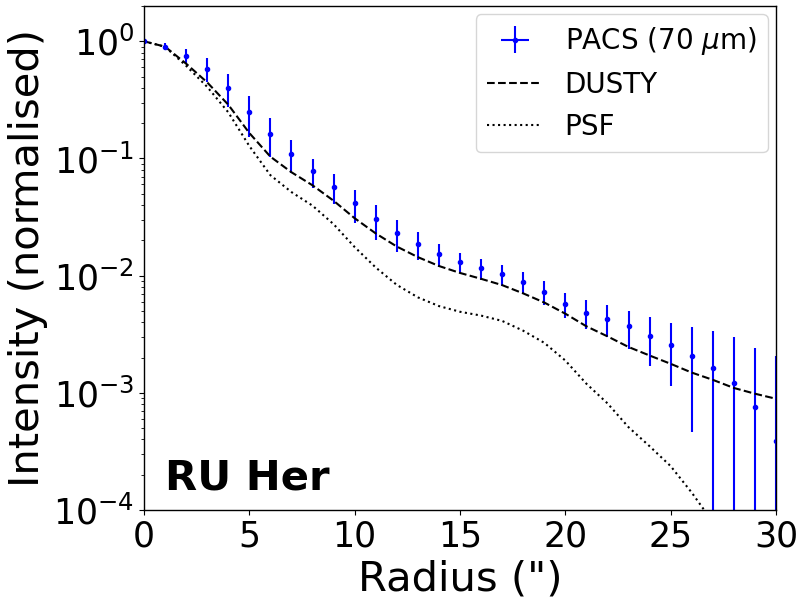}
     \end{subfigure}
     \begin{subfigure}[b]{0.49\linewidth}
         \centering
         \includegraphics[width=\linewidth]{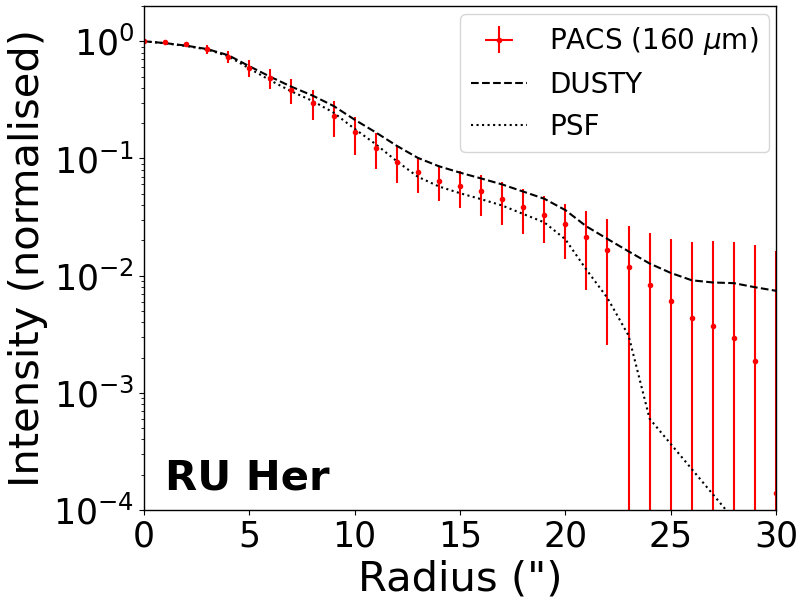}
     \end{subfigure}
     \begin{subfigure}[b]{0.49\linewidth}
         \centering
         \includegraphics[width=\linewidth]{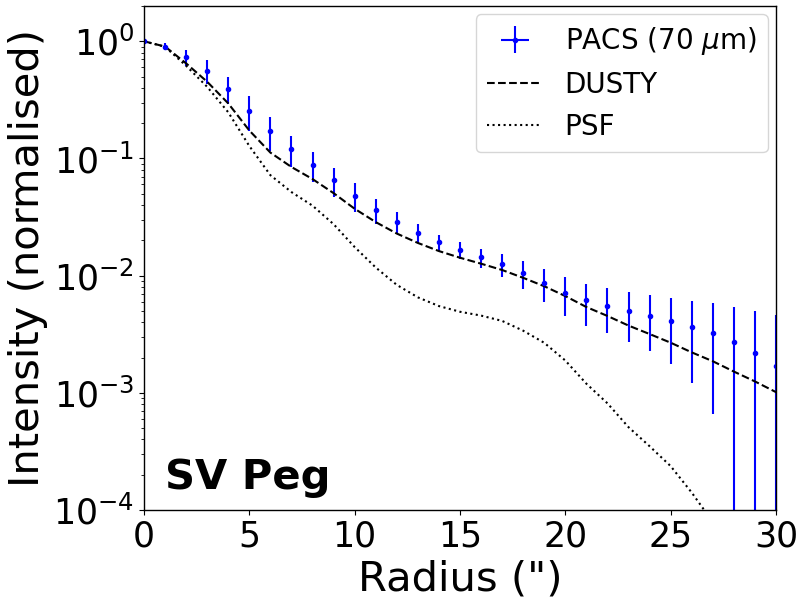}
     \end{subfigure}
     \begin{subfigure}[b]{0.49\linewidth}
         \centering
         \includegraphics[width=\linewidth]{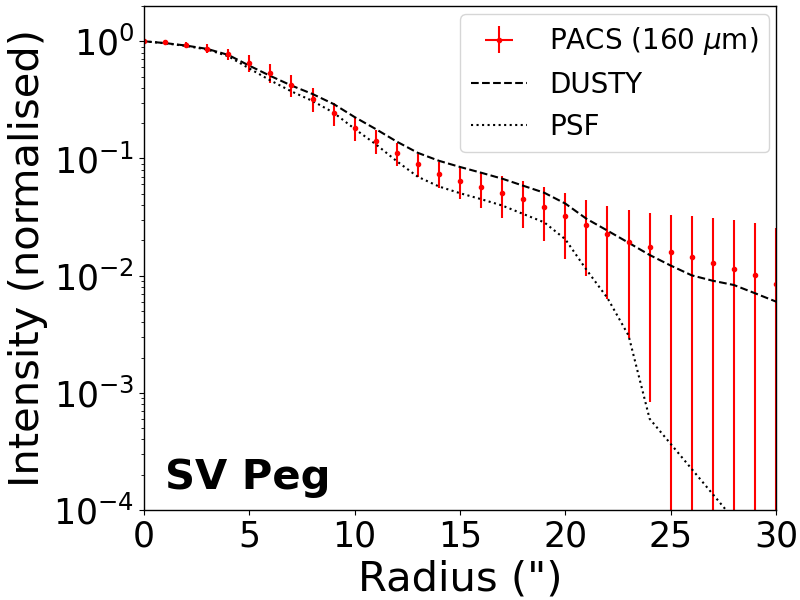}
     \end{subfigure}
     \begin{subfigure}[b]{0.49\linewidth}
         \centering
         \includegraphics[width=\linewidth]{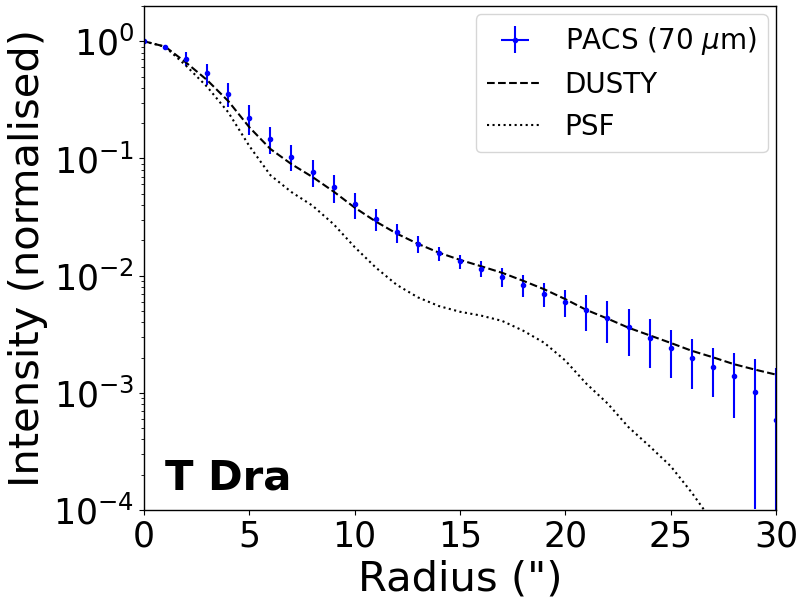}
     \end{subfigure}
     \begin{subfigure}[b]{0.49\linewidth}
         \centering
         \includegraphics[width=\linewidth]{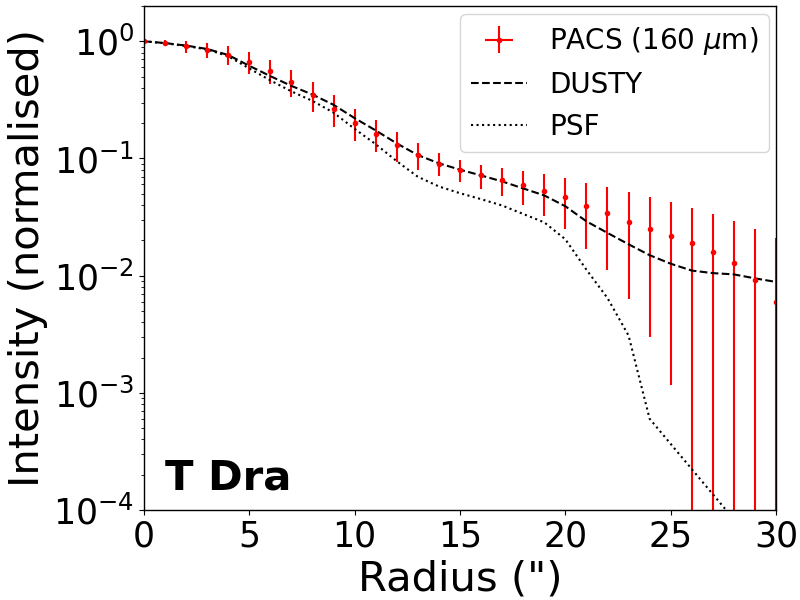}
     \end{subfigure}
     \begin{subfigure}[b]{0.49\linewidth}
         \centering
         \includegraphics[width=\linewidth]{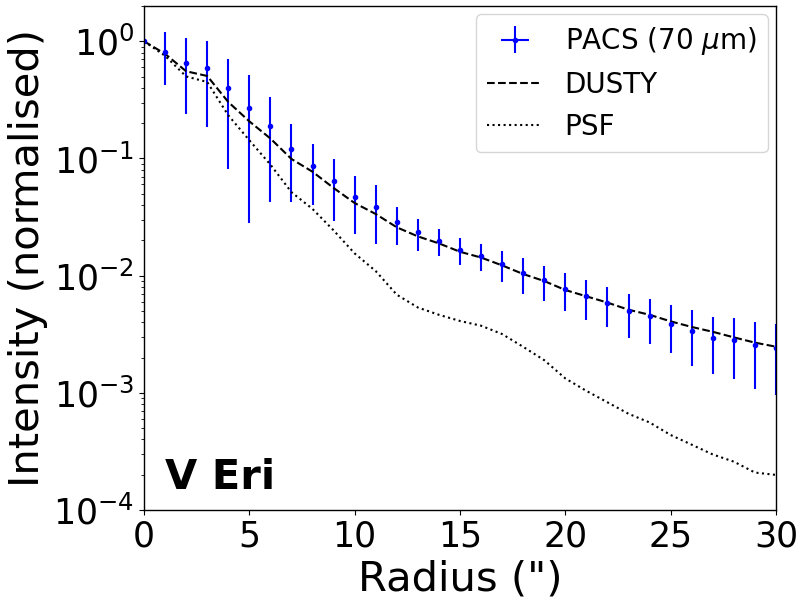}
     \end{subfigure}
     \begin{subfigure}[b]{0.49\linewidth}
         \centering
         \includegraphics[width=\linewidth]{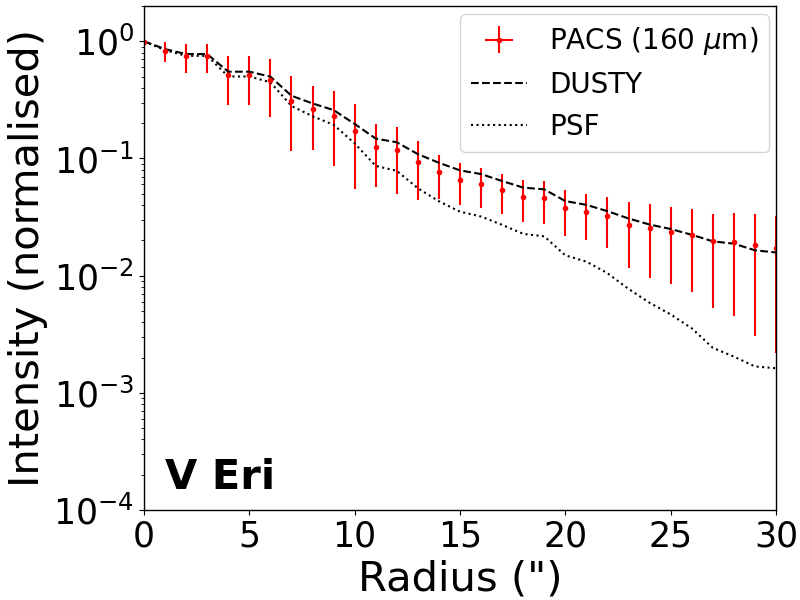}
     \end{subfigure}
     \par\bigskip
     \caption{Radial surface brightness profiles at 70 \mum\ ({\it left}) and 160 \mum\ ({\it right}) for the four semi-extended sources in the sample (from {\it top} to {\it bottom}: RU\,Her, SV\,Peg, T\,Dra and V\,Eri). Solid circles correspond to {\it Herschel}/PACS radial surface brightness profiles, dashed lines to DUSTY synthetic images convolved with their respective PSFs and dotted lines to the PSFs to represent point-source radial surface brightness.}
     \label{radial_profiles}
\end{figure}

Fig.~\ref{radial_profiles} shows radial surface brightness profiles for the four semi-extended sources presented in this study (RU\,Her, SV\,Peg, T\,Dra and V\,Eri). In addition, we created DUSTY synthetic images for their best-fit models in the effective wavelength of each {\it Herschel}/PACS filter and convolved them with their respective PSFs. Then, we radially averaged the surface brightness, in both the observed and the synthetic images, in radial annuli of 1\arcsec and compared them. The uncertainties were estimated as the sum of the standard deviation of the flux and the average instrumental error in each annulus.

The best-fit DUSTY models (see Sect.~\ref{fit_results}) reproduce correctly the angular sizes after convolving with the correspondent PSF. In the four cases the difference between the angular sizes and the {\it Herschel}/PACS PSFs can be appreciated better on the B filter (70 \mum) due to its higher angular resolution.

\section{VY\,UMa detached shell}\label{VY-UMA}

We present a more detailed analysis of the SED modelling for the particular case of VY\,UMa, which has a large-scale detached shell in addition to a compact, bright central emission component that represents the present-day mass-loss rate. This extended emission was previously identified in the {\it Herschel}/PACS images \citep{Cox_2012, van_Marle_2014} and manifested in the SED as the large excesses in the far-infrared (see the case of VY\,UMa in Fig.~\ref{SED_fitting}). 

The contribution of the detached shell is only noticed in the SED in the far-infrared because it is composed of very cold dust ($\sim$10--100K), whose emission is concentrated in the $\sim$30--300 $\mu$m region, and it does not modify the SED at shorter wavelengths. Therefore, we used the {\it Herschel}/PACS fluxes, which isolates the present-day mass-loss (black filled circles in Fig.~\ref{SED_fitting}) and the extended emission from the detached shell (black filled diamonds in Fig.~\ref{SED_fitting}), and rejected IRAS and AKARI because they included partially the emission form the detached shell.

\begin{figure}[h!]
    \centering
    \includegraphics[width=\linewidth]{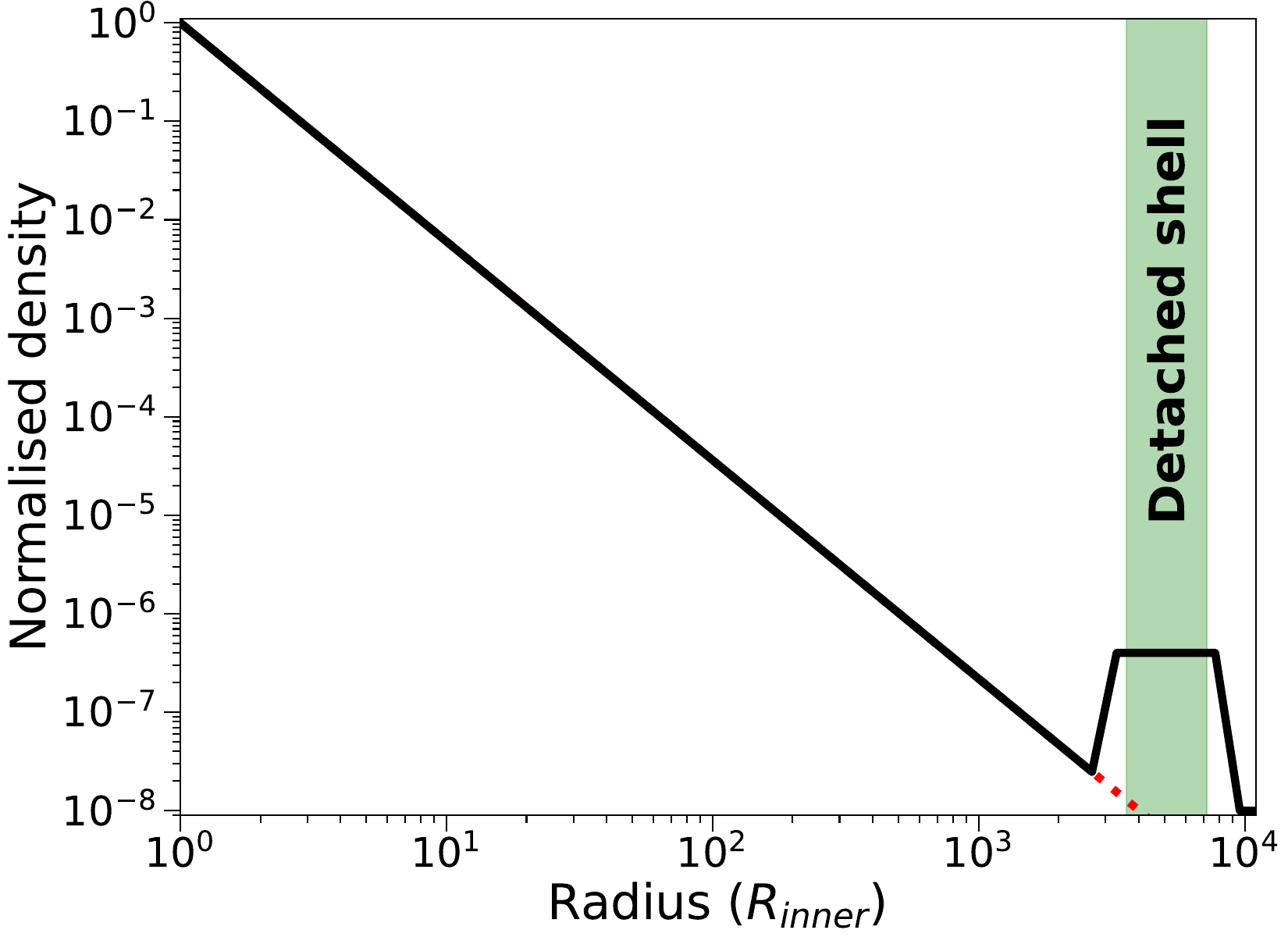}
    \caption{Adopted density profile for VY\,UMa. The solid black line represents the modified density distribution to fit the detached shell. The dotted red line represents the smooth density profile from the present-day mass-loss. The green area represent the location of the detached shell in the {\it Herschel}/PACS images.}
    \label{fig:VY-UMA_density}
\end{figure}

To correctly fit the SED, we followed a similar methodology to that presented in \cite{Mecina_2014, Mecina_2020} to separate the contribution from the present-day mass-loss and the detached shell. We started by fitting the SED of the present-day mass-loss (including {\it Herschel}/PACS fluxes and all the photometric observations at $\lambda$<\,30 $\mu$m) following the same procedure as in the rest of the sources. Once the best-fit solution for the present-day mass-loss was found, we modified the normalised density distribution including a peak at 3200--8000$\times$$R_{\rm inn}$, which is equivalent to 31--78\arcsec\ considering the source distance and in agreement with the location of the detached shell (see Fig.~\ref{fig:VY-UMA_density}).

This model reproduces satisfactorily both the FIR excesses and the {\it Herschel}/PACS radial surface brightness profiles. It allows us to estimate an overall density and dust mass of the extended detached shell. However, we note that this single 1D model is not able to reproduce the asymmetrical shape of the detached shell, which would require a more advanced 3D radiative transfer model.

\begin{figure*}[h!]
     \centering
    \begin{subfigure}[b]{0.45\linewidth}
         \centering
         \includegraphics[width=\linewidth]{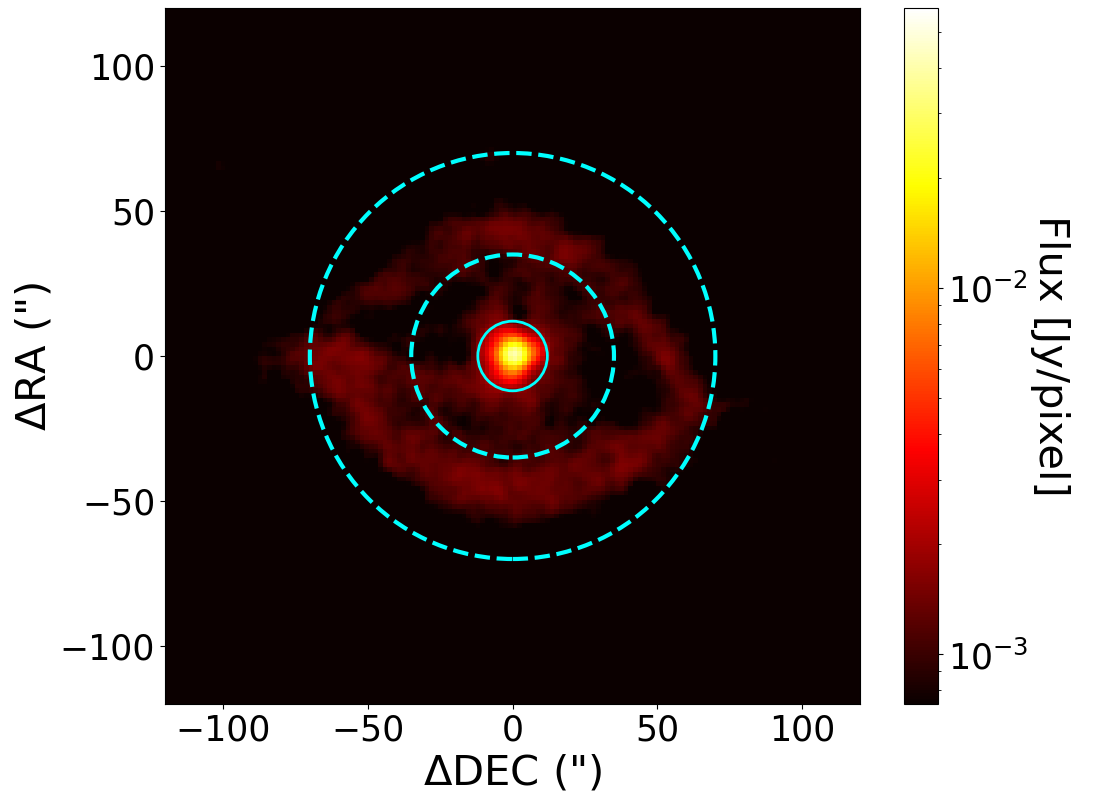}
     \end{subfigure}
     \begin{subfigure}[b]{0.45\linewidth}
         \centering
         \includegraphics[width=\linewidth]{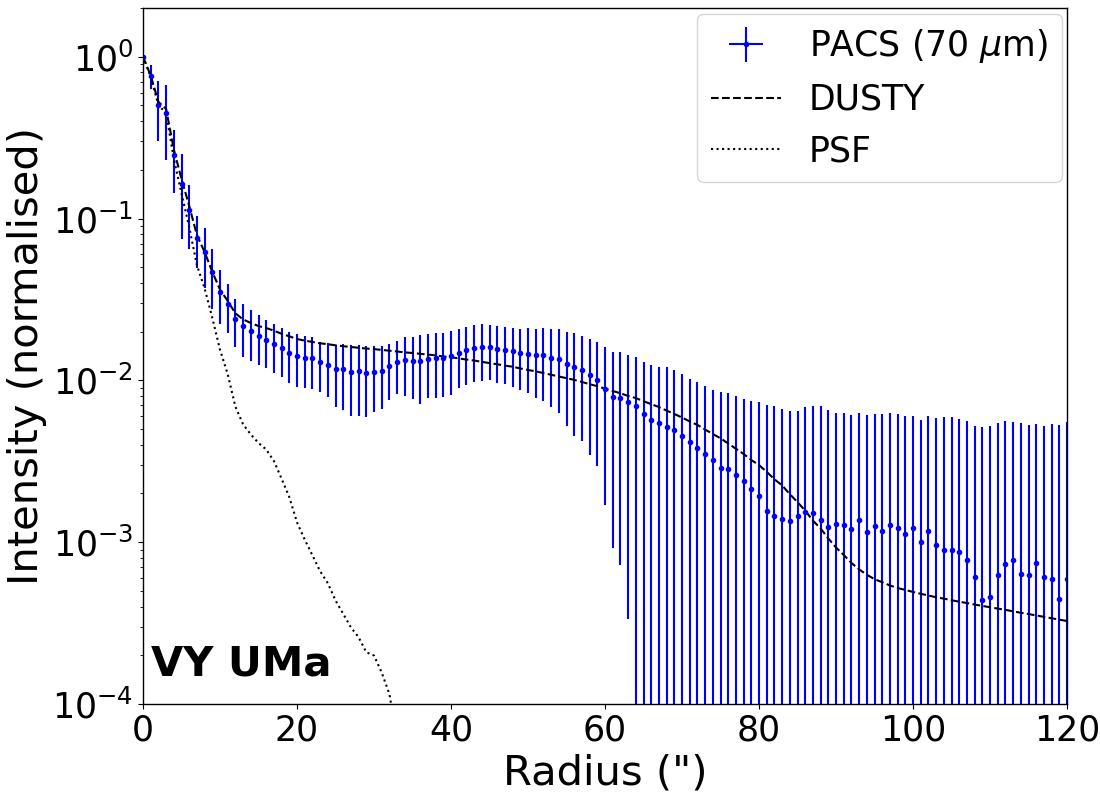}
     \end{subfigure}
     \par\bigskip
     \begin{subfigure}[b]{0.45\linewidth}
         \centering
         \includegraphics[width=\linewidth]{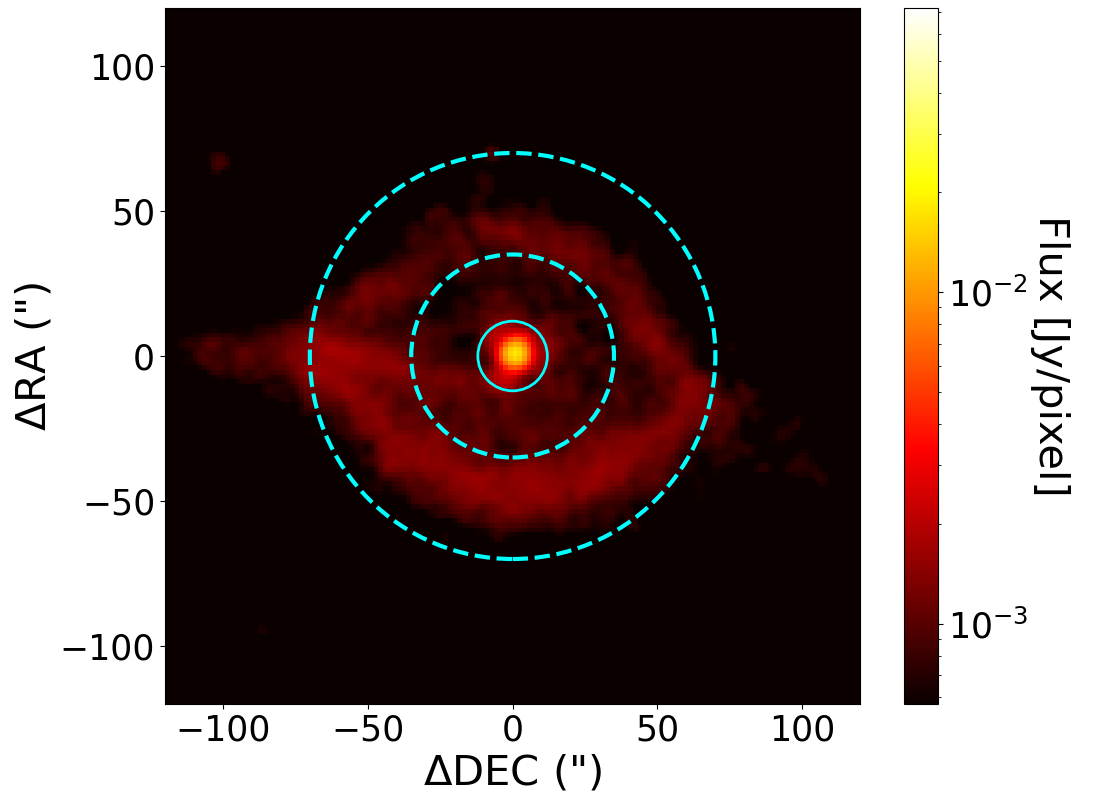}
     \end{subfigure}
     \begin{subfigure}[b]{0.45\linewidth}
         \centering
         \includegraphics[width=\linewidth]{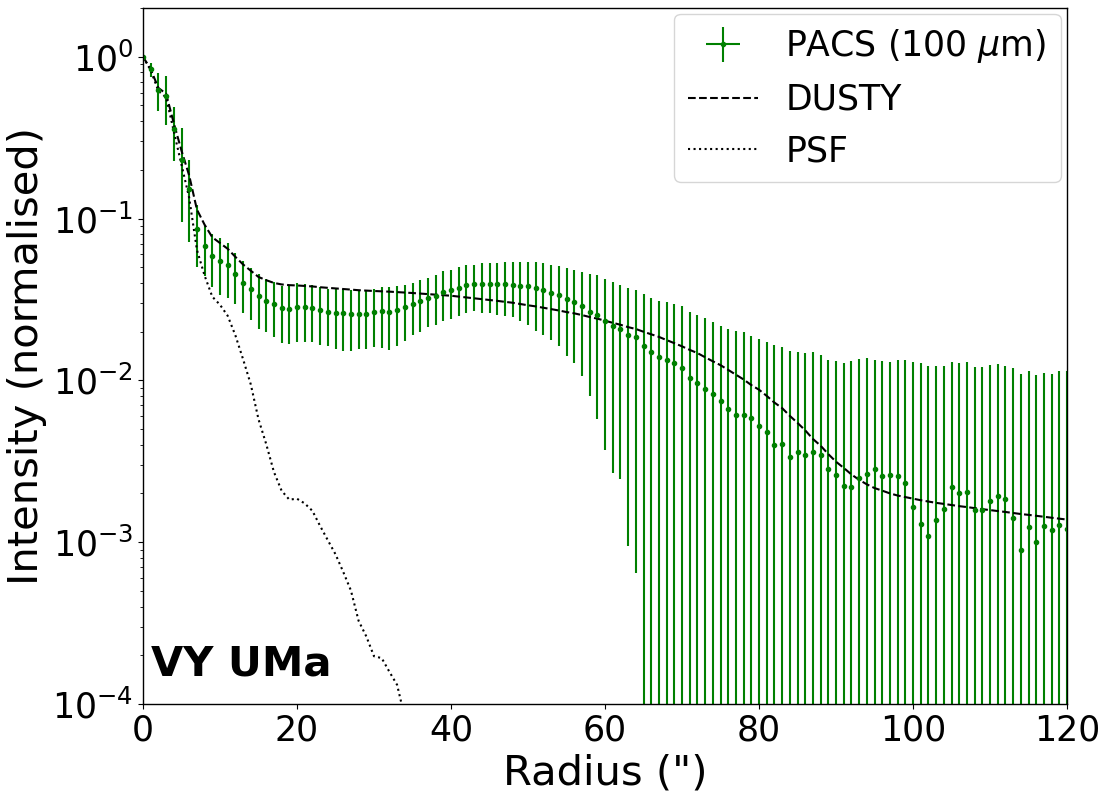}
     \end{subfigure}
     \par\bigskip
     \begin{subfigure}[b]{0.45\linewidth}
         \centering
         \includegraphics[width=\linewidth]{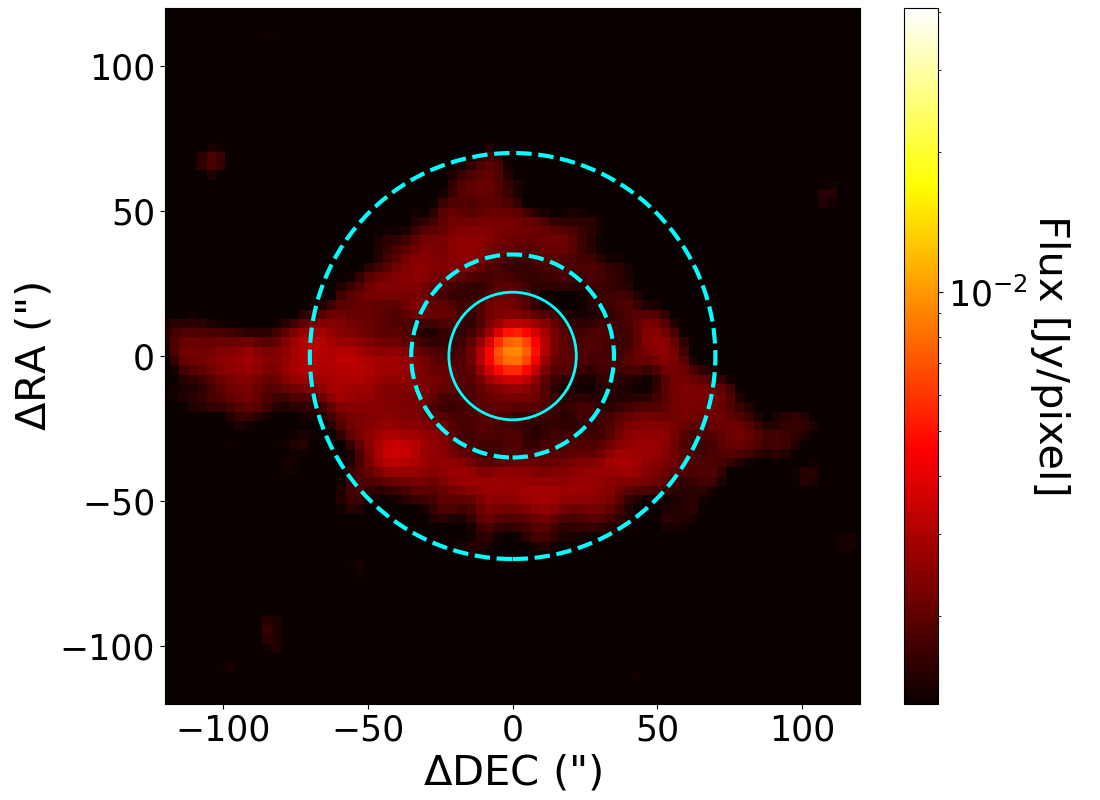}
     \end{subfigure}
     \begin{subfigure}[b]{0.45\linewidth}
         \centering
         \includegraphics[width=\linewidth]{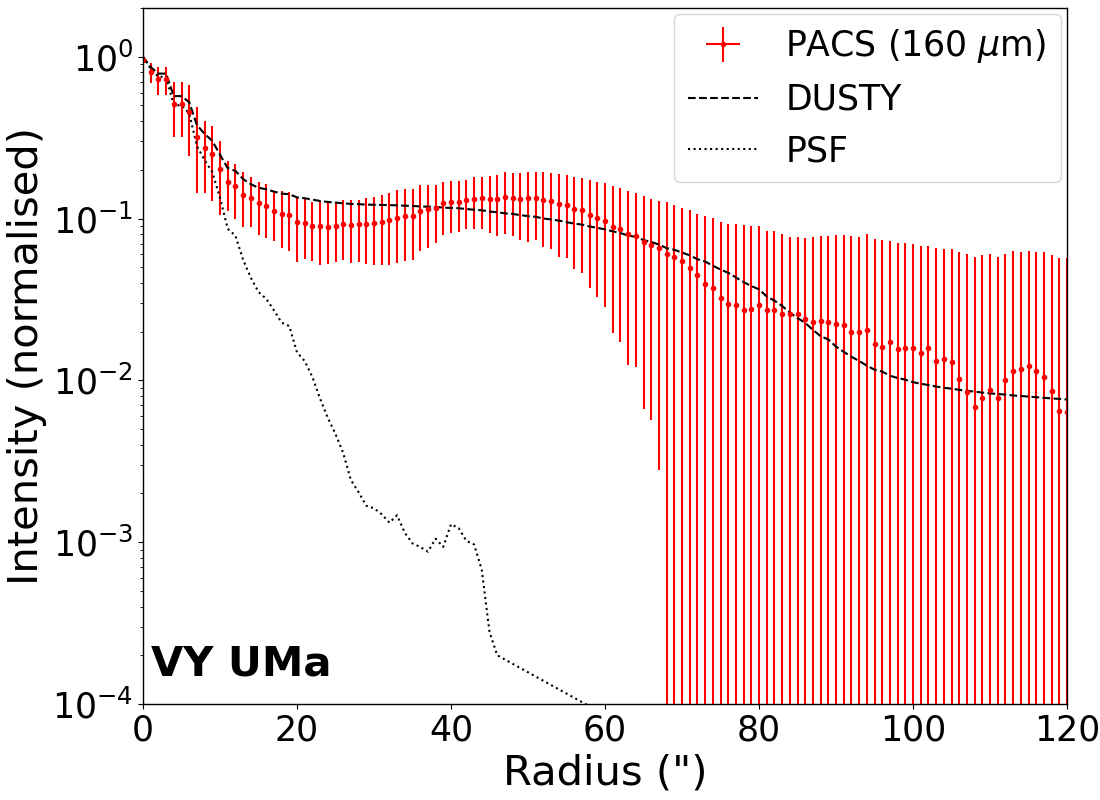}
     \end{subfigure}
        \caption{{\it Left}: {\it Herschel}/PACS images of VY\,UMa. The solid cyan represents the aperture used to isolate the central component (representing the present-day mass-loss), and the dashed cyan lines the aperture used to measure the extended detached shell. {\it Right}: Radial surface brightness profiles, the markers indicate the surface brightness from the images, the dashed line the surface brightness from DUSTY models convolved with their respective {\it Herschel}/PACS PSFs, and the dotted lines the {\it Herschel}/PACS PSFs.}
        \label{fig:VY-UMA_Radial}
\end{figure*}



\section{UV excesses and dust attenuation}\label{UV_opacities}

\begin{table*}[h!]

\renewcommand{\arraystretch}{1.0}
\small
\centering

\caption{UV emission parameters derived from GALEX observations and SED modelling} 
\label{tab:UV_excesses}

\begin{adjustbox}{max width=\textwidth}
\begin{threeparttable}[b]

\begin{tabular}{l >{\centering\arraybackslash}p{1.30cm} >{\centering\arraybackslash}p{1.30cm} >{\centering\arraybackslash}p{1.30cm} >{\centering\arraybackslash}p{1.30cm} >{\centering\arraybackslash}p{1.30cm} >{\centering\arraybackslash}p{1.30cm} >{\centering\arraybackslash}p{1.20cm} >{\centering\arraybackslash}p{1.00cm} >{\centering\arraybackslash}p{1.00cm} >{\centering\arraybackslash}p{1.20cm}}
\hline\hline 
Source & NUV & FUV  & NUV  & FUV  & NUV  & FUV  & $R_{\rm FUV/NUV}$  & $\tau_{\rm NUV}$ & $\tau_{\rm FUV}$ & $R^{\rm corr}_{\rm FUV/NUV}$\\  
 & (Obs.) & (Obs.)  & (SED) & (SED) &  (excess) & (excess) & (Obs.)  &  &  & \\  
    & ($\mu$Jy) & ($\mu$Jy) & ($\mu$Jy) & ($\mu$Jy) &   & ($\times$$10^{3}$) &  & &   \\ 
\hline 

AT\,Dra  & 2190$^{+40}_{-50}$ & 177$^{+13}_{-14}$  &  1700  & 7.6e$\times$$10^{-2}$ & 1.29$^{+0.02}_{-0.03}$ & 2.30$^{+0.17}_{-0.18}$ & 0.080$^{+0.007}_{-0.007}$ & -- & -- & 0.080$^{+0.007}_{-0.007}$\\ 

BC\,CMi$^{(*)}$  & 1013$^{+16}_{-16}$ &  55$^{+5}_{-5}$ &    720 & 2.9$\times$$10^{-2}$ &  1.41$^{+0.02}_{-0.02}$ & 1.90$^{+0.17}_{-0.17}$ & 0.054$^{+0.006}_{-0.005}$ & -- & -- & 0.054$^{+0.006}_{-0.005}$\\ 

BD\,Cam$^{(*)}$  & 5970$^{+40}_{-40}$ & 2330$^{+30}_{-30}$  &  4200  & 3.2$\times$$10^{-1}$ & 1.42$^{+0.01}_{-0.01}$ & 7.28$^{+0.9}_{-0.9}$ & 0.390$^{+0.008}_{-0.007}$ & -- & -- & 0.390$^{+0.008}_{-0.007}$\\ 

BY\,Boo  & 3140$^{+130}_{-110}$ & 214$^{+10}_{-8}$ &   2100  & 8.3$\times$$10^{-2}$ & 1.50$^{+0.06}_{-0.05}$ & 2.58$^{+0.12}_{-0.10}$ & 0.068$^{+0.004}_{-0.005}$ & -- & -- & 0.068$^{+0.004}_{-0.005}$\\ 

CG\,UMa  & 1498$^{+16}_{-30}$ &  108$^{+9}_{-9}$ &  1200  &  4.9$\times$$10^{-2}$ & 1.25$^{+0.01}_{-0.03}$ & 2.20$^{+0.18}_{-0.18}$ & 0.072$^{+0.008}_{-0.007}$ & -- & -- & 0.072$^{+0.008}_{-0.007}$\\ 

DF\,Leo  & 590$^{+90}_{-130}$ &  36$^{+4}_{-8}$ & 1000   & 7.6$\times$$10^{-2}$  & 0.59$^{+0.09}_{-0.13}$ & 0.47$^{+0.05}_{-0.10}$ & 0.061$^{+0.017}_{-0.011}$ & -- & -- & 0.061$^{+0.017}_{-0.011}$\\ 

EY\,Hya & 110$^{+30}_{-30}$ & 82$^{+15}_{-20}$  & 23  & 2.8$\times$$10^{-4}$  &  4.8$^{+1.3}_{-1.3}$ & 290$^{+50}_{-70}$ & 0.740$^{+0.018}_{-0.007}$ & 0.99 & 1.67 & 1.461$^{+0.04}_{-0.014}$\\ 

FH\,Vir  & 400$^{+40}_{-50}$ & 49$^{+16}_{-10}$  &  370  & 1.4$\times$$10^{-2}$ &  1.08$^{+0.11}_{-0.13}$ & 3.5$^{+1.1}_{-0.7}$ & 0.120$^{+0.04}_{-0.010}$ & 0.14 & 0.24 & 0.133$^{+0.04}_{-0.011}$\\ 

IN\,Hya  & 620$^{+40}_{-60}$ & 44$^{+6}_{-6}$ & 1200   & 8.6$\times$$10^{-2}$ &  0.52$^{+0.03}_{-0.05}$ & 0.51$^{+0.7}_{-0.7}$ & 0.071$^{+0.017}_{-0.007}$ & 0.11 & 0.19 & 0.077$^{+0.018}_{-0.008}$\\ 

ome\,Vir$^{(*)}$  & 2950$^{+30}_{-30}$ & 179$^{+12}_{-12}$  &  2100  & 9.3$\times$$10^{-2}$ &  1.40$^{+0.01}_{-0.01}$ & 1.92$^{+0.13}_{-0.13}$ & 0.061$^{+0.004}_{-0.005}$ & -- & -- & 0.061$^{+0.004}_{-0.005}$\\ 

R\,LMi$^{(b)}$  & 40$^{+20}_{-30}$ &  2.4$^{+0.7}_{-0.7}$ & 7.6   & 5.2$\times$$10^{-5}$ & 5$^{+3}_{-4}$  & 46$^{+13}_{-13}$ & 0.040$^{+0.013}_{-0.012}$ & 4.50 & 7.63 & 0.9$^{+0.3}_{-0.3}$\\ 

R\,UMa  & 90$^{+30}_{-40}$ & 50$^{+12}_{-7}$  & 30   & 5.3$\times$$10^{-4}$ & 3.0$^{+1.0}_{-1.3}$ & 94$^{+20}_{-13}$ & 0.6$^{+0.5}_{-0.2}$ & 1.13 & 1.91 & 1.3$^{+1.1}_{-0.4}$\\ 

RR\,Eri  & 430$^{+70}_{-60}$ & 28$^{+4}_{-4}$  &  300  & 8.4$\times$$10^{-3}$ &  1.4$^{+0.2}_{-0.2}$ & 3.3$^{+0.5}_{-0.5}$ & 0.064$^{+0.013}_{-0.016}$ & 0.08 & 0.14 & 0.068$^{+0.014}_{-0.017}$\\ 

RR\,UMi  & 7200$^{+700}_{-800}$ & 1870$^{+90}_{-110}$ & 4100 & 1.8$\times$$10^{-1}$ & 1.76$^{+0.17}_{-0.2}$ & 1.04$^{+0.05}_{-0.06}$ & 0.26$^{+0.04}_{-0.02}$ & -- & -- & 0.26$^{+0.04}_{-0.02}$\\ 

RT\,Cnc  & 190$^{+30}_{-30}$ & 32$^{+14}_{-15}$  &  170  & 3.7$\times$$10^{-3}$ &  1.12$^{+0.18}_{-0.18}$ & 9$^{+4}_{-4}$ & 0.19$^{+0.09}_{-0.05}$ & 0.76 & 1.29 & 0.32$^{+0.15}_{-0.08}$\\ 

RU\,Her$^{(b)}$  & 80$^{+8}_{-6}$ & $<$5.75  & 27 & 5.8$\times$$10^{-4}$ & 3.0$^{+0.3}_{-0.2}$ & $<$10 & $<$0.08 & 0.56 & 0.95 & $<$0.12\\ 

RW\,Boo  & 410$^{+50}_{-60}$ & 23$^{+8}_{-11}$  &  180  & 5.3$\times$$10^{-3}$ & 2.3$^{+0.3}_{-0.3}$ & 4.3$^{+1.5}_{-2}$ & 0.057$^{+0.012}_{-0.03}$ & 0.17 & 0.29 & 0.064$^{+0.014}_{-0.03}$\\ 

RZ\,UMa$^{(b)}$  & 18$^{+6}_{-3}$ & 8$^{+4}_{-4}$  &  22  & 3.6$\times$$10^{-4}$ & 0.82$^{+0.3}_{-0.14}$  & 22$^{+11}_{-11}$ & 0.4$^{+0.3}_{-0.2}$ & 0.51 & 0.86 & 0.6$^{+0.4}_{-0.3}$\\ 

ST\,UMa$^{(*)}$ & 994$^{+14}_{-14}$ & 168$^{+13}_{-13}$  & 680 & 2.7$\times$$10^{-2}$ & 1.46$^{+0.02}_{-0.02}$  & 6.2$^{+0.5}_{-0.5}$ & 0.17$^{+0.01}_{-0.02}$ & 0.25 & 0.43 & 0.20$^{+0.01}_{-0.02}$\\ 

SV\,Peg$^{(b)}$  & 29$^{+5}_{-1}$ &  $<$5.75 &  39  & 4.6$\times$$10^{-4}$ & 0.74$^{+0.13}_{-0.03}$ & $<$13 & $<$0.21 & 0.55 & 0.91 & $<$0.30\\ 

T\,Dra$^{(a, b)}$  & 32$^{+3}_{-3}$ & 7.9$^{+1.1}_{-1.1}$  &  2.6  & 4.2$\times$$10^{-5}$ & 12.3$^{+1.2}_{-1.2}$ & 190$^{+30}_{-30}$ & 0.25$^{+0.06}_{-0.06}$ &  7.76 & 8.05 & 0.33$^{+0.08}_{-0.08}$\\ 

TU\,And$^{(b)}$  & 15$^{+9}_{-9}$ &  $<$5.75 &  19  & 4.3$\times$$10^{-4}$ & 0.8$^{+0.5}_{-0.5}$  & $<$13 & $<$0.96 &  0.98 & 1.67 & $<$1.91\\ 

UY\,Leo  & 150$^{+20}_{-30}$ & 27$^{+5}_{-8}$  &  29  & 7.6$\times$$10^{-4}$ &  5.2$^{+0.7}_{-1.0}$ & 36$^{+7}_{-11}$ & 0.18$^{+0.05}_{-0.04}$ &  0.56 & 0.95  & 0.27$^{+0.07}_{-0.06}$\\ 

V\,Eri  & 100$^{+50}_{-50}$ &  67$^{+15}_{-30}$ & 34   & 4.0$\times$$10^{-4}$  &  2.9$^{+1.5}_{-1.5}$ & 170$^{+40}_{-80}$ & 0.67$^{+0.6}_{-0.14}$ &  1.41 & 2.39 & 1.8$^{+1.6}_{-0.4}$\\ 

VY\,UMa  & 140$^{+100}_{-70}$ & 8$^{+6}_{-3}$  &  110  & 3.0$\times$$10^{-3}$ &  1.3$^{+0.9}_{-0.6}$ & 2.7$^{+2}_{-1.0}$ & 0.06$^{+0.10}_{-0.04}$ &  0.69 & 0.72 & 0.06$^{+0.10}_{-0.04}$\\ 

W\,Peg  & 180$^{+30}_{-30}$ &  13$^{+3}_{-3}$ & 41 & 5.8$\times$$10^{-4}$ & 4.4$^{+0.7}_{-0.7}$  & 22$^{+5}_{-5}$ & 0.065$^{+0.017}_{-0.017}$ &  0.98 & 1.67 & 0.13$^{+0.03}_{-0.03}$\\ 

Y\,CrB$^{(b)}$  & 12$^{+7}_{-4}$ & $<$5.75  &  16  & 3.3$\times$$10^{-4}$ & 0.8$^{+0.4}_{-0.3}$ & $<$17 & $<$0.67 &  0.70 & 1.19 & $<$1.09 \\ 

Y\,Gem  & 6000$^{+6000}_{-4000}$ & 5000$^{+6000}_{-4000}$  &4.6  & 1.0$\times$$10^{-3}$  & 1300$^{+1300}_{-900}$ & 5000$^{+6000}_{-4000}$ & 0.86$^{+0.10}_{-0.3}$ &  0.08 & 0.14 & 0.91$^{+0.11}_{-0.3}$\\ 

Z\,Cnc  & 73$^{+4}_{-6}$ & 12$^{+3}_{-4}$  & 9.2   & 2.4$\times$$10^{-3}$ & 7.9$^{+0.4}_{-0.6}$ & 5.0$^{+1.3}_{-1.7}$ & 0.17$^{+0.04}_{-0.07}$ &  0.28 & 0.48 & 0.21$^{+0.05}_{-0.09}$\\ 

\hline 

\end{tabular} 
\begin{tablenotes}
\item \normalsize \textbf{Notes.} Column (1): Source, Col. (2): Averaged observed NUV flux, Col. (3): Averaged observed FUV flux, Col. (4): NUV flux expected from SED modelling, Col. (5): FUV flux expected from SED modelling, Col. (6): NUV excess, Col. (7): FUV excess, Col. (8): Averaged ratio between FUV and NUV fluxes, Col. (9): Optical depth at NUV effective wavelength, Col. (10): Optical depth at FUV effective wavelength, Col. (11): Averaged ratio between FUV and NUV fluxes corrected from dust attenuation.
\end{tablenotes}

\end{threeparttable}
\end{adjustbox}

\renewcommand{\arraystretch}{1.0}

\end{table*}

We computed synthetic stellar photometry on the GALEX bands and compared them with the observed fluxes to estimate the UV excesses \citep[similarly to][]{Sahai_2008, Ortiz_2016}. We noted that GALEX NUV and FUV photometric bands fall outside the wavelength range covered by the COMARCS models used in this work (see Sect.~\ref{modelling}). Therefore, the stellar fluxes were estimated by extrapolating the input spectra following a blackbody distribution and convolving it with the transmission curves of their respective filters. Table~\ref{tab:UV_excesses} shows the observed and modelled properties of the UV emission for the sources used in this sample.

We acknowledge that the real stellar spectrum differs from a blackbody approximation mainly due to the presence of (i) molecular absorption bands, and (ii) emission-line features due to possible chromospheric emission, that can produce a significant effect on the shape of the stellar spectra in the UV. The features in (ii) are expected to have a much stronger effect in the FUV (compared to the NUV) where the photospheric contribution is relatively much weaker. Note that the sole purpose of this simple approximation is to explore whether $R_{\rm FUV/NUV}$ is a reasonable proxy for the FUV excess. 

\begin{figure}[h!]
    \centering
    \includegraphics[width=1.0\linewidth]{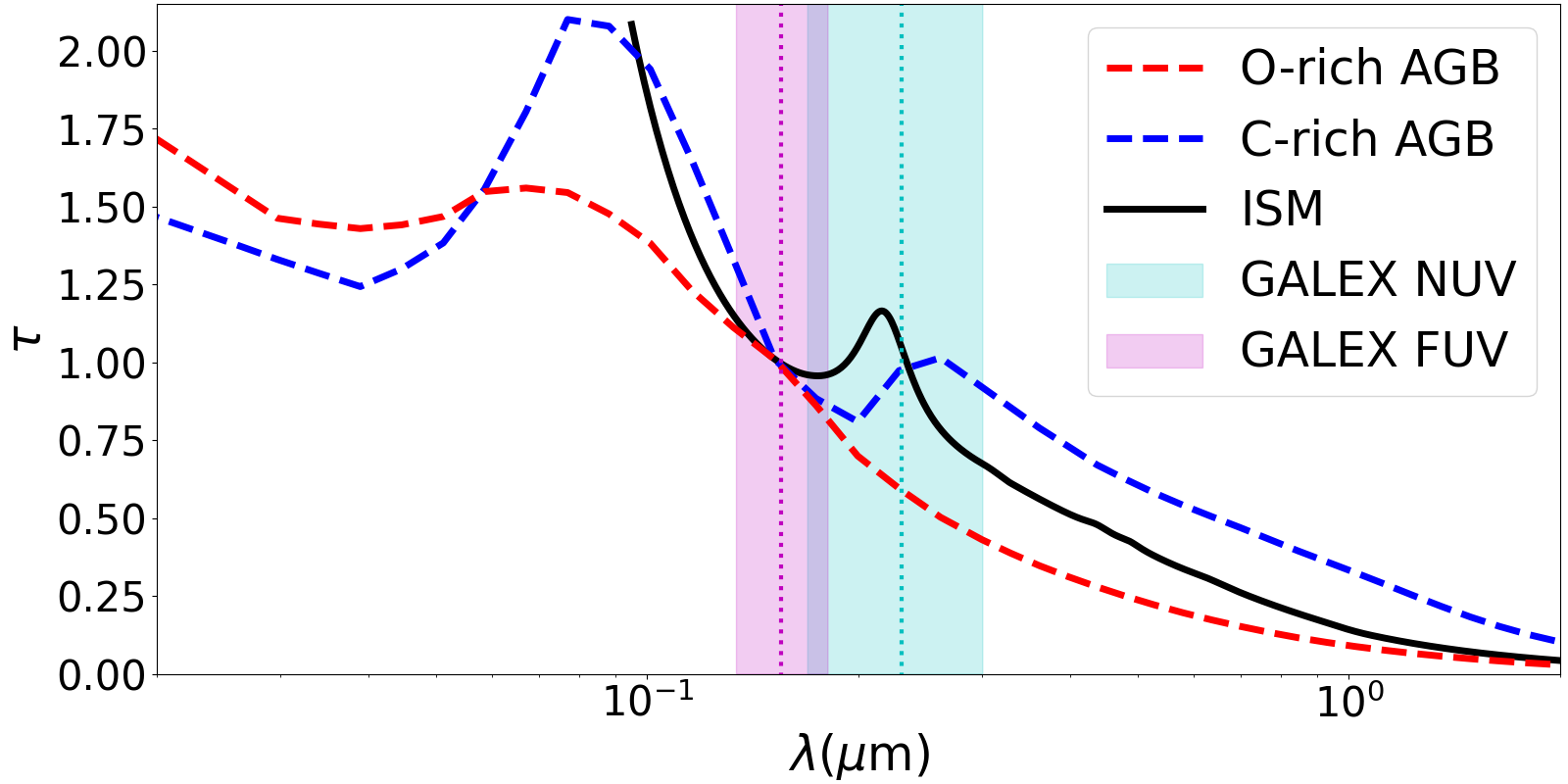}
    \caption{Dashed red: dust extinction curve for O-rich AGB stars (provided by DUSTY). Dashed blue: dust extinction curve for C-rich AGB stars (provided by DUSTY). Solid black: ISM extinction law from \cite{Gordon_2023}. The effective wavelengths and filter widths of the GALEX bands are represented as dotted lines and shadowed areas (cyan and magenta respectively). The three curves are scaled to match in the GALEX FUV band.}
    \label{fig:DUSTY_opacities}
\end{figure}

\begin{figure*}[h!]
     \centering
     \begin{subfigure}[b]{0.45\linewidth}
         \centering
         \includegraphics[width=\linewidth]{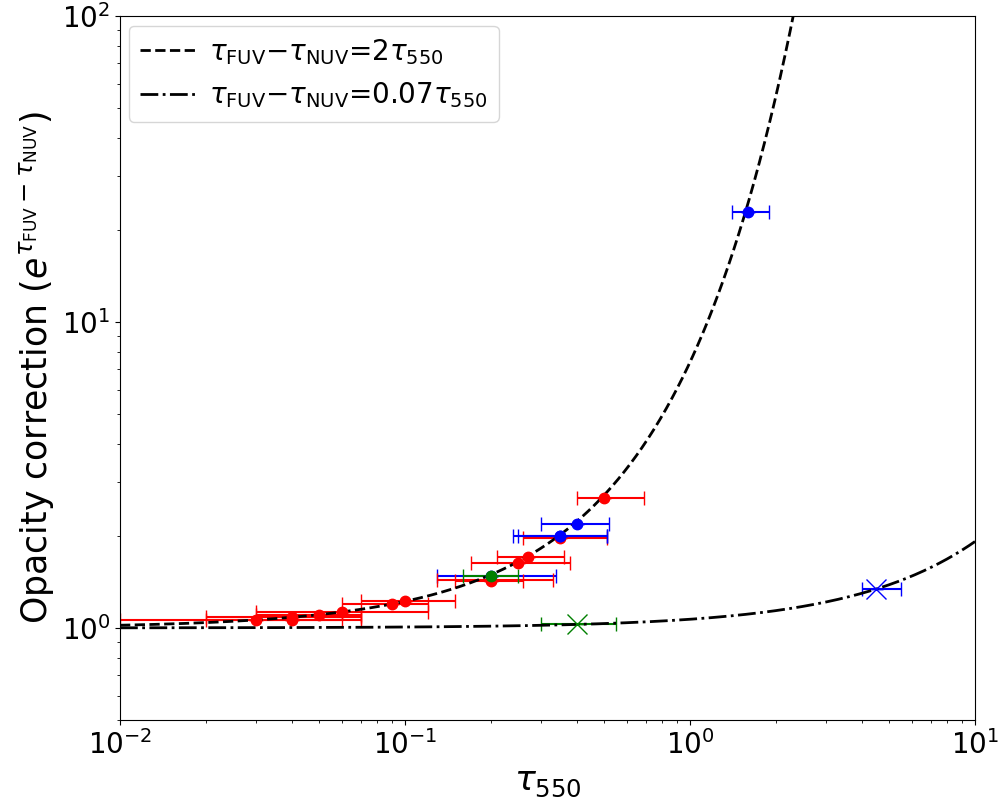}      
     \end{subfigure}
     \begin{subfigure}[b]{0.45\linewidth}
         \centering
         \includegraphics[width=\linewidth]{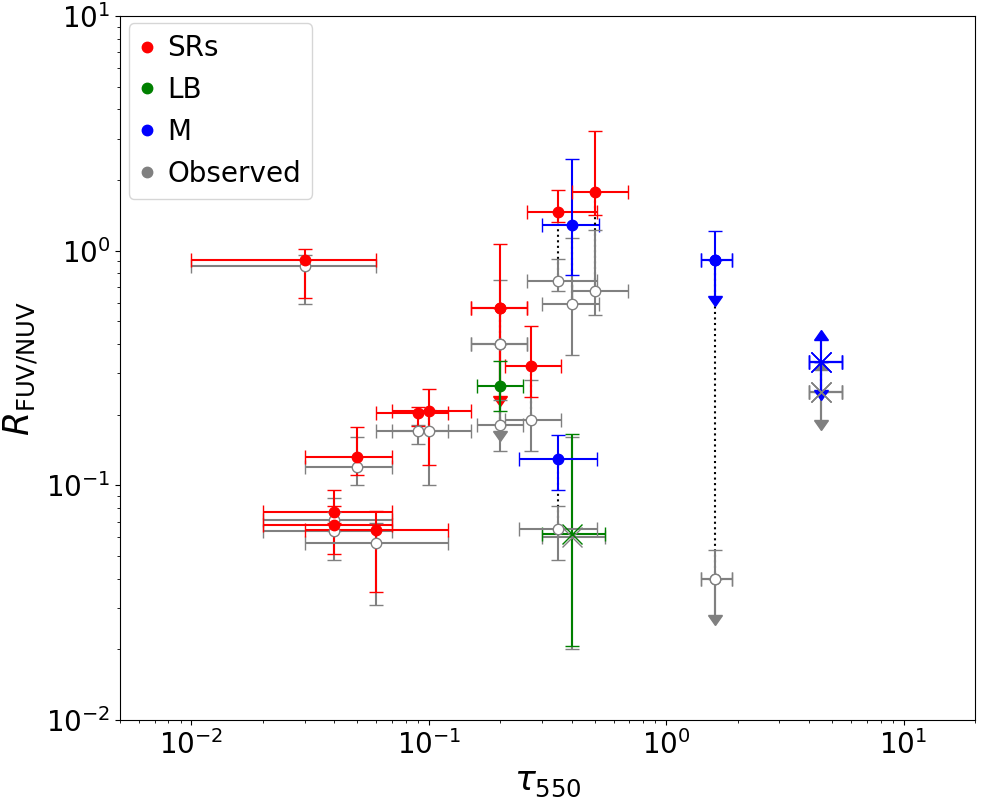}
     \end{subfigure}
        \caption{{\it Left}: Comparison between the opacity correction factor for the $R_{\rm FUV/NUV}$ ratio and the opacity at the fiducial wavelength. The dashed and dash-dotted lines represents the correlations of $e^{(\tau_{\rm FUV}-\tau_{\rm NUV})}$$\simeq$2$\tau_{\rm 550}$ for O-rich AGB stars and $e^{(\tau_{\rm FUV}-\tau_{\rm NUV})}$$\simeq$0.025$\tau_{\rm 550}$ for C-rich AGB stars respectively. {\it Right}: Comparison between $R_{\rm FUV/NUV}$ and the opacity at the fiducial wavelength. The colour and shape of the markers represent the stellar variability and chemistry type as in Fig.~\ref{fig:HR_diagram}. Grey markers are those without correction from dust attenuation. The dotted lines link the markers for the same sources before and after the dust attenuation correction.}
        \label{fig:opacity_vs_tau}
\end{figure*}

\begin{figure*}[h!]
     \centering
     \begin{subfigure}[b]{0.45\linewidth}
         \centering
         \includegraphics[width=\linewidth]{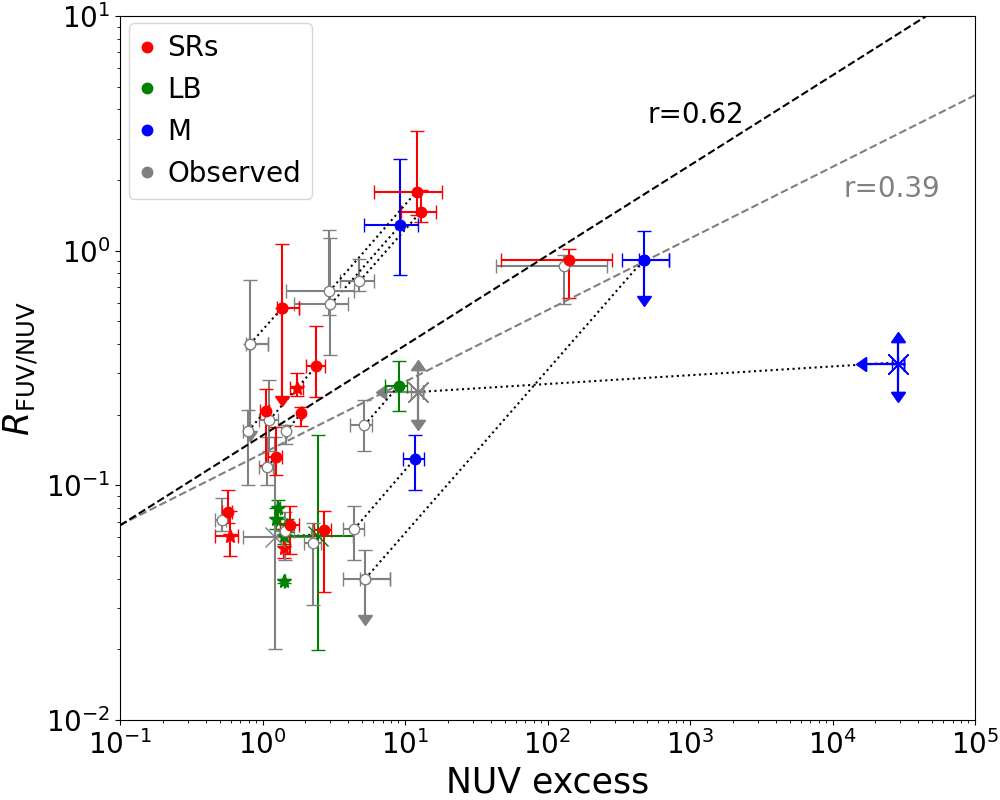}
     \end{subfigure}
     \begin{subfigure}[b]{0.45\linewidth}
         \centering
         \includegraphics[width=\linewidth]{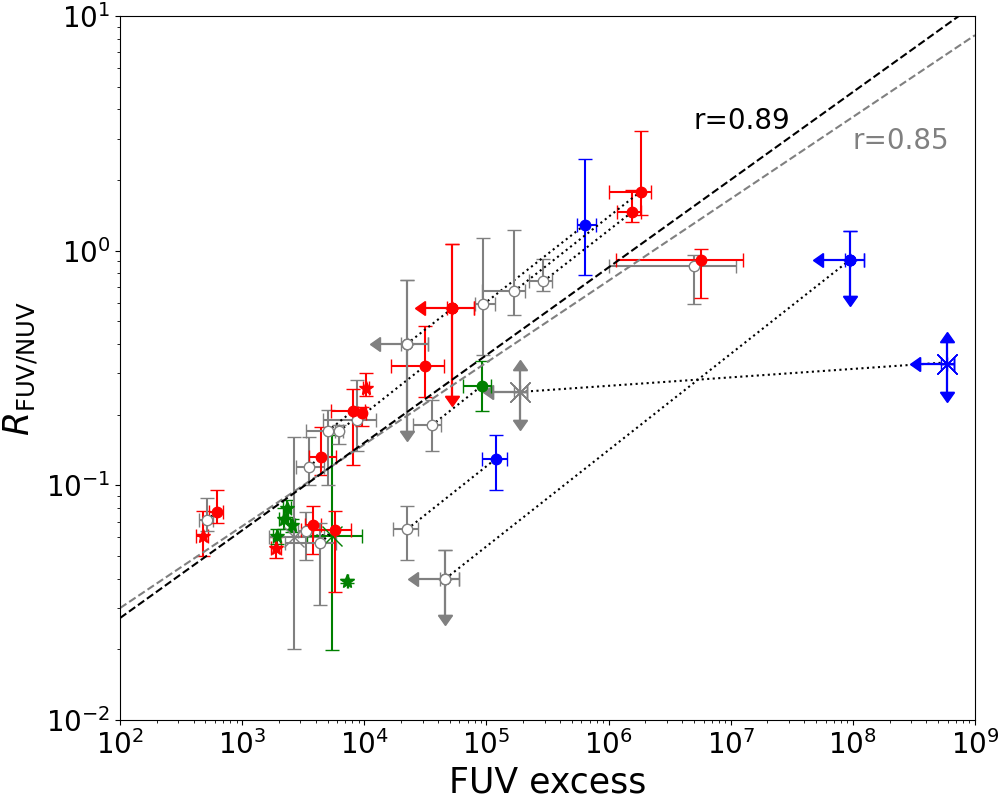}
     \end{subfigure}
        \caption{Correlations between $R_{\rm FUV/NUV}$ and the NUV and FUV excesses ({\it left} and {\it right} respectively). The colour and shape of the markers represent the stellar variability and chemistry type as in Fig.~\ref{fig:HR_diagram}. Grey markers are those without correction from dust attenuation. The dotted lines link the markers for the same sources before and after the dust attenuation correction. The dashed lines represents linear fits to the points.}
        \label{fig:UV_excesses}
\end{figure*}

In addition, the UV spectral range is specially affected by circumstellar dust attenuation, which might imply a large difference between the intrinsic and the observed UV emission, therefore, affecting the estimation of the UV excesses. Moreover, the wavelength dependency of the opacity produces a variation of the $R_{\rm FUV/NUV}$.

Fig.~\ref{fig:DUSTY_opacities} shows the opacity curves for the generic dust composition of C-rich AGB stars and O-rich AGB stars described in Sect.~\ref{DUSTY} provided by DUSTY as well as the ISM from \cite{Gordon_2023}. In O-rich AGB stars the dust opacity is dominated by silicates and oxides, resulting in $\tau_{\rm FUV} \simeq 1.7 \tau_{\rm NUV}$. In C-rich AGB stars the dust opacity is dominated by amorphous carbon, which shows the $\sim$2200\AA\, NUV bump, resulting in $\tau_{\rm FUV} \simeq 1.02 \tau_{\rm NUV}$. Finally, the opacity curve of the ISM also displays the NUV bump, although with a different overall shape, resulting in $\tau_{\rm FUV} \simeq 0.95 \tau_{\rm NUV}$.

The observed UV fluxes were corrected for the dust attenuation in order to estimate their intrinsic values, assuming that the UV source is located inside the dust envelope and, therefore, the flux is attenuated by the full radial optical depth of the dust envelope. This assumption is reasonable considering that the UV emission is expected to be produced by the presence of an interacting close stellar companion, located at a few stellar radii and inside the dust formation zone.

Fig.~\ref{fig:opacity_vs_tau} shows the dependence of the opacity correction factor with the optical depth and its effect on $R_{\rm FUV/NUV}$. The optical depth is larger in the FUV band than in the NUV band, leading to intrinsic $R_{\rm FUV/NUV}$ larger than observed. Furthermore, the opacity correction factor $e^{(\tau_{\rm FUV}-\tau_{\rm NUV})}$ increases exponentially with the optical depth. The effect on O-rich AGB stars is significant for large opacities, whereas in C-rich AGB stars the effect remains very small in this opacity range, due to the differences between the opacity on NUV and FUV bands.

Fig~\ref{fig:UV_excesses} shows the comparison between $R_{\rm FUV/NUV}$ and the UV excesses. It also shows linear fits between these parameters. There is a weak correlation ($r$=0.39) between $R_{\rm FUV/NUV}$ and the NUV excess, which becomes stronger ($r$=0.62) after correcting for the dust attenuation. In contrast, the correlation between $R_{\rm FUV/NUV}$ and the FUV excess is strong ($r$=0.85) and becomes even stronger ($r$=0.89) after the correction. This correlation shows that $R_{\rm FUV/NUV}$ is a good tracer of UV excesses and can be used independently to explore trends.

This result is expected because small UV excesses indicate a larger contribution of the stellar component to $R_{\rm FUV/NUV}$, and stellar photospheric emission is characterised by low $R_{\rm FUV/NUV}$. As described in Sect.~\ref{intro}, both UV excesses and $R_{\rm FUV/NUV}$ are related to the nature of the UV emission: small UV excesses and $R_{\rm FUV/NUV}$ indicate an intrinsic origin (i.e. chromospheric activity), whereas large UV excesses and $R_{\rm FUV/NUV}$ indicate an extrinsic origin (i.e. accretion activity). We highlight that these correlations can be seen despite the scatter produced by UV emission variability, and they become more pronounced when corrected for dust attenuation.

\section{SEDs and best-fit models for the rest of dusty AGB stars}\label{SED_appendix}

\begin{figure*}[ h!]

     \begin{subfigure}[b]{0.245\linewidth}
         \centering
         \includegraphics[width=\linewidth]{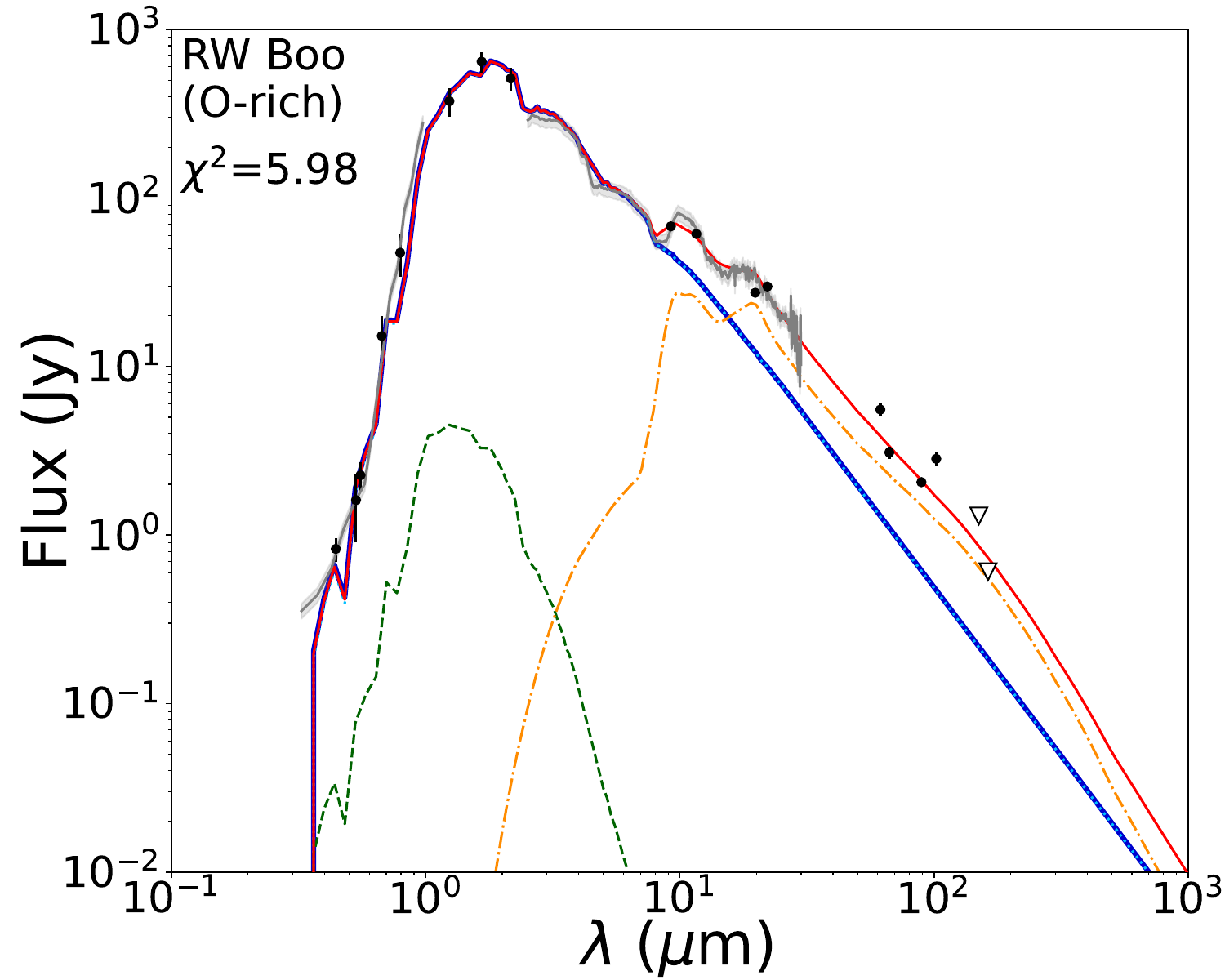}
     \end{subfigure}
     \begin{subfigure}[b]{0.245\linewidth}
         \centering
         \includegraphics[width=\linewidth]{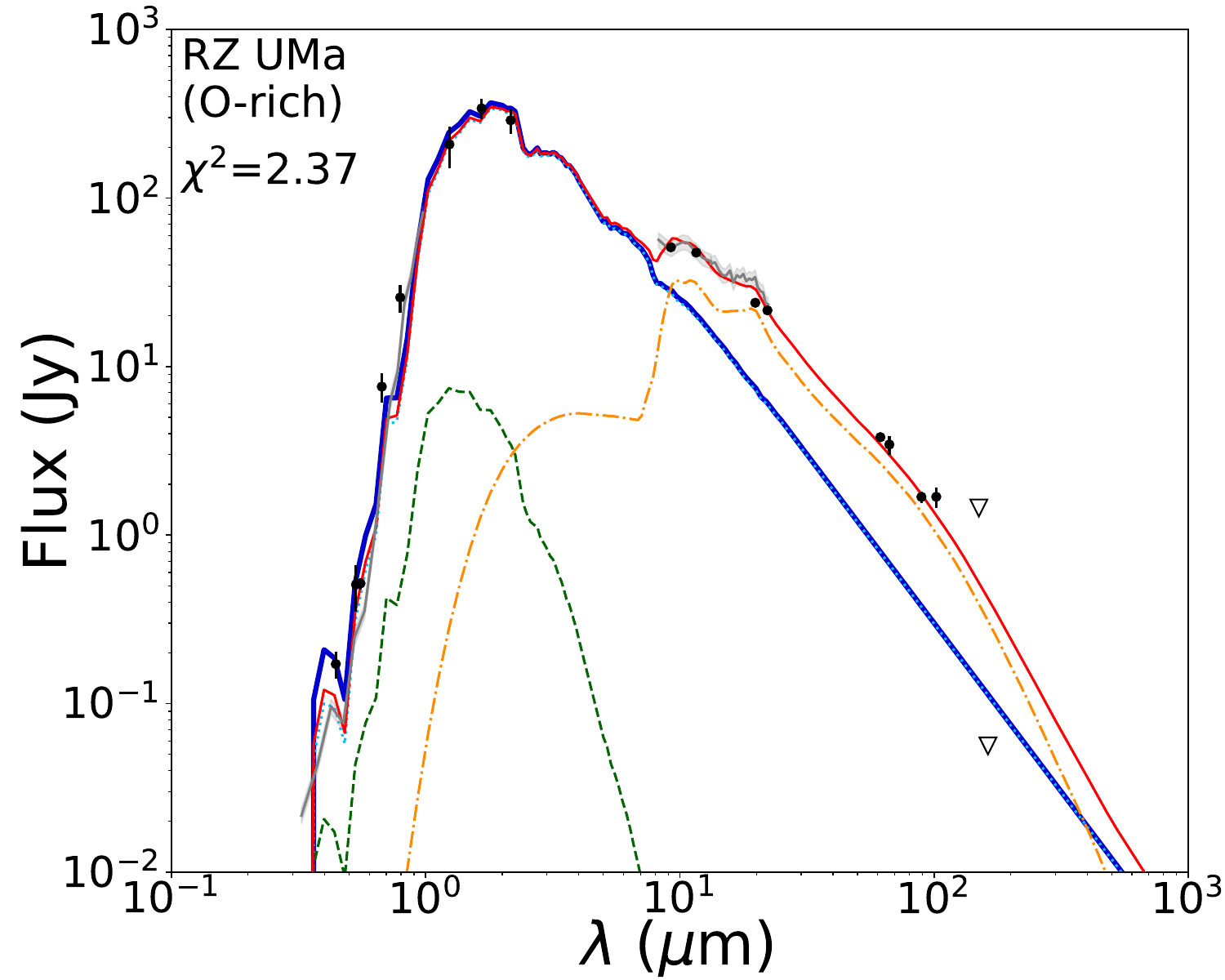}
     \end{subfigure}
     \begin{subfigure}[b]{0.245\linewidth}
         \centering
         \includegraphics[width=\linewidth]{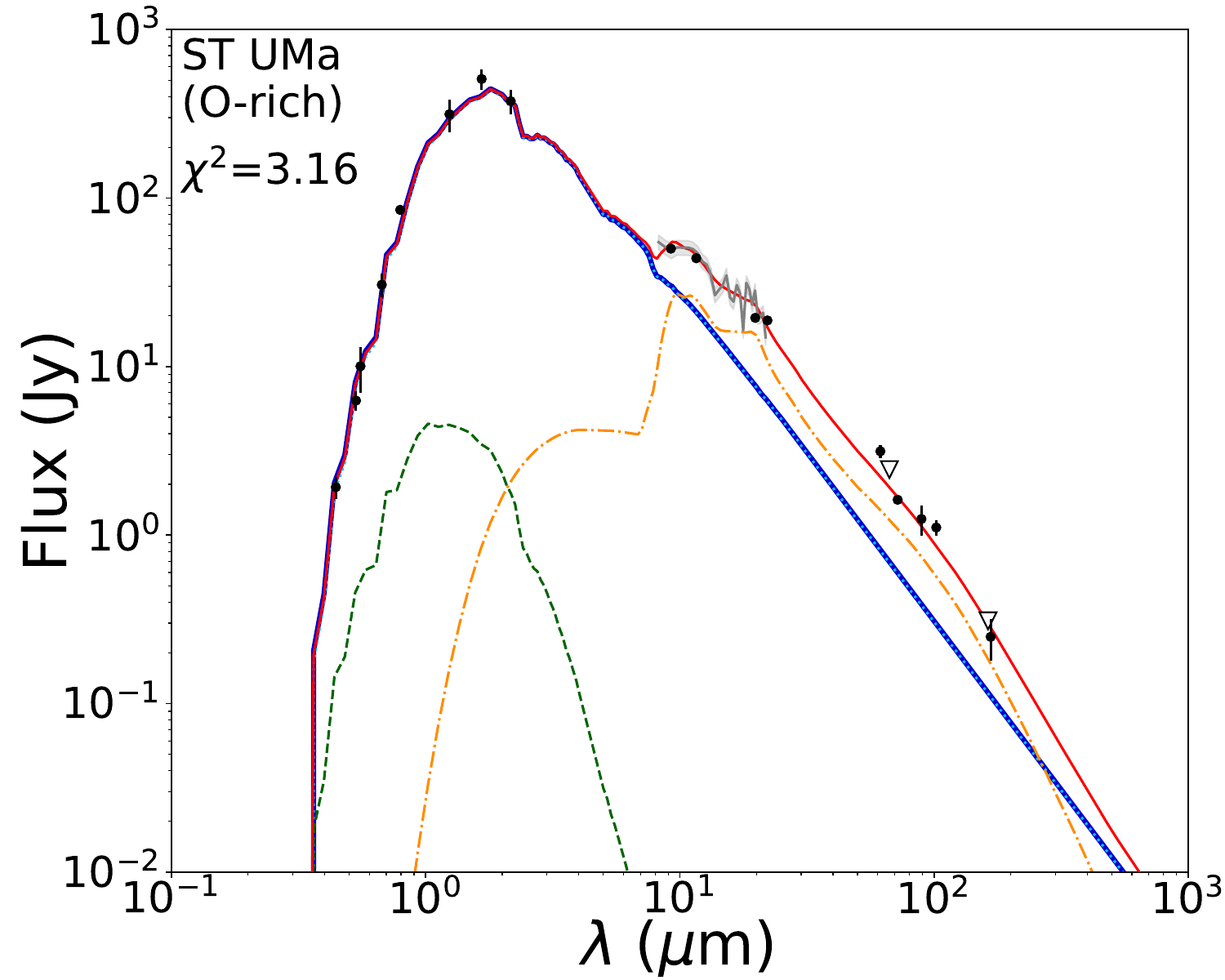}
     \end{subfigure}
     \begin{subfigure}[b]{0.245\linewidth}
         \centering
         \includegraphics[width=\linewidth]{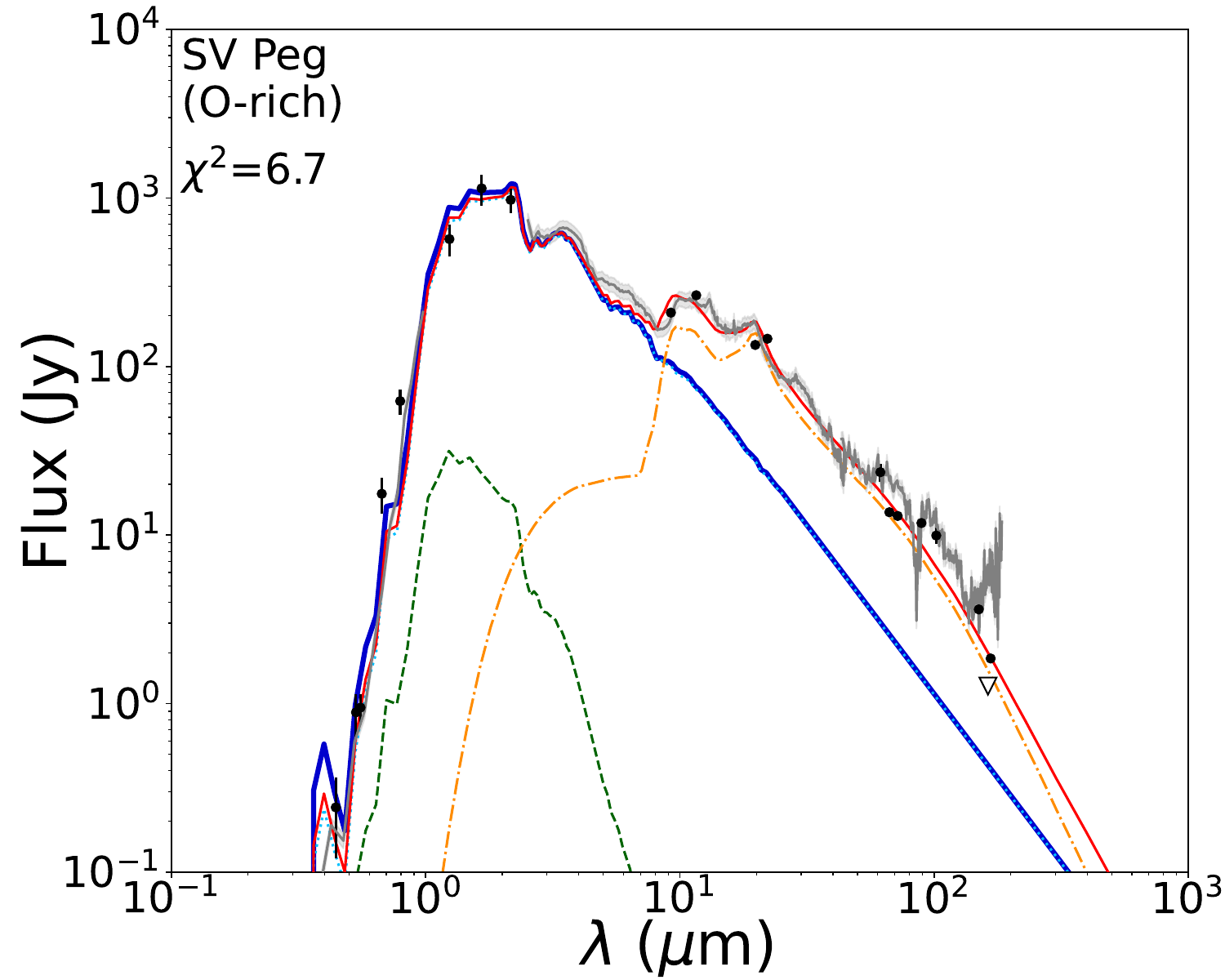}
     \end{subfigure}

     \begin{subfigure}[b]{0.245\linewidth}
         \centering
         \includegraphics[width=\linewidth]{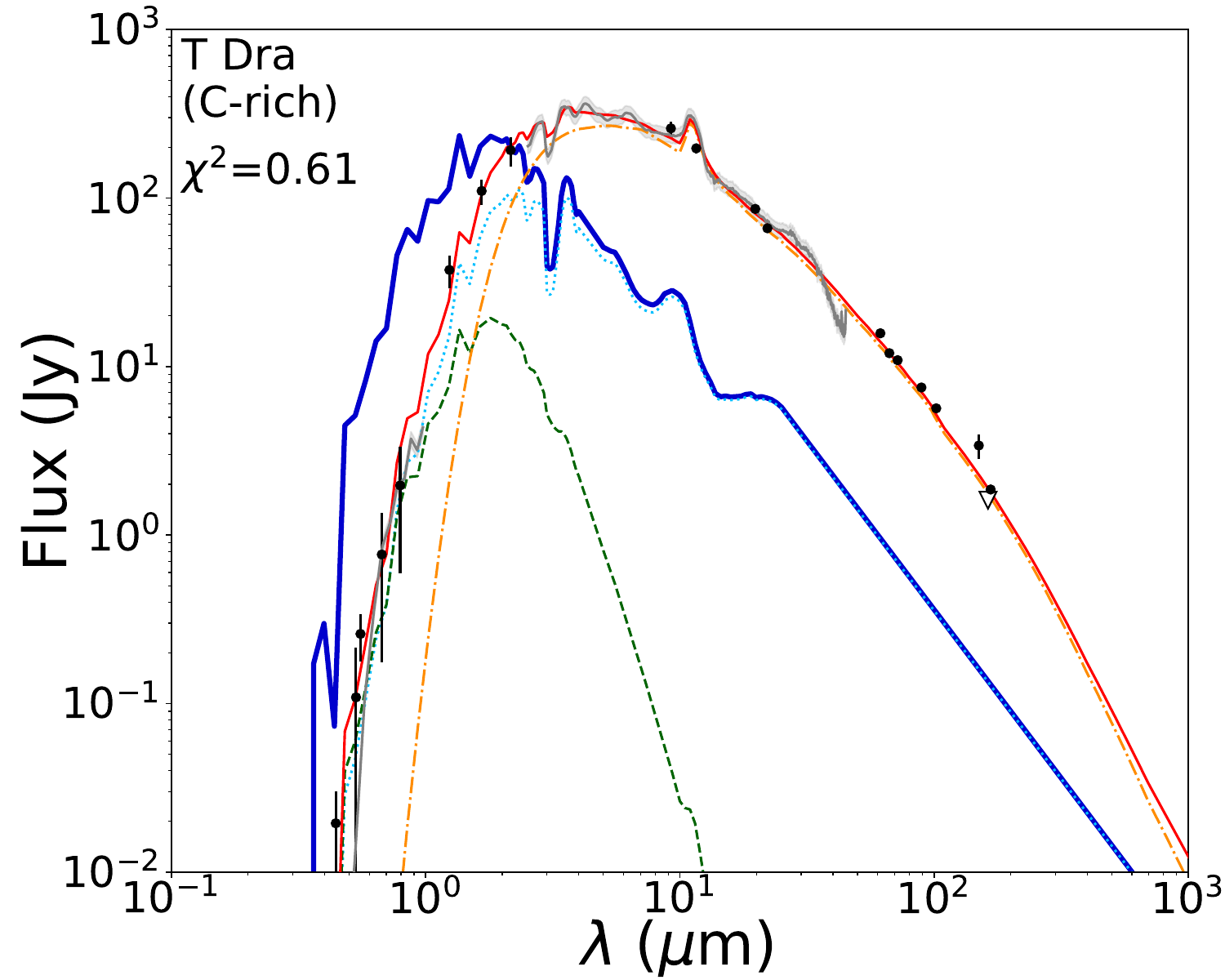}
     \end{subfigure}
     \begin{subfigure}[b]{0.245\linewidth}
         \centering
         \includegraphics[width=\linewidth]{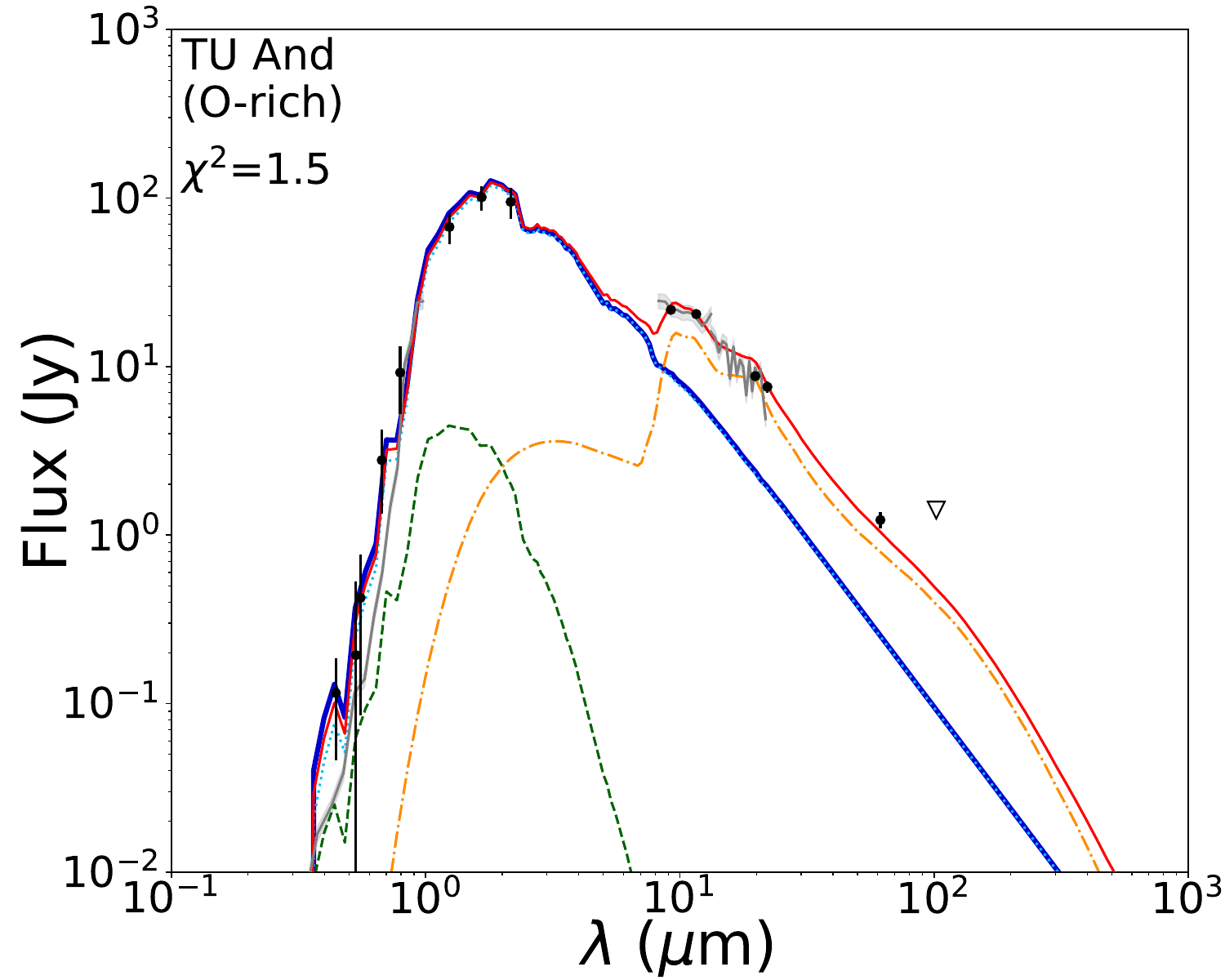}
     \end{subfigure}
     \begin{subfigure}[b]{0.245\linewidth}
         \centering
         \includegraphics[width=\linewidth]{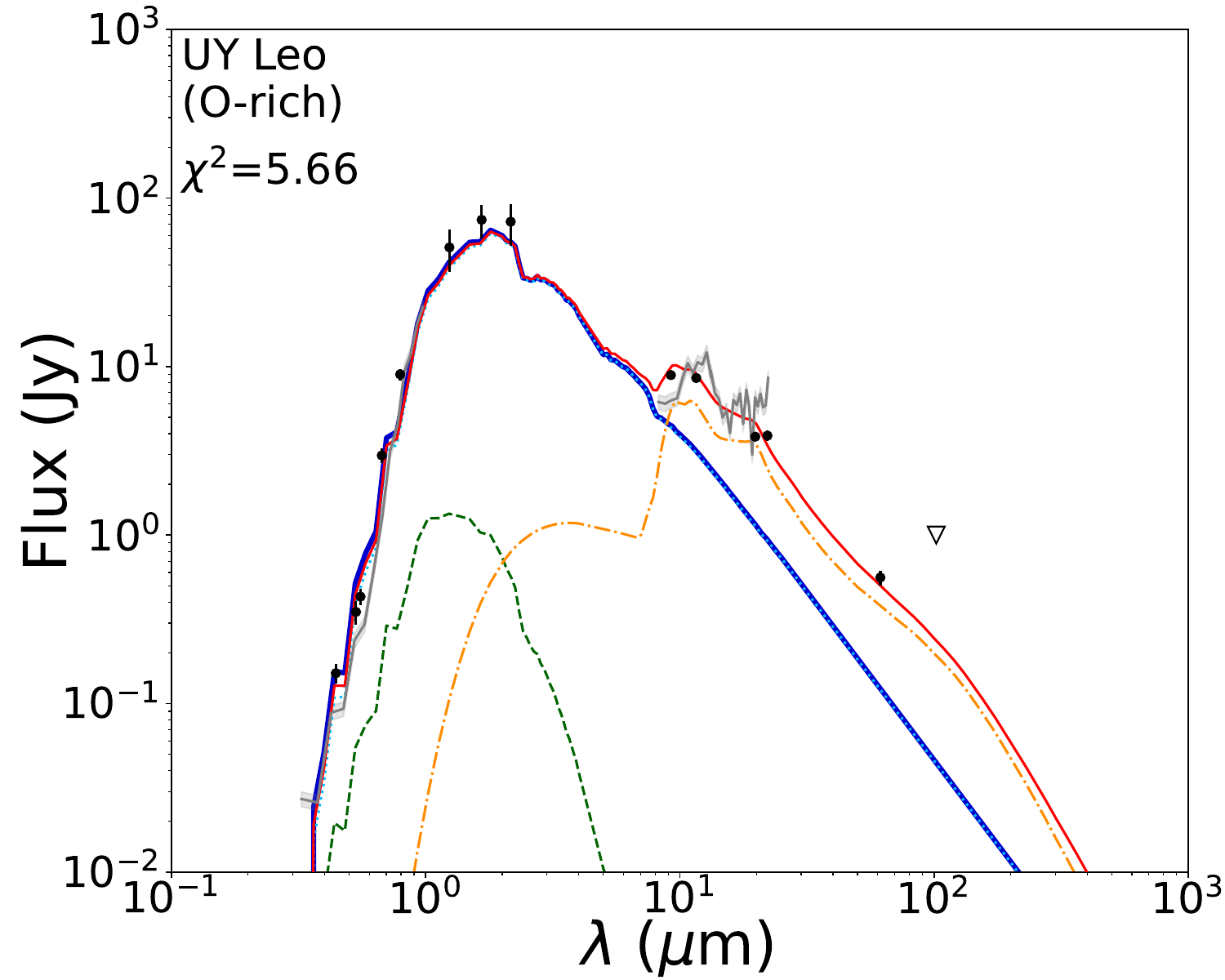}
     \end{subfigure}
     \begin{subfigure}[b]{0.245\linewidth}
         \centering
         \includegraphics[width=\linewidth]{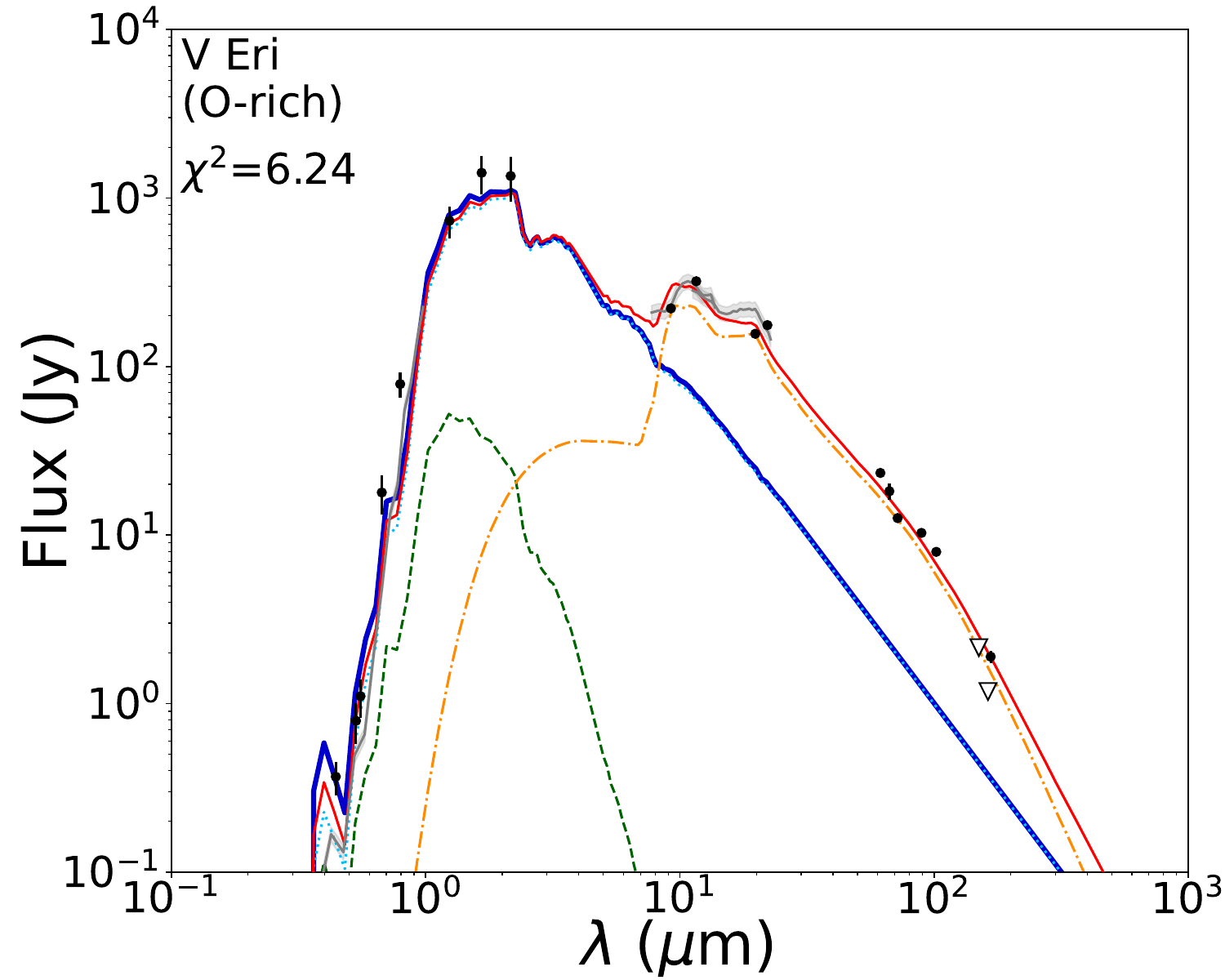}
     \end{subfigure}
     
     \begin{subfigure}[b]{0.245\linewidth}
         \centering
         \includegraphics[width=\linewidth]{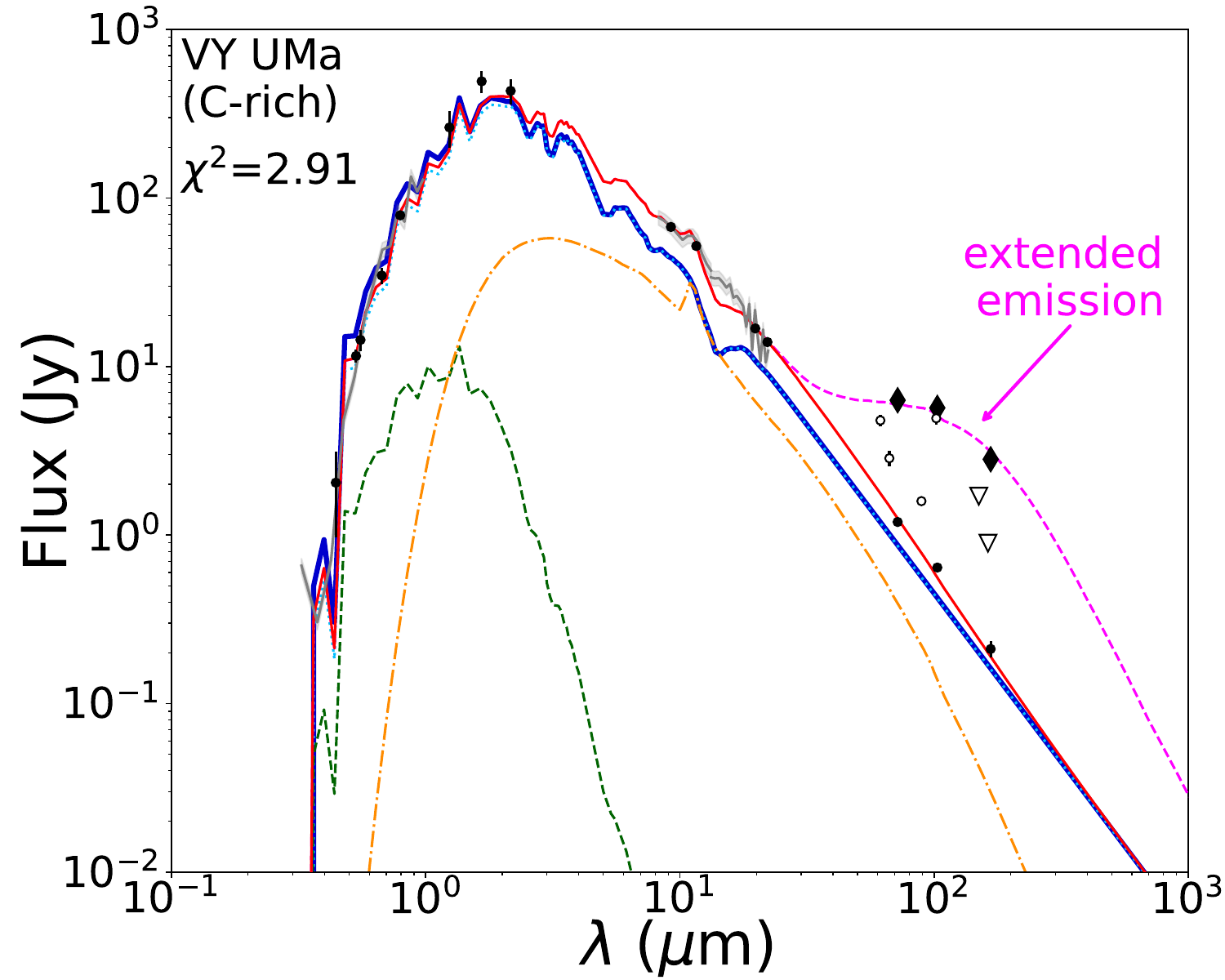}
     \end{subfigure}
     \begin{subfigure}[b]{0.245\linewidth}
         \centering
         \includegraphics[width=\linewidth]{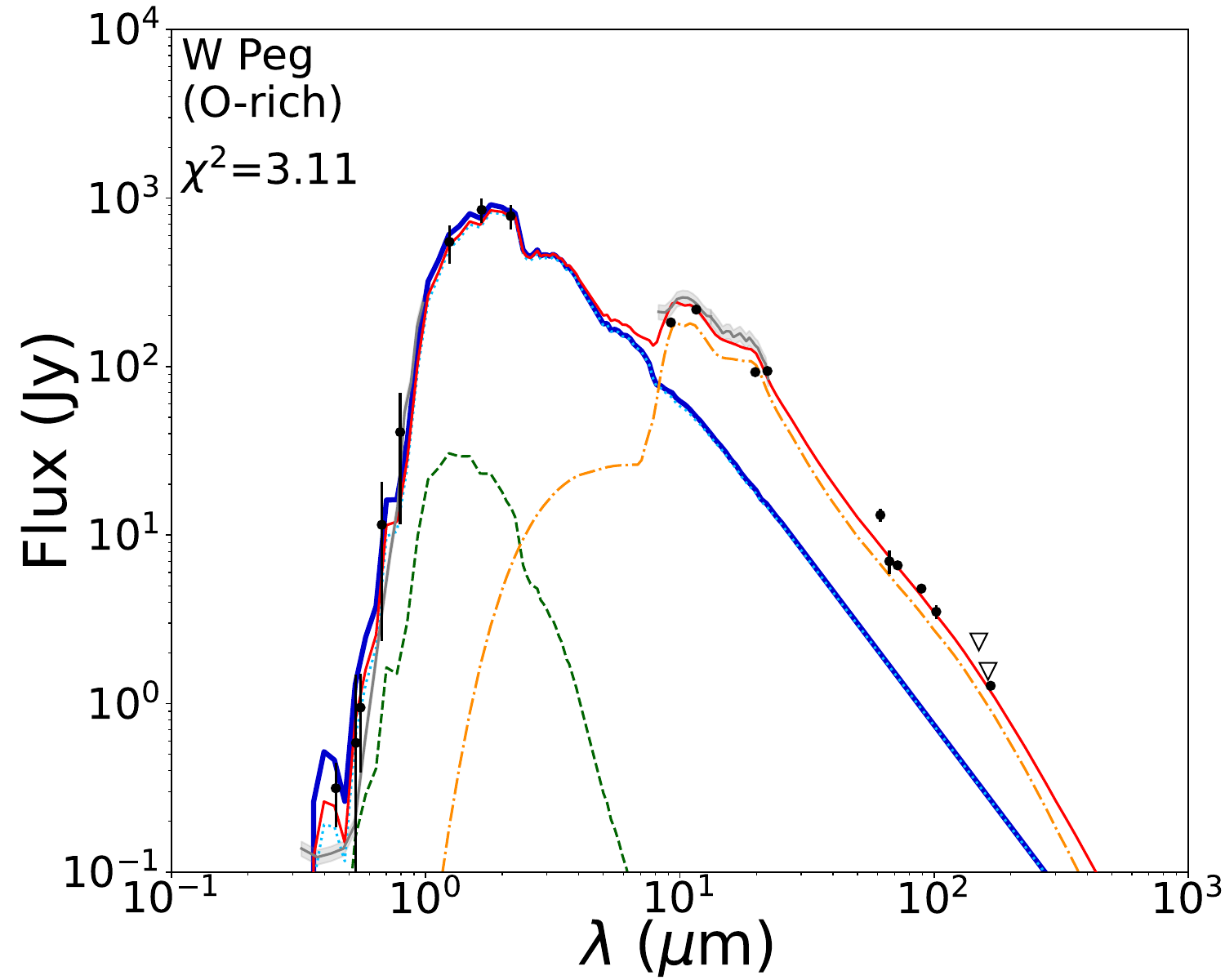}
     \end{subfigure}
     \begin{subfigure}[b]{0.245\linewidth}
         \centering
         \includegraphics[width=\linewidth]{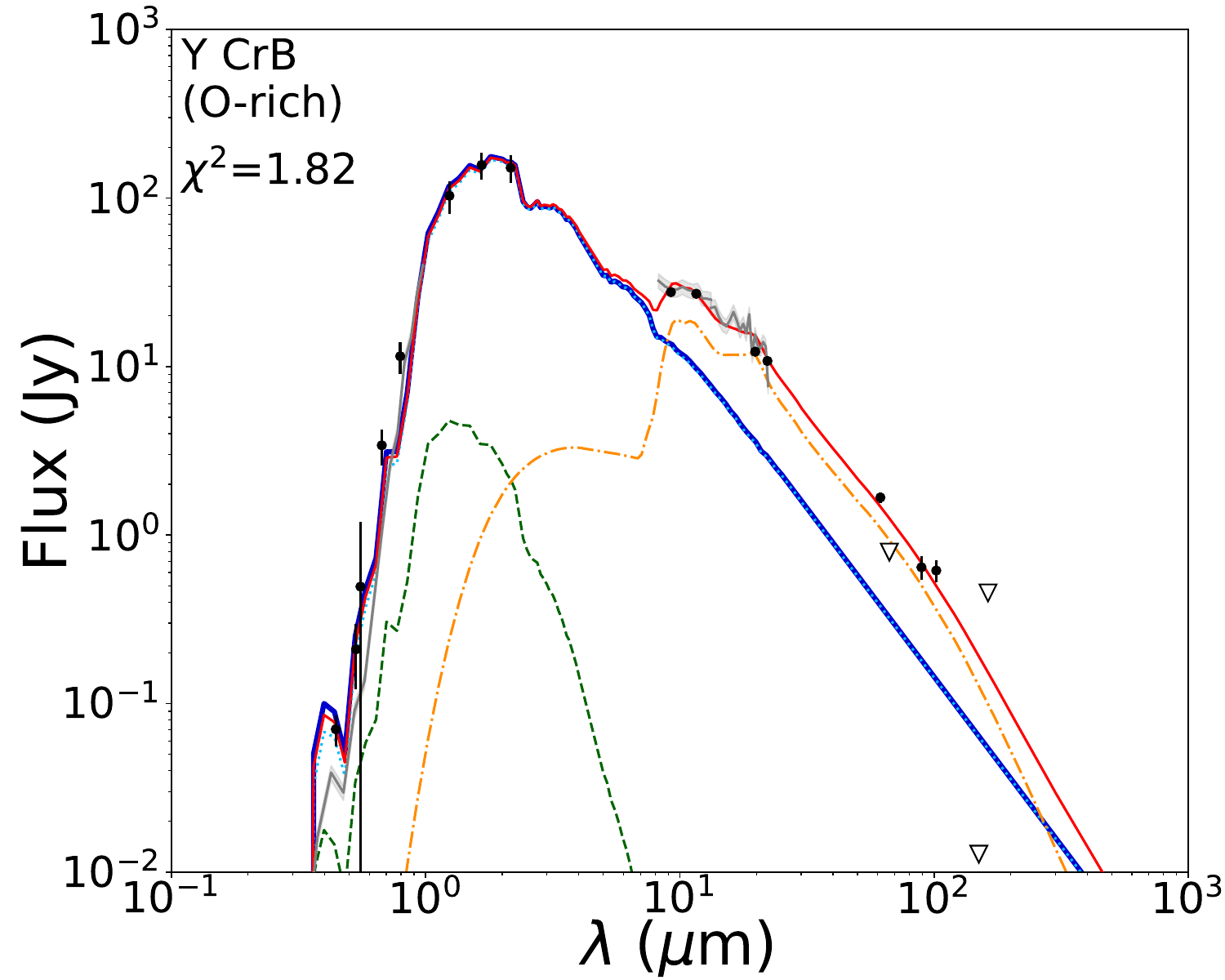}
     \end{subfigure}
     \begin{subfigure}[b]{0.245\linewidth}
         \centering
         \includegraphics[width=\linewidth]{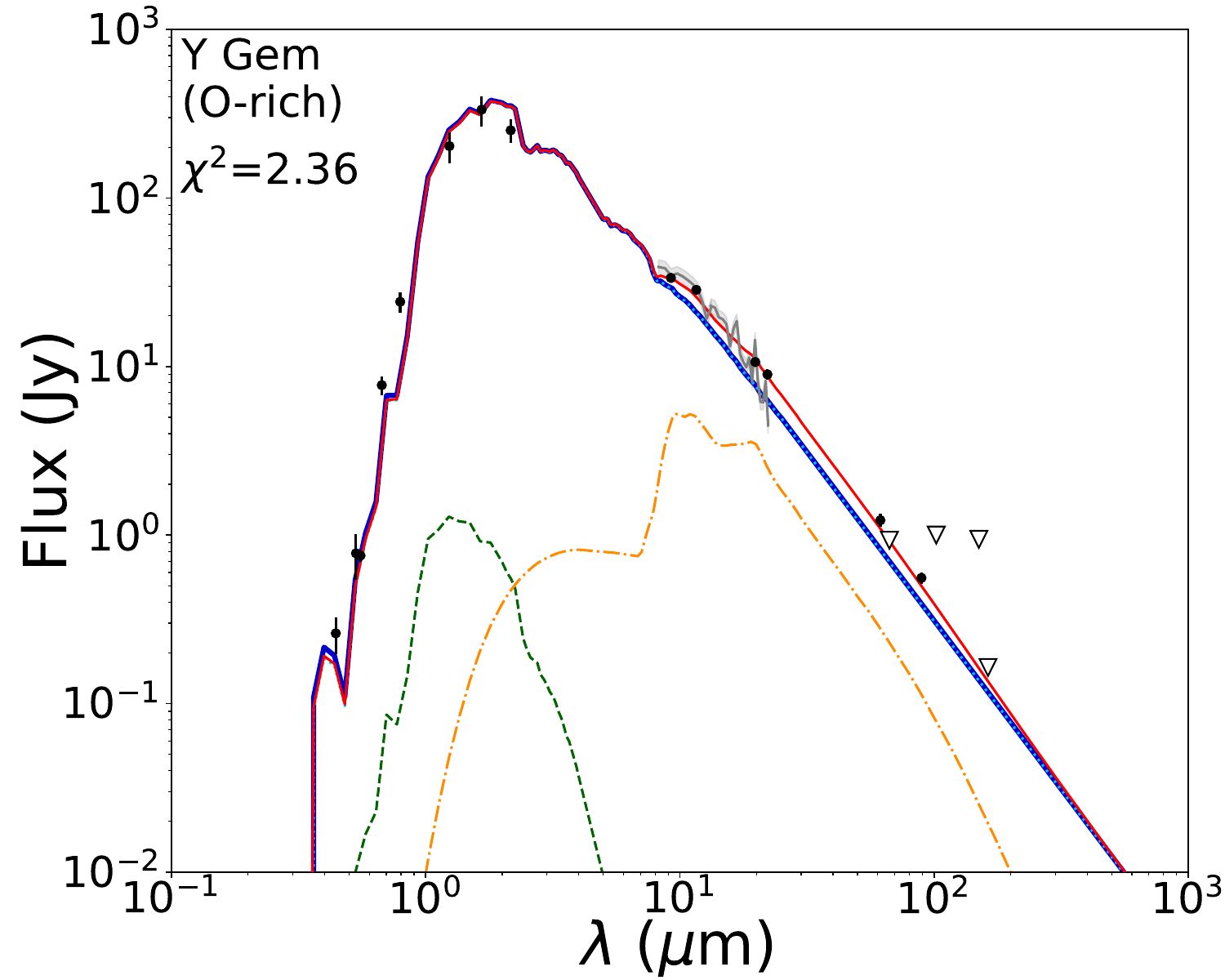}
     \end{subfigure}
     \begin{subfigure}[b]{0.245\linewidth}
         \centering
         \includegraphics[width=\linewidth]{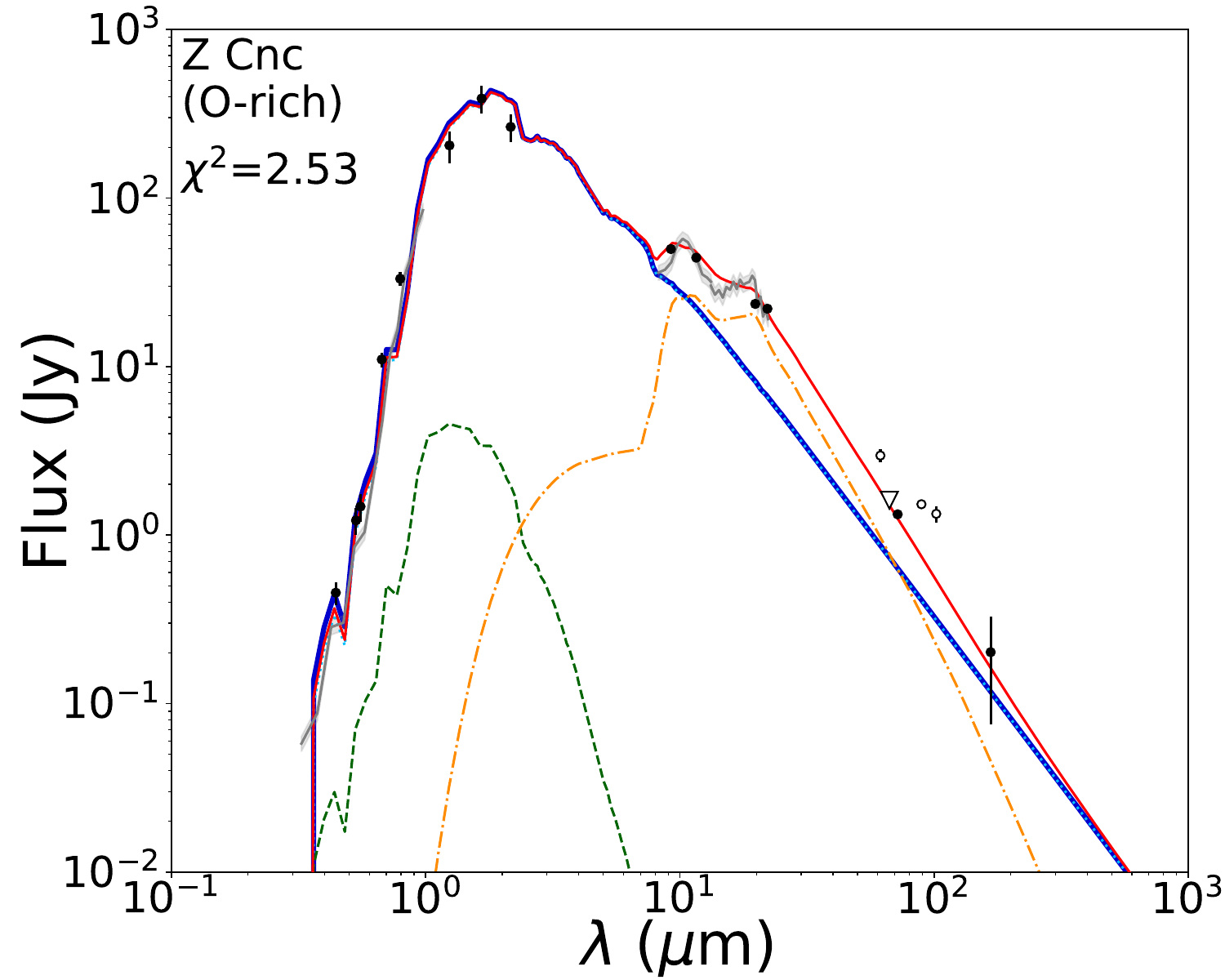}
     \end{subfigure}
        \caption{Same as Fig.~\ref{SED_fitting} for the rest of dusty AGB stars in the sample. In the case of VY\,UMa, the solid diamonds and the dashed magenta line represent respectively the fluxes and the DUSTY model for the detached shell.}
        \label{SED_fitting_2}
\end{figure*}

Fig.~\ref{SED_fitting_2} shows the SEDs and their best-fit DUSTY models for the rest of dusty AGB stars to complement the sources already shown in Fig.~\ref{SED_fitting}.

\section{Parameter uncertainties and correlations from the SED modelling}\label{correlations}

We present a detailed description of the procedure used to estimate the uncertainties of the free parameters used in the SED modelling and identify correlations between them. Single parameter uncertainties were estimated as 68\% confidence intervals from the normalised log-likelihood distribution  $ln(\mathcal{L}(x))$. We defined the $\chi^{2}$ likelihood function as:

\begin{equation}
    \mathcal{L}(x) \propto e^{-\chi^{2}(x)/2},
\end{equation}
\noindent

where $x$ are the free parameters. On the other hand, we explore degeneracies between pairs of parameters from their $\chi^{2}$ distributions. This method allowed us to identify which parameters are degenerated and visualise the degree to which these degeneracies affect the $\chi^{2}$ value.

\begin{figure*}[h!]
    \centering
    \includegraphics[width=0.9\linewidth]{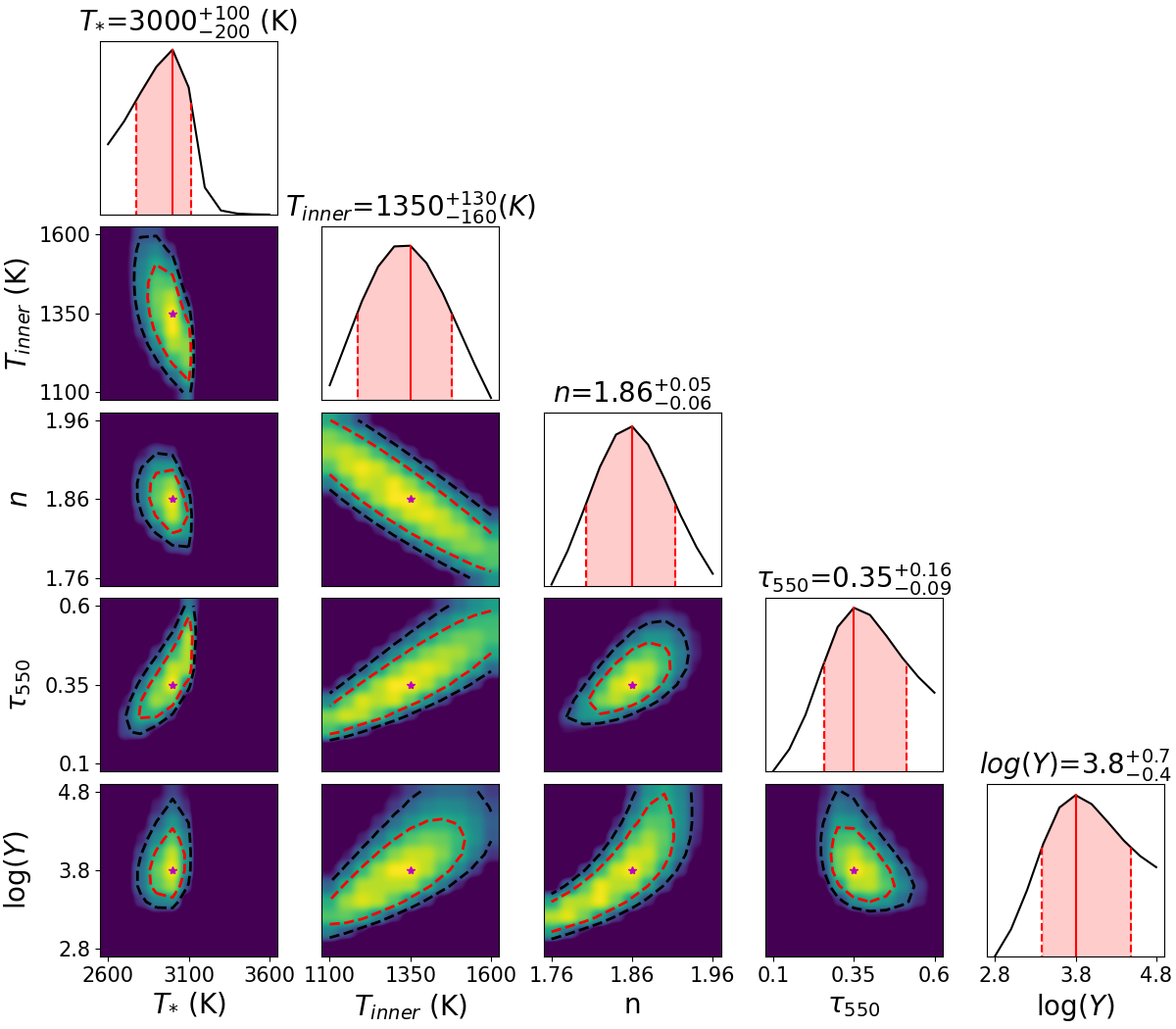}
    \caption{Corner plot of EY\,Hya to exemplify correlations between the different free parameters. The 1D curves display the log-likelihood functions, the solid red line indicates the most likely value and the dashed red lines and shadowed region the $\pm$1$\sigma$ values. The 2D histograms display the $\chi^{2}$ distributions, the dashed red and black lines indicate the areas with values lower than $\chi^{2}$+1 and $\chi^{2}$+2 respectively.}
    \label{fig:corner_plot}
\end{figure*}

Fig.~\ref{fig:corner_plot} shows the corner plot obtained for EY\,Hya as an example of the estimation of the uncertainties associated with the free parameters from the SED modelling and the identification of the correlations between them. We found the following sets of positive correlations: $T_{*}$--$\tau$, $T_{\rm inn}$--$\tau$, $T_{\rm inn}$--$Y$, $n$--$\tau$, and $n$--$Y$. On the other hand, we found the following set of anticorrelations: $T_{*}$--$T_{\rm inn}$, $T_{\rm inn}$--$n$, and $\tau$--$Y$. 

These correlations are quite linear in these ranges of the parameters, although there are two exceptions: (i) log(Y) that presents non-linearities with $n$ and $\tau_{550}$ and (ii) $T_{*}$ that presents sharp upper limits, this limit is related with an emission excess in optical wavelengths that cannot be compensated by the rest of parameters. Similar correlations were found for the rest of dusty sources and can be generalised for the modelling procedure.

\section{Comparison with NESS sample}\label{NESS_comparison}

We present a complementary comparison with the Nearby Evolved Stars Survey \citep[NESS,][]{Scicluna_2022}, which contains 485 nearby AGB stars. We first compared the CO($J$=2-1) and IRAS 60 \mum\, fluxes between our sample and NESS \citep[obtained from][]{Wallstrom_2025} to check whether uvAGBs have statistically lower CO intensity at similar IRAS 60 \mum\,as suggested by \cite{alonso-hernandez_2024} using the CO($J$=1-0).

\begin{figure*}[h!]
     \centering
     \begin{subfigure}[b]{0.45\linewidth}
         \centering
         \includegraphics[width=\linewidth]{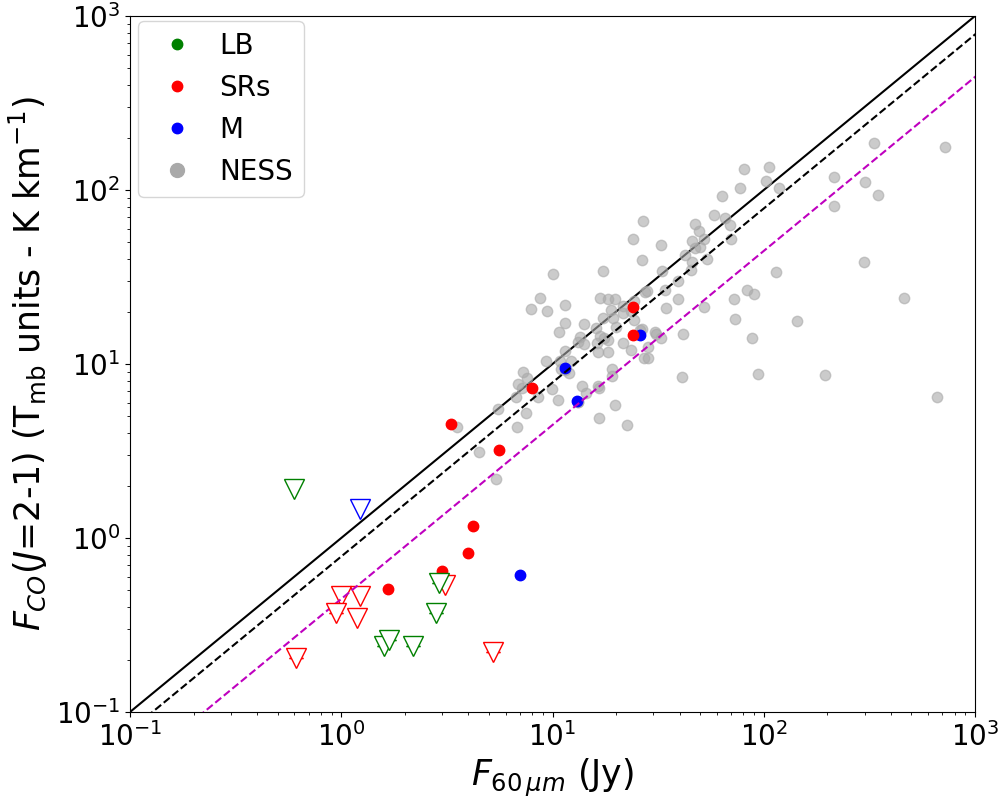}
     \end{subfigure}
     \begin{subfigure}[b]{0.45\linewidth}
         \centering
         \includegraphics[width=\linewidth]{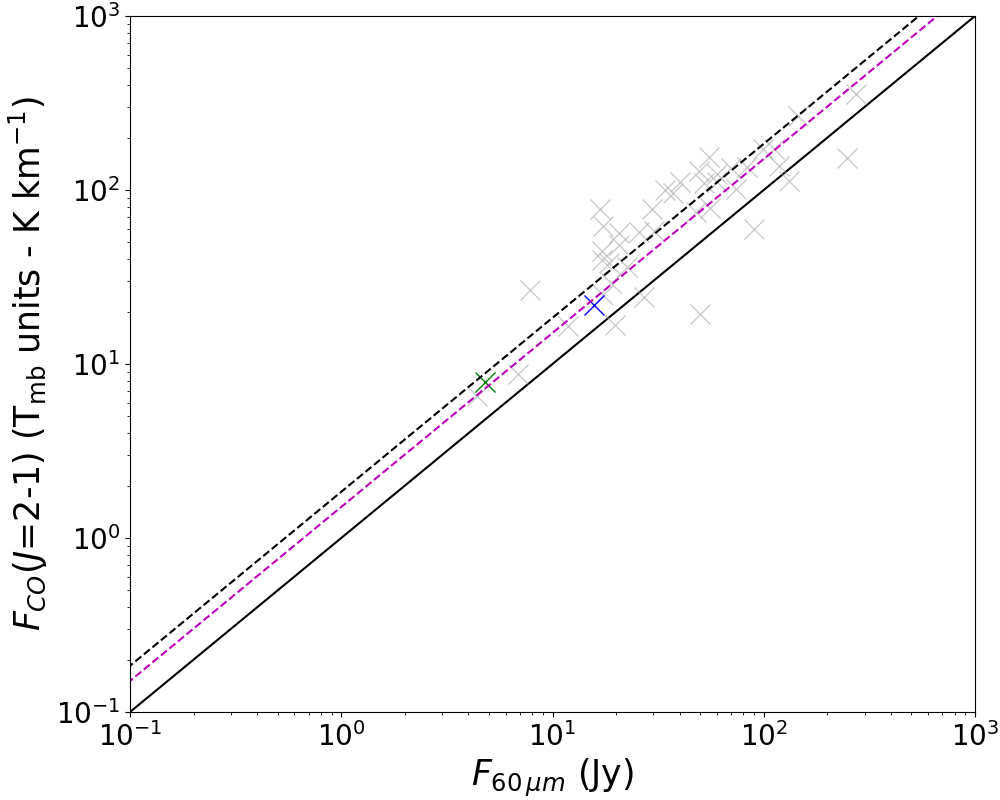}
     \end{subfigure}
        \caption{Comparison of CO($J$=2-1) with IRAS 60 \mum\, for O-rich and C-rich AGB stars ({\it left} and {\it right} respectively). The colour and shape of the markers represent the stellar variability and chemistry type as in Fig.~\ref{fig:HR_diagram}. Empty triangles are upper limits for the CO($J$=2-1) intensities. Grey markers represent NESS sources. The solid black lines indicates the equality relationship between CO($J$=2-1) intensity and IRAS 60 \mum\, fluxes. The dashed lines represents linear fits to the points, solid and purple for our sample and NESS respectively.}
        \label{fig:NESS_CO_vs_IRAS}
\end{figure*}

Fig.~\ref{fig:NESS_CO_vs_IRAS} shows the comparison between CO($J$=2-1) integrated intensity and IRAS 60 \mum\, flux. We used the CO($J$=2-1) intensities as they are available for both our sample and NESS. We scaled NESS CO($J$=2-1) intensities to match our intensities for the three commonly detected in CO in both samples (R\,LMi, SV\,Peg and W\,Peg).

In the case of unresolved and optically thin envelopes the CO intensity is proportional to the infrared flux (i.e. slope of 1 in logarithmic scale). Nevertheless, we noted that some scattering in this relationship is expected due to optically depth, partially resolved sources and ISM pollution. We found that our uvAGBs with the largest IRAS 60 \mum\, fluxes in our sample overlap with the trend described by NESS sources. However, our sources with lowest IRAS 60 \mum\, fluxes are systematically under this trend, although they fall outside the IRAS 60 \mum\, range covered by NESS. The resulting fits indicate that our sources have indeed lower CO($J$=2-1) intensities as well as CO($J$=1-0), as previously inferred by \cite{alonso-hernandez_2024}.

We also compared the correlations that we found in Fig.~\ref{fig:delta_vs_UV} for the gas-to-dust ratio with (i) the ratio between CO and IRAS 60 \mum\, and (ii) with $R_{\rm FUV/NUV}$. Our $\delta$ estimates are in the upper part of those covered by NESS, although this parameter depends in the methodology to estimate the mass-loss rates and it is sensitive to systematic effects. In the case of $R_{\rm FUV/NUV}$, we used the values without dust attenuation correction because we do not have the dust optical depths for the NESS sources.

\begin{figure*}[h!]
     \centering
     \begin{subfigure}[b]{0.45\linewidth}
         \centering
         \includegraphics[width=\linewidth]{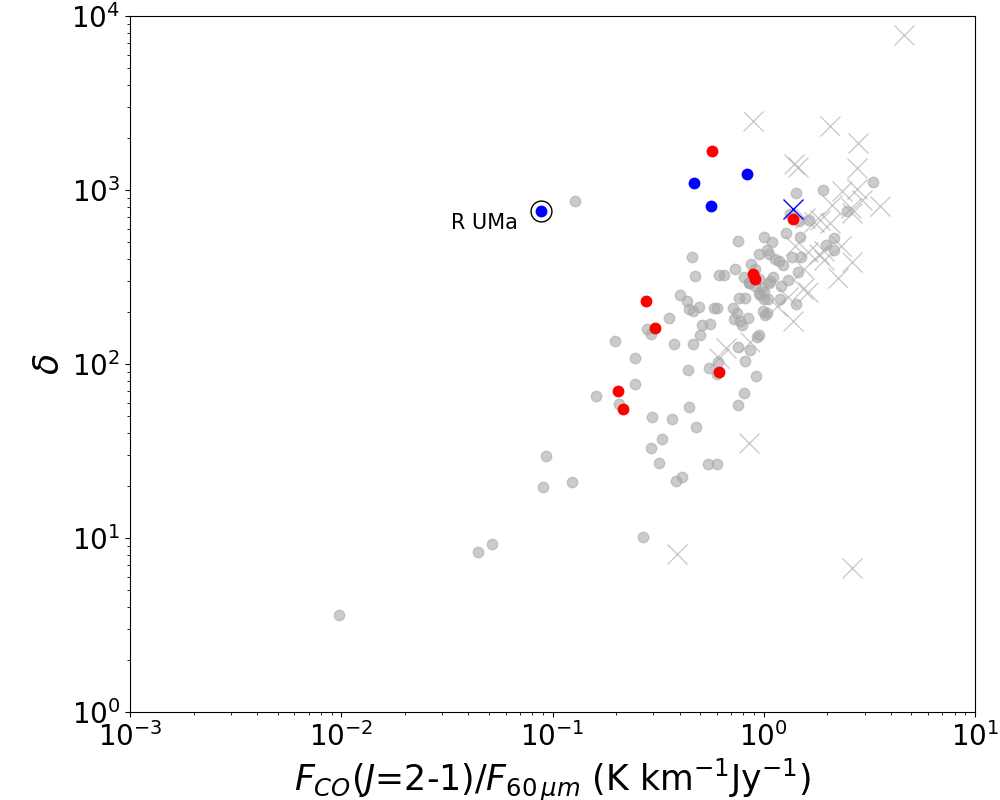}
     \end{subfigure}
     \begin{subfigure}[b]{0.45\linewidth}
         \centering
         \includegraphics[width=\linewidth]{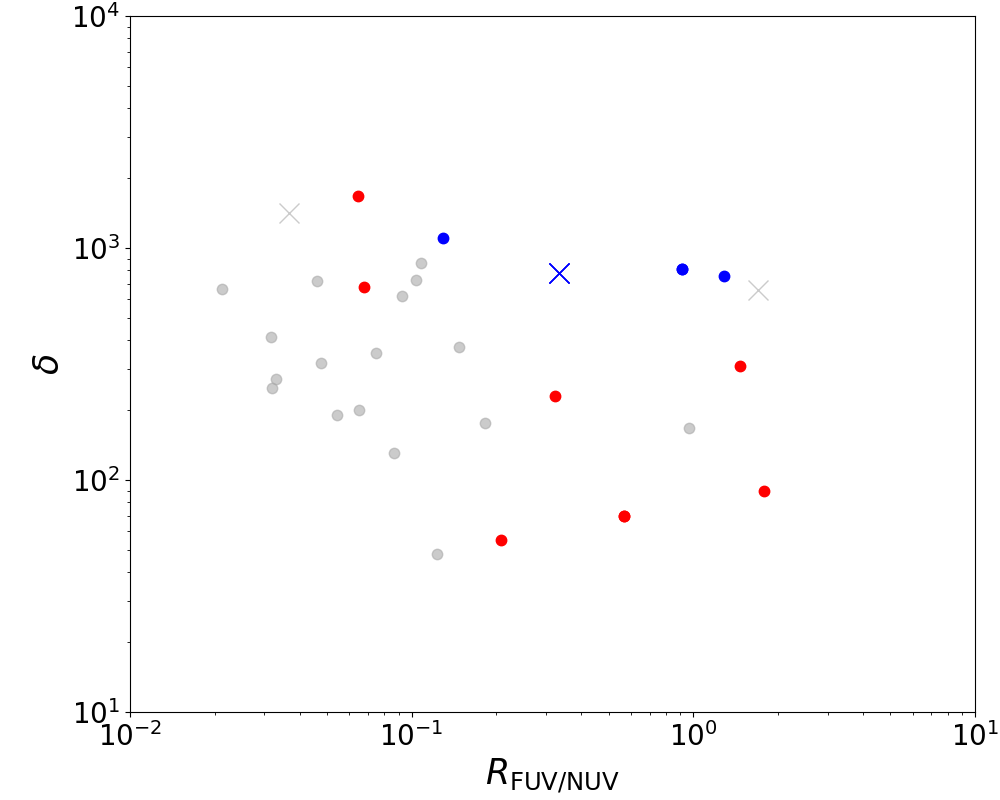}
     \end{subfigure}
        \caption{Comparison of the gas-to-dust ratio with the ratio between CO($J$=2-1) and IRAS 60 \mum\, and $R_{\rm FUV/NUV}$ ({\it left} and {\it right} respectively). The colour and shape of the markers represent the stellar variability and chemistry type as in Fig.~\ref{fig:HR_diagram}. Empty triangles are upper limits for the CO($J$=2-1) intensities. Grey markers represent NESS sources. The values of $R_{\rm FUV/NUV}$ shown in this figure were not corrected from dust attenuation. It can be appreciated a correlation and an anticorrelation in the figures on the {\it left} and {\it right} respectively.}
        \label{fig:NESS_delta}
\end{figure*}

Fig.~\ref{fig:NESS_delta} shows the comparison between the gas-to-dust ratio and the ratio between CO and IRAS 60 \mum\, and $R_{\rm FUV/NUV}$. It is shown that $\delta$ correlates with the ratio between CO and infrared emission, although some outliers (likely related with pollution in the IRAS 60 \mum\, images) are present. We noted that $\delta$ is not a empirical value and depends on the employed assumptions and methodology. Therefore, it is likely the presence of some systematic differences between our $\delta$ estimates and NESS. On the other hand, $\delta$ anticorrelates with $R_{\rm FUV/NUV}$, although it presents some scatter that can be related with UV emission variability and it is not corrected from dust attenuation.

Finally, we crossmatched the NESS catalogue with GALEX and found that among the 485 AGB stars in NESS, 163 were detected in the NUV band and 95 were also detected in the FUV band. These percentages illustrate that large AGB star samples also contain some uvAGBs.

\end{appendix}

\end{document}